\documentclass[iicol, pdflatex, sn-basic]{sn-jnl}

\usepackage{graphicx}%
\usepackage{multirow}%
\usepackage{amsmath,amssymb,amsfonts}%
\usepackage{amsthm}%
\usepackage{mathrsfs}%
\usepackage[title]{appendix}%
\usepackage{xcolor}%
\usepackage{textcomp}%
\usepackage{manyfoot}%
\usepackage{booktabs}%
\usepackage{algorithm}%
\usepackage{algorithmicx}%
\usepackage{algpseudocode}%
\usepackage{listings}%
\usepackage{threeparttable}
\usepackage{makecell}
\usepackage{siunitx}
\usepackage{subcaption}
\usepackage{comment}
\usepackage{lmodern}
\usepackage{hyperref, float}
\hypersetup{
    colorlinks,
    linkcolor={red!50!black},
    citecolor={blue!50!black},
    urlcolor={blue!80!black}
}

\begin{document}
\title[Rapid Rotation in Be stars]{Predicted observational effects of rapid rotation for Be stars}

\author*[1]{\fnm{Rina G.} \sur{Rast}}\email{krast@uwo.ca}

\author[1]{\fnm{Carol E.} \sur{Jones}}\email{cejones@uwo.ca}

\author[1]{\fnm{Mark W.} \sur{Suffak}}\email{msuffak@uwo.ca}

\author[2,3]{\fnm{Jonathan} \sur{Labadie-Bartz}}\email{jbartz@udel.edu}

\author[4]{\fnm{Asif} \sur{ud Doula}}\email{asif@psu.edu}

\author[5]{\fnm{Alex C.} \sur{Carciofi}}\email{carciofi@astro.iag.usp.br}

\author[1]{\fnm{Peter} \sur{Quigley}}\email{pquigley@uwo.ca}

\author[2]{\fnm{Coralie} \sur{Neiner}}\email{coralie.neiner@obspm.fr}

\author[6]{\fnm{Jeremy J.}\sur{Drake}}\email{jeremy.1.drake@lmco.com}

\affil*[1]{\orgdiv{Physics and Astronomy}, \orgname{The University of Western Ontario}, \orgaddress{\street{1151 Richmond Street}, \city{London}, \postcode{N6A 3K7}, \state{Ontario}, \country{Canada}}}

\affil[2]{\orgdiv{LIRA, Paris Observatory}, 
    \orgname{CNRS, PSL University, Universit\'e Paris Cité, Sorbonne University, CY Cergy University}, 
    \orgaddress{\street{5 place Jules Janssen}, \city{Meudon}, \postcode{92195}, \country{France}}}

\affil[3]{\orgdiv{DTU Space}, 
    \orgname{Technical University of Denmark}, 
    \orgaddress{\street{Elektrovej 327}, \city{Kgs., Lyngby}, \postcode{2800}, \country{Denmark}}} 

\affil[4]{\orgname{Penn State Scranton}, 
    \orgaddress{\street{120 Ridge View Drive}, \city{Dunmore}, \postcode{18512}, \state{PA}, \country{USA}}} 

\affil[5]{\orgdiv{Instituto de Astronomia, Geof\'ica e Ci\^encias Atmosf\'ericas}, \orgname{Universidade de S\~ao Paulo}, \orgaddress{\city{ S\~ao Paulo}, \postcode{94304}, \state{SP}, \country{Brazil}}} 

\affil[6]{\orgdiv{Advanced Technology Center}, \orgname{Lockheed Martin}, \orgaddress{\street{3251 Hanover St}, \city{Palo Alto}, \postcode{94304}, \state{CA}, \country{USA}}} 

\abstract{We conduct a systematic study on the effects of rapid rotation on predicted Be star observables. We use the three-dimensional Monte Carlo radiative transfer code, \textsc{hdust}, to model a comprehensive range of Be star subtypes at varying rotation rates. Using these models, we predict $V$ magnitude and photometric color, H$\alpha$ line profiles, and polarization at UV wavelengths as well as in the $V$-band for Be stars from B0 to B8. For each spectral subtype, we investigate the effects of disk density on the produced observables. We find that reddening and brightening effects of gravity darkening may cause rapidly-rotating stars to appear more evolved than they truly are. Rotational effects on the H$\alpha$ line profile shape may reduce line intensity for Be stars viewed at low inclinations and increase line intensity for those viewed at high inclinations. Additionally, rapid rotation can significantly impact the measured equivalent width of the line produced by a star with a moderate to high density disk, especially at high inclinations. When the star-disk system is viewed near edge-on, gravity darkening can result in stronger H$\alpha$ emission than would otherwise be expected for a disk of a given density. We also find that the competing effects of rapid rotation and H\,\textsc{i} opacity cause the slope of the polarized continuum (the polarization color) to be very sensitive to changes in the stellar rotation rate. This quantity offers a strong diagnostic for the rotation rate of Be stars.}


\maketitle

\section{Introduction}\label{sec1}

Be stars are often described using the definition by \citet{col87} which characterizes these objects as rapidly rotating, main sequence or slightly evolved stars that have exhibited hydrogen emission at some time. These stars represent roughly one in five B-type stars in the Galaxy \citep{bod20} and show variations in photometric, spectroscopic, polarimetric and interferometric observations. For a comprehensive review of Be stars, see \citet{riv13}. 

The spectral emission originates in a geometrically thin, disk-like distribution of gas orbiting the equatorial region of the Be star in Keplerian fashion. This disk is built and sustained by material ejected from the stellar surface. While the mass-loss mechanisms involved are not well understood, it is likely that rapid rotation and nonradial pulsation both play a role \citep{baa20, nei20, lab22}. Recent work has shown that the outflowing material may be launched from a localized azimuthal range along the stellar surface near the equator, initially orbiting near the star and eventually taking the shape of an axisymmetric disk \citep{car25, lab25}.

Rapid rotation can distort Be stars into oblate spheroids, since the stellar equatorial radius grows with rotation rate. This phenomenon results in a redistribution of the internal flux away from the equator and toward the poles, where the effective gravity is higher, and is referred to as gravity darkening. Often the von Zeipel Law \citep{von24}, which states that the radiative flux is proportional to the effective gravity, is used to approximate the variation in temperature between the stellar poles and equator. This law says that the stellar effective temperature, $T_{\rm{eff}}$, varies with the local effective gravity $g_{\rm{eff}}$ at a given latitude according to $T_{\rm{eff}} \propto g_{\rm{eff}}^{\beta}$, with the parameter $\beta$ set at $1/4$, independent of the stellar rotation rate. However, interferometry has shown that this law may over-estimate the temperature variation, particularly for the most rapidly rotating stars. More recently, a formalism developed by \citet{esp11} has been widely adopted for rapid rotators, where $\beta$ scales directly with the stellar rotation rate. 

Based on a study of He and Mg line profiles, \citet{tow04} argued that the gravity darkening effect is often overlooked and suggested the possibility that Be star rotation rates have been systematically underestimated. However, \citet{zor16} accounted for rotationally-induced gravity darkening in their critical analysis of the rotational velocities of a large sample of Be stars, and found that on average Be stars rotate no greater than $\sim$80\% of the critical rate. This suggests that either Be stars as a class rotate substantially sub-critically (prompting questions as to how material is able to achieve orbit in the form of a disk), or that Be stars do in fact rotate very close to critical and there remains some deficiency in the treatment of spectral synthesis at the highest rotation rates \citep{lab25}. 

Studying the effects of gravity darkening in rapidly rotating stars has had a long history. \citet{col91} modeled atmospheres of rotating B stars by expanding on the work of \citet{son77}, \citet{sle80} and \citet{sle92}. \citet{sle80} added metal line blanketing at ultraviolet (UV) wavelengths and the number of lines considered was increased. Photospheric models produced by \citet{col91} indicated that rotational effects are present in the spectra of B-type stars at UV wavelengths, despite the fact that the inclusion of metallic line blanketing results in a decrease of UV flux by several tenths of a magnitude. They show that rotational effects can increase the source function gradient and enhance the UV polarization that is produced by rotational distortion of the star itself. They also suggest the possibility that Be stars could be rotating critically. Furthermore, \citet{col91} demonstrated that an increase in the polarization produced by the oblate star is evident in the wings of narrow absorption lines and suggested that this effect could be even stronger in the UV. This would add complexity to interpreting observations correctly in rapid rotators.

While gravity darkening clearly has a major impact in the observables originating from the star, the disks around Be stars are also sensitive to the stellar gravity darkening. \citet{mcg13} studied the effects of gravity darkening on the thermal structure of Be star disks.
Not surprisingly, these authors found that the reduction in the inner disk temperature at the midplane as a result of increasing disk density was more significant than the decrease in disk temperature from moderate changes in rotation. However, they also found that increasing the rotation rate from 0 to 99\% of critical (defined below) can produce the same effect as changing the disk density by a factor of 2.5-7.5 times. \citet{mcg13} also demonstrated that the disk becomes less isothermal with increasing stellar rotation. In turn, this means that observations will be affected since disk temperature is directly related to the state of the gas and the  observations produced.

In this study, we systematically vary the rotation rates of Be stars to examine the rotational effects on observables for a full range of spectral subtypes. Specifically, we apply nonlocal thermodynamic equilibrium (NLTE) radiative transfer models to Be stars with low, moderate, and high density disks, and compare them to purely photospheric models with no disks. We explore a range of spectral subtypes from B0 to B8, producing synthetic observables for each system including $V$-band magnitudes and $B-V$ color, H$\alpha$ line profiles and equivalent widths (EWs), and UV and $V$-band polarization degree.

This is the first systematic parameter study dedicated to understanding the effects of rotation on predicted observations for Be stars with disks. We demonstrate that considering stellar rotation is a necessary ingredient to interpreting predictions correctly. We also discuss the value of having UV observations to infer stellar rotation rate. This paper is organized as follows: in Section~\ref{sec2} we describe our methods, in Section~\ref{sec3} we provide our results and finally we provide a discussion in Section~\ref{sec4} which highlights advances in our study that could be gained with UV spectropolarimetry. We summarize our findings in Section~\ref{sec5}.

\section{Methods}\label{sec2}

We use the 3D NLTE Monte Carlo radiative transfer code, \textsc{hdust} \citep{car06a,car08a}. This code generates a Be disk model based on stellar parameters such as mass, radius, luminosity, $T_{\rm{eff}}$, and rotation rate,  as well as disk properties such as radial extent, maximum density, and density distribution. Given these, the code simulates the propagation of photons through the disk to create a 3D map of ionization levels and temperature. \textsc{hdust} uses this information to predict observables such as the spectral energy distribution, emission line profiles, and intrinsic polarization levels. It has been used to model individual Be stars and predict their observables at specific moments in the system's evolution (for example, see \citealt{suf20, mar22}). \textsc{hdust} has also been used by \citet{suf23, suf24} to predict the properties and observables of Be disks with complex or asymmetrical geometries.

\begin{table*}[ht] 
\caption{Stellar and disk parameters used in the \textsc{hdust} models, taken from Figure 8(a) of \citet{vie17} and roughly consistent with those chosen by \citet{suf23}. Note that the $\rho_0$ values are for $W=0$ and were scaled with the rotation rate, as described in the text.}
\label{tab:parameter_summary}
    \centering
    \begin{threeparttable}
    \begin{tabular}{c c c c c c}
    \hline
    Spectral Type & $M$ (M$_{\odot}$) & $R_\mathrm{p}$ (R$_{\odot}$) & $T_\mathrm{eff}$ (K) & $L$ (L$_{\odot}$) & $\rho_0$ (\SI{e-11}{\gram\per\cubic\cm}) \\ [0.5ex] 
    \hline
    B0 & 17.5 & 7.4  & 30,000 & 39,740 & 0.1/1/10 \\
    B2 & 9.11 & 5.33 & 21,000 & 4,950 & 0.1/1/10 \\ 
    B5 & 5.9  & 3.9  & 15,000 & 690    & 0.05/0.5/5 \\ 
    B8 & 3.8  & 3.0  & 12,000 & 167    & 0.01/0.05/0.5  \\ 
    \hline
    Parameter &  \multicolumn{5}{c}{Range} \\ [0.5ex] 
    \hline
    $W$ & \multicolumn{5}{c}{0.1, 0.5, 0.6, 0.7, 0.75, 0.8, 0.9, 0.95, 0.99} \\
    $i$ & \multicolumn{5}{c}{5$^{\circ}$, 25$^{\circ}$, 45$^{\circ}$, 70$^{\circ}$, 85$^{\circ}$, 89$^{\circ}$} \\
    \hline
    \end{tabular}
\end{threeparttable}
\end{table*}

To test the effects of gravity darkening over a range of stellar properties, we created a computational grid with spectral types from B0 to B8. This grid is shown in \autoref{tab:parameter_summary}. The stellar properties for each spectral subtype are taken from \citet{cox00} and \citet{sil10} and are consistent with those used in \citet{suf23}. For each spectral subtype, we set the initial assumed disk temperature to 60\% of the stellar $T_{\rm{eff}}$ \citep{mil99a, mil99b, car06a}, and the code determines a self-consistent solution for the temperature structure of the disk assuming radiative equilibrium. The energy balance is dominated by radiative processes; while \citet{anu25b} showed that shear heating can become significant in Be stars of type A2 and later, it is negligible for earlier-type Be stars and therefore would not affect the stars modeled in this study \citep{anu25a}. The disk is modeled as azimuthally symmetric, with its density distribution following the equation
\begin{equation}
    \rho(r,z) = \rho_{0} \left(\frac{r}{R_{\rm{eq}}}\right)^{-n} \exp \left(- \frac{z^2}{2H^2}\right)\,,
    \label{eq:power_law}
\end{equation}
where $r$ and $z$ describe the radial and vertical positions in the disk, respectively, $R_{\rm{eq}}$ represents the equatorial stellar radius, $\rho_0$ is the density where $r$ = $R_{\rm{eq}}$ and $z = 0$, $H$ is the scale height, and $n$ is a constant that governs how quickly the density drops off with increasing $r$ \citep{bjo97, bjo05}. The scale height is given by 
\begin{equation}
    H(r) = \frac{c_s R_{\rm{eq}}}{v_{\rm{orb}}} \left( \frac{r}{R_{\rm{eq}}} \right)^{3/2} \\,
\end{equation}
where $c_s$ is the sound speed in the disk and $v_{\rm{orb}} = \left( GM_\star / R_{\rm{eq}} \right)^{1/2}$ is the Keplerian circular orbital velocity at the stellar equator \citep{car06a}. We assume a steady state disk with $n=3.5$ \citep{lee91, por99} and vary $\rho_0$ in Eq. \ref{eq:power_law} to test low, moderate, and high disk densities. Since the range of densities seen in Be disks varies with spectral subtype, we chose values for $\rho_0$ based on Figure 8(a) in \citet{vie17} which, for each stellar $T_{\rm{eff}}$, are roughly consistent with those chosen by \citet{suf23} and \citet{rub23}. 

We model the Be stars as rigid rotators and for each spectral type and density, vary the stellar rotation rate $W$, defined by 
\begin{equation}
    W =  \frac{v_{\rm{rot}}}{v_{\rm{crit}}} 
\end{equation}
where $v_{\rm{rot}}$ is the equatorial rotational velocity of the star and $v_{\rm{crit}}=v_{\rm{orb}}$ is defined above \citep{riv13}. Here, critical rotation would correspond to $W=1$. In a sample of 233 Be stars, \citet{zor16} found rotation rates from 30\% to 95\% of the critical linear velocity at the equator. Here, we test values of $W$ from 0.1 to 0.99, as shown in \autoref{tab:parameter_summary}. The lower limit of 0.1, which is rare for B-type stars in general \citep{hua10} and well below the lower limit for Be stars suggested by \citet{zor16}, provides a basis for comparison against larger rotation rates. We use the von Zeipel law for gravity darkening, with the gravity darkening exponent constant at $\beta = 0.25$ \citep{von24}. In total, our final sample shown in Table~\ref{tab:parameter_summary} included 108 models. 

We hold the disk radius constant at 25 $R_{\rm{eq}}$ for all simulations. The H$\alpha$ line is formed within the first $\sim10-20$ $R_{\star}$ of the Be disk \citep{car11}, depending on disk density and stellar properties, while continuum emission at visual wavelengths is formed within a smaller area very close to the stellar surface \citep{car06a, hau12}, and the polarization is contained within 10 $R_{\star}$ \citep{car11}. Therefore, extending the tested disk radius beyond this point would not affect the dependence of each observable we tested on the stellar rotation rate. Since $W$ can be related to the equatorial radius of a Be star according to the relation \citep{rim18},
\begin{equation}
    W = \sqrt{2\left( \frac{R_{\rm{eq}}}{R_{\rm{pole}}} -1 \right)}\,,
\label{eq:radius_ratio}
\end{equation}
the equatorial radius of the star scales as $(W^2/2) +1$. As a result, the radial extent of the disk ($R_{\rm{disk}}$) changes significantly across the parameter space investigated in this work. For example, a star rotating at $W=0.99$ was modeled with a disk that extended nearly 1.5 times further than a star rotating at $W=0.1$. Since the models used the same number of gridcells regardless of the rotation rate, the mass contained within a cell also increased significantly with rotation rate, if no adjustments were made to correct the density or volume of the cell. This directly affects the observables which depend on the mass contained within the emitting region. In order to better constrain the mass within a cell, we adjusted the base density $\rho_0$ listed in Table~\ref{tab:parameter_summary} by the same factor that the equatorial radius increased. While this method did not perfectly account for the increase in cell volume with $W$, and the disk mass within a given radius (in R$_{\odot}$) increased by a factor of about seven times between $W=0.1$ and $W=0.99$, this was half the mass increase that would occur if no density adjustments were applied. 

We find the H$\alpha$ EW, the average polarization degree across the $V$-band, and the photometric magnitude in the $V$-band for all models. We also compare the H$\alpha$ emission profile for all models to the photospheric H$\alpha$ absorption profile. We selected key inclinations $i$ from 5$^{\circ}$ to 89$^{\circ}$ in order to test a full range of possible viewing angles, also shown in Table~\ref{tab:parameter_summary}. Our models therefore represent a comprehensive set of Be stars at different spectral subtypes, disk densities, and inclinations. 

\section{Results}\label{sec3}

We present our findings organized by each predicted observable. We show the results for the predicted $V$-band magnitude and color in Section~\ref{sec:Vband_results}, our results for the H$\alpha$ EW in Section~\ref{sec:ha_ew_results} and those for the $V$-band polarization in Section~\ref{sec:pol_results}.

\subsection{Photometry}\label{sec:Vband_results}

\begin{figure*}[!ht]
\centering
   \begin{subfigure}{0.49\textwidth}
        \includegraphics[width=\textwidth]{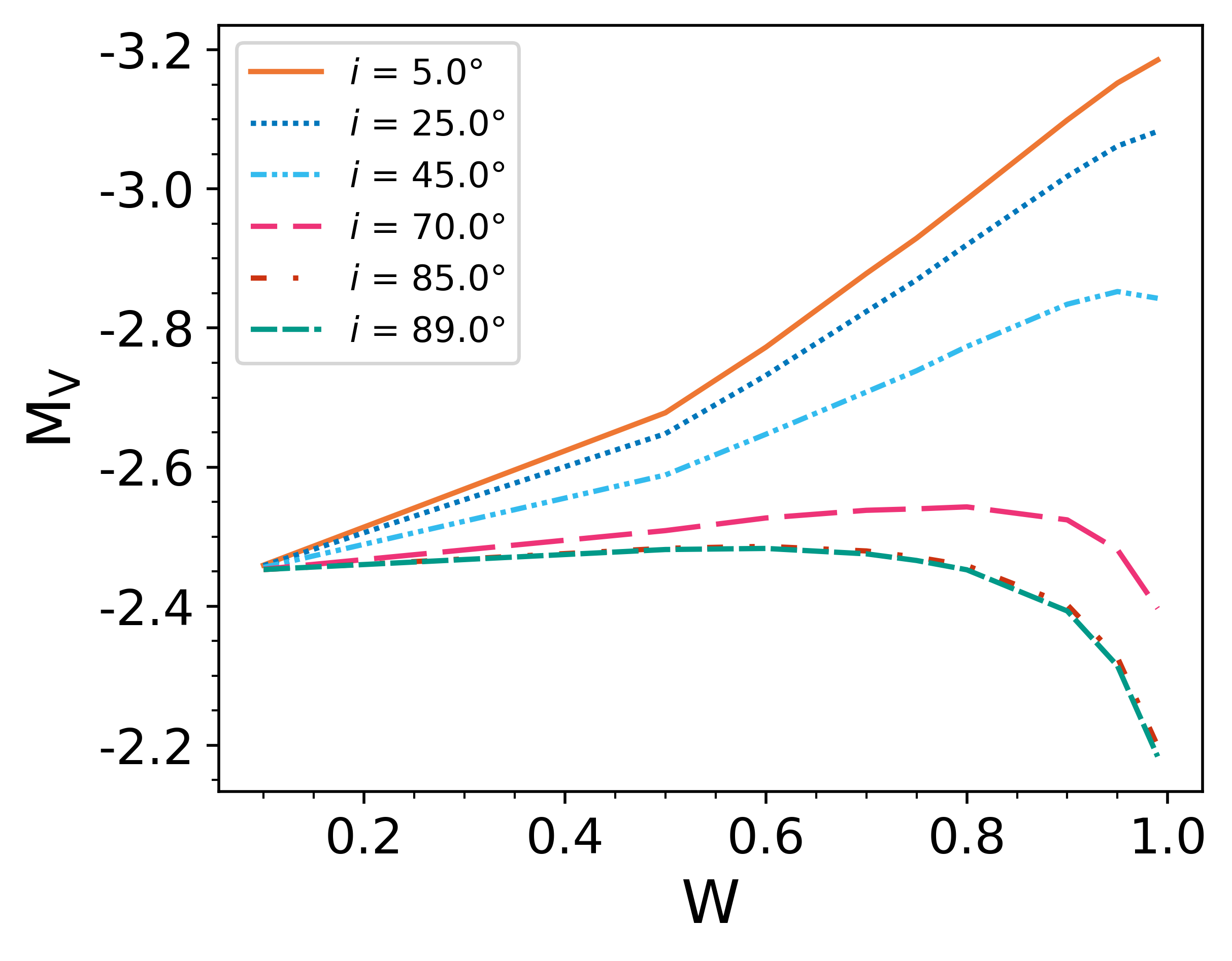}
    \end{subfigure}
    \begin{subfigure}{0.49\textwidth}
        \includegraphics[width=\textwidth]{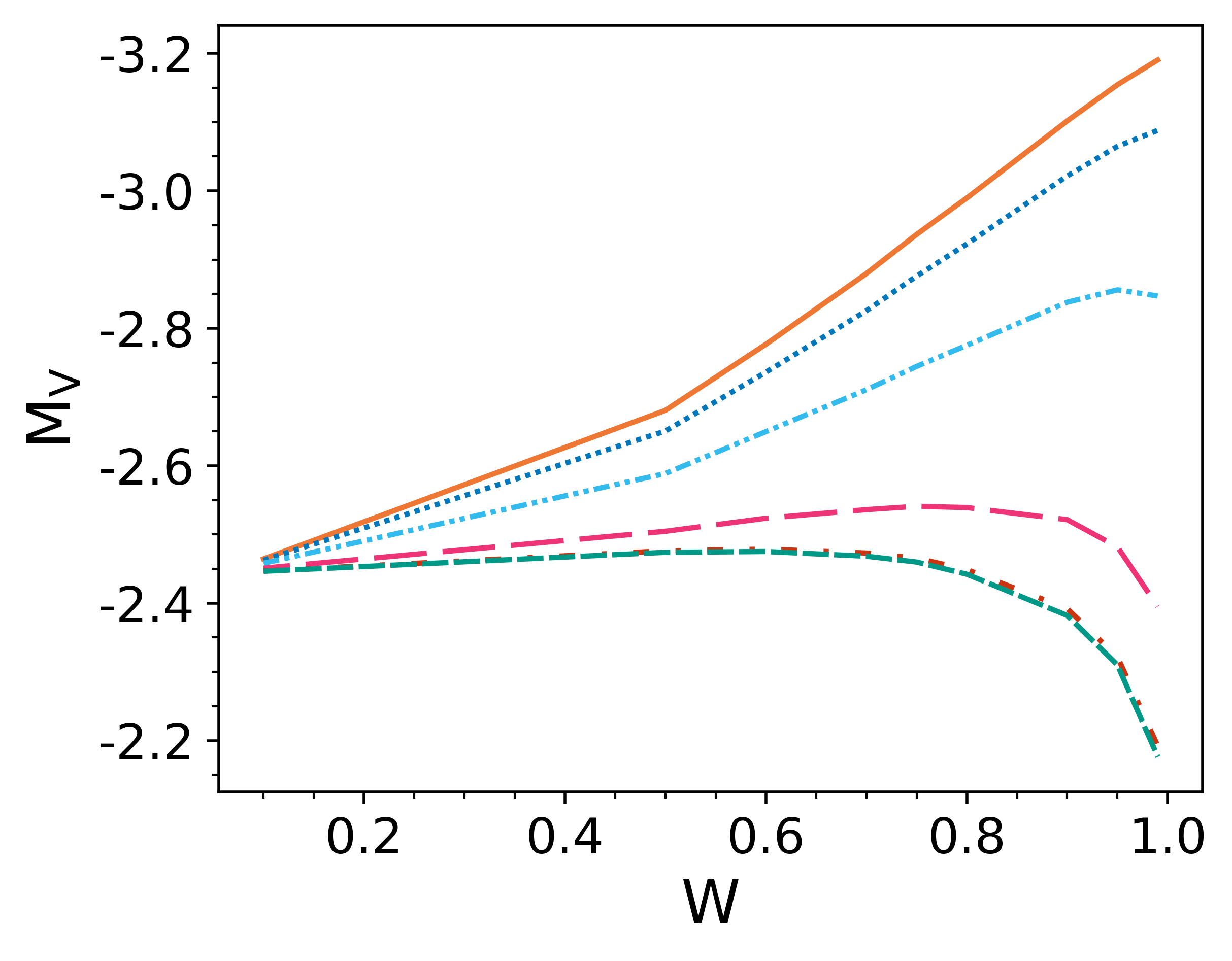}
    \end{subfigure}
    \begin{subfigure}{0.49\textwidth}
        \includegraphics[width=\textwidth]{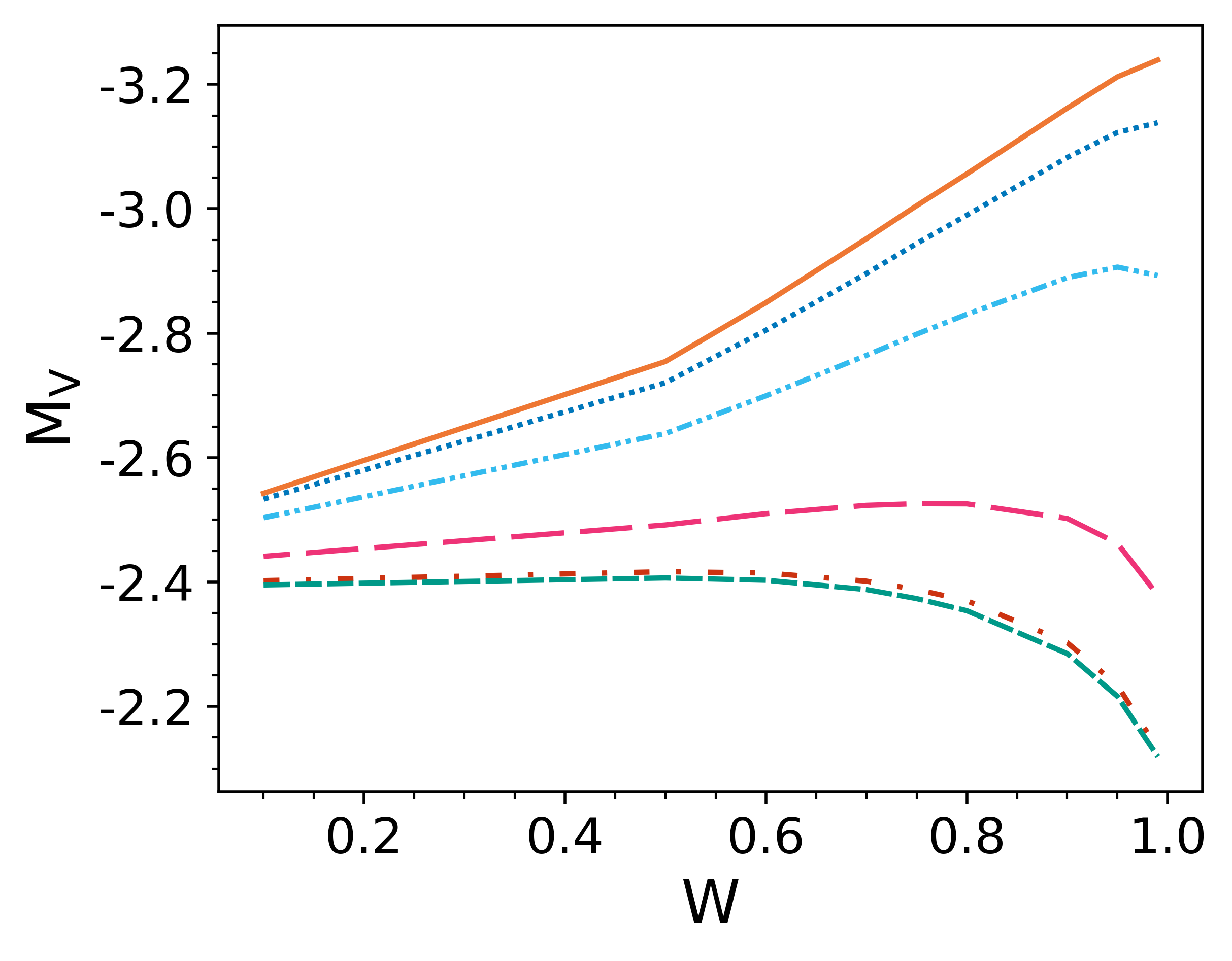}
    \end{subfigure}
    \begin{subfigure}{0.49\textwidth}
        \includegraphics[width=\textwidth]{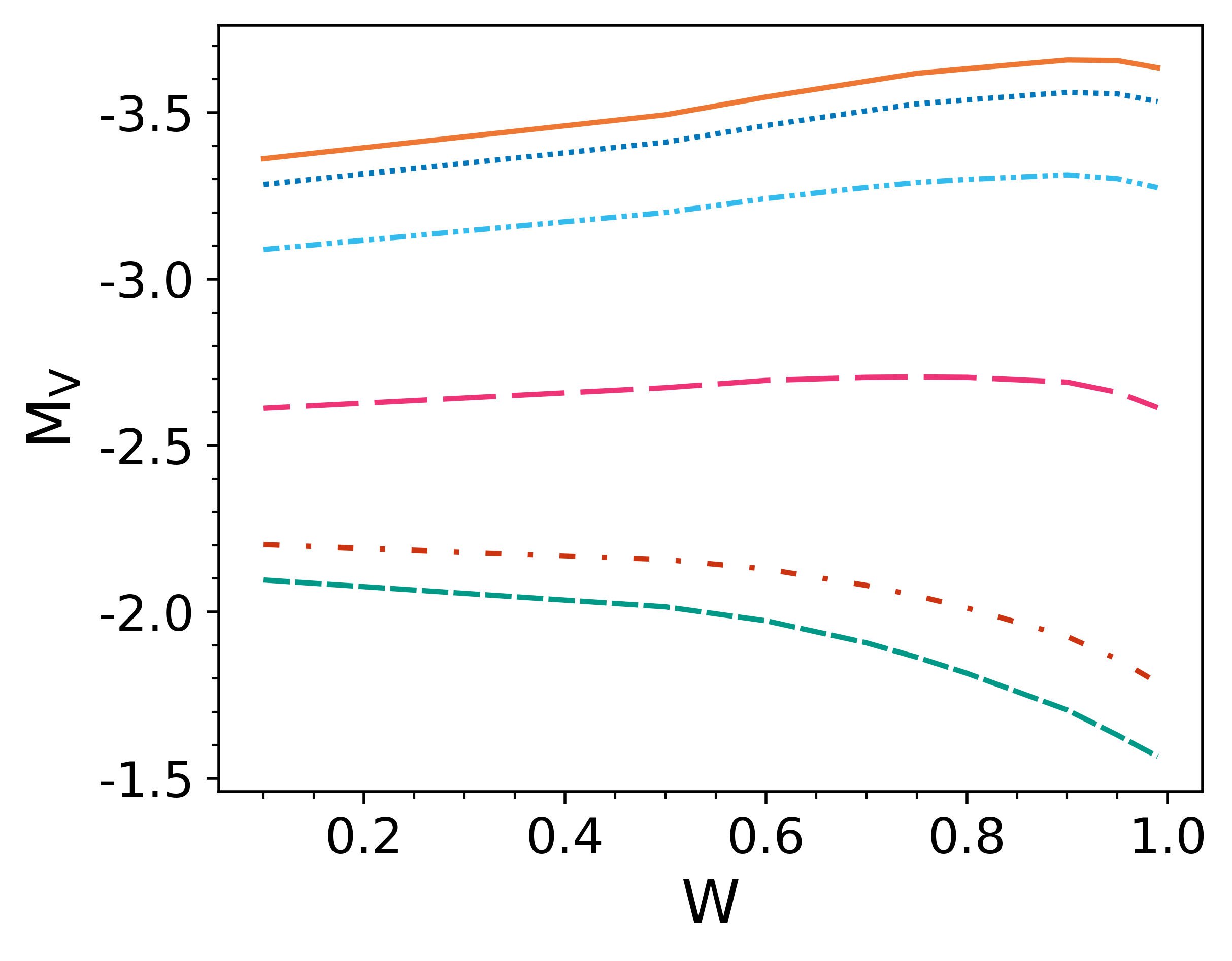}
    \end{subfigure}
\caption{Absolute $V$-band magnitude trends with increasing $W$ for B2 models with no disk (top left), low density disk (top right), moderate density disk (bottom left) and high density disk (bottom right). The inclination angles are indicated in the legend in the top left panel.}
    \label{fig:b2_vmag}
\end{figure*}

\begin{figure*}[!ht]
\centering
   \begin{subfigure}{0.49\textwidth}
        \includegraphics[width=\textwidth]{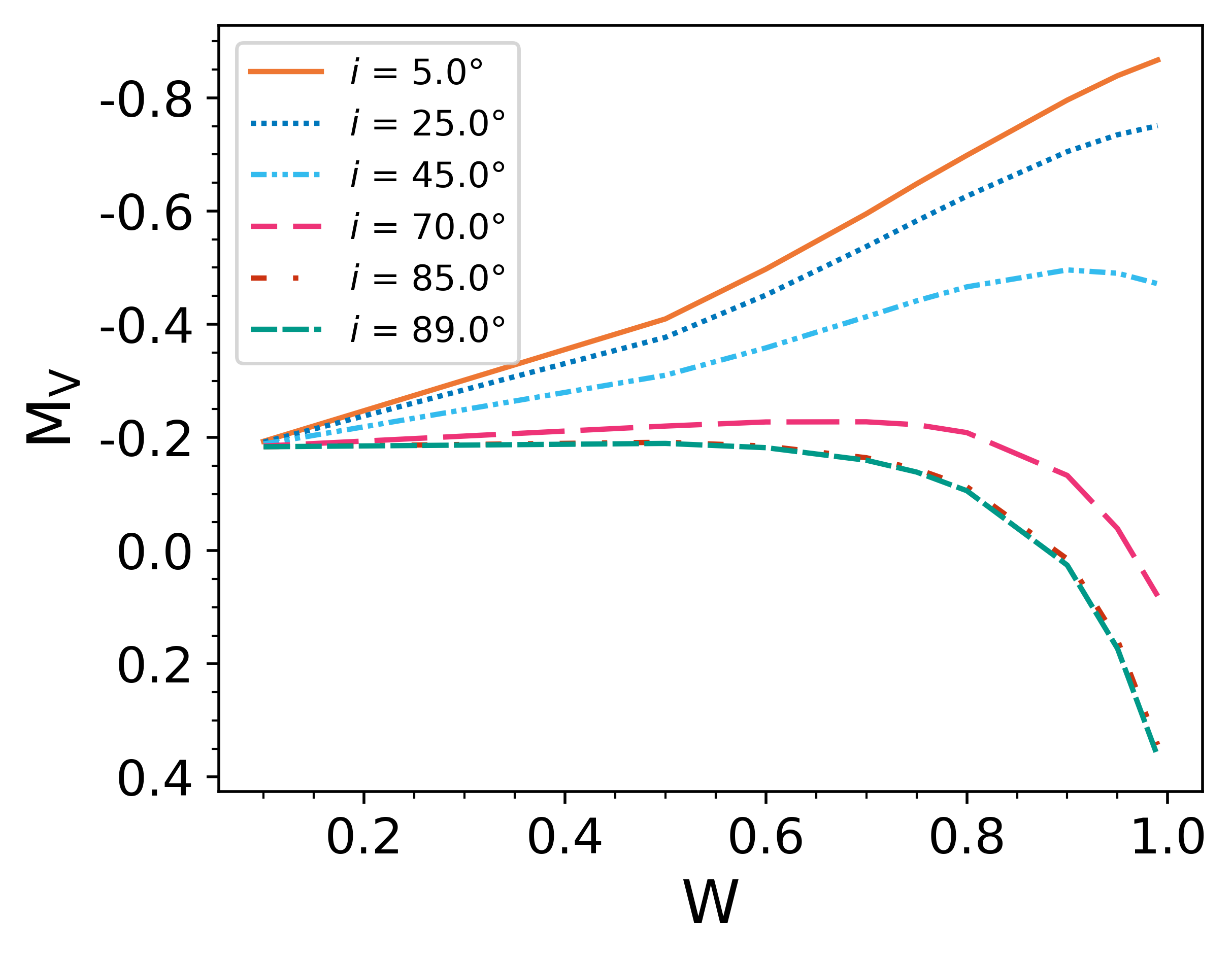}
    \end{subfigure}
    \begin{subfigure}{0.49\textwidth}
        \includegraphics[width=\textwidth]{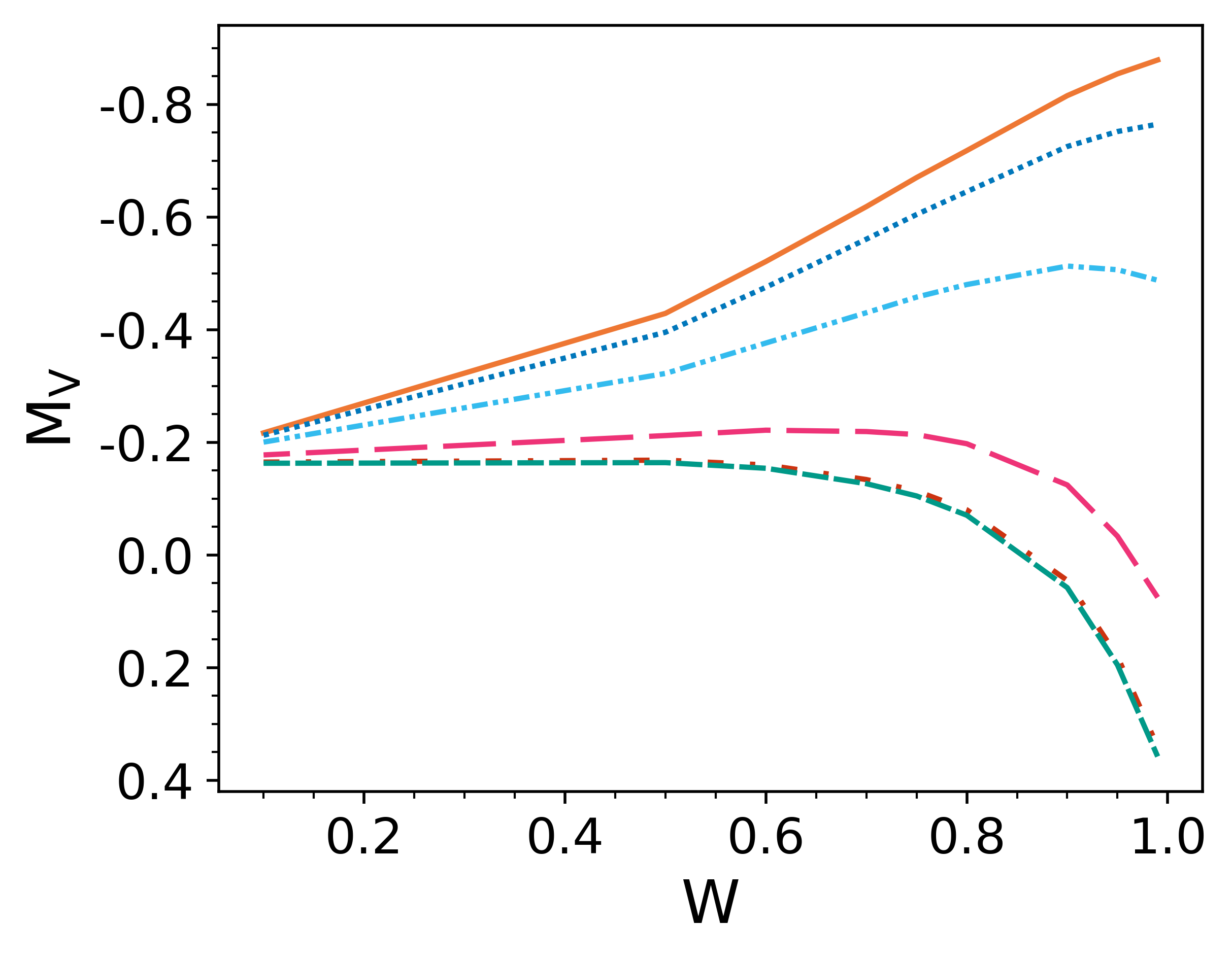}
    \end{subfigure}
\caption{Absolute $V$-band magnitude for the B8 diskless (left) and high density disk (right) models. The inclination angles are indicated in the legend in the left panel.}
    \label{fig:b8_vmag}
\end{figure*}

We computed the predicted absolute $B$ and $V$ magnitude as well as the $B-V$ index over the full range of tested rotation rates and spectral subtypes. Figure~\ref{fig:b2_vmag} shows the $V$-band magnitude as a function of $W$ over a range of inclination angles for all B2 models. It is representative of the B0, B2 and B5 models, which all show the same trends with changing rotation rate. The figures for the B0 and B5 models are included in Appendix~\ref{secA1}. For these subtypes, low density disks have no significant impact on the brightness of the system. Both diskless stars and those with low density disks show that the effect of increasing the rotation rate (overall brightening or dimming) depends on the inclination angle, $i$, except for the smallest rotation rate. Stars viewed at low or moderate inclinations (i.e., closer to pole-on) brighten as the rotation rate increases, primarily due to an increase in polar temperatures and the larger projected stellar area. Meanwhile, stars viewed at high inclinations (i.e., more equator-on) remain roughly the same brightness for rotation rates lower than $W=0.8$, but dim by several tenths of a magnitude at the highest rates as $W$ approaches critical. 

For moderate and high density disks, varying inclination produces different magnitude values even for the lowest tested rotation rate. The high density disk models show the largest departure from the diskless trends. When the systems with the densest disks are viewed near face-on (small inclinations), the rotational effects on brightness are less significant than for the diskless models. By contrast, when the same systems are viewed near edge-on (large inclinations) and the rotation rate is increased, the system dims more substantially than when there is no disk. 

The trends for the coolest subtype in our models, B8, are distinct from the earlier subtypes. These models are characterized by a lack of dependence on disk density, and even the systems with the densest disks are nearly indistinguishable from the stars with no disks. This is illustrated by the comparison of the diskless model and the high density model in Figure~\ref{fig:b8_vmag} (the remaining B8 models are shown in Appendix~\ref{secA1}). The disk for the B8 models has no significant contribution to the brightness of the system at any rotation rate, and the star behaves as if it had no disk at all. 

\begin{figure*}[!ht]
\centering
    \begin{subfigure}{0.49\textwidth}
        \includegraphics[width=\textwidth]{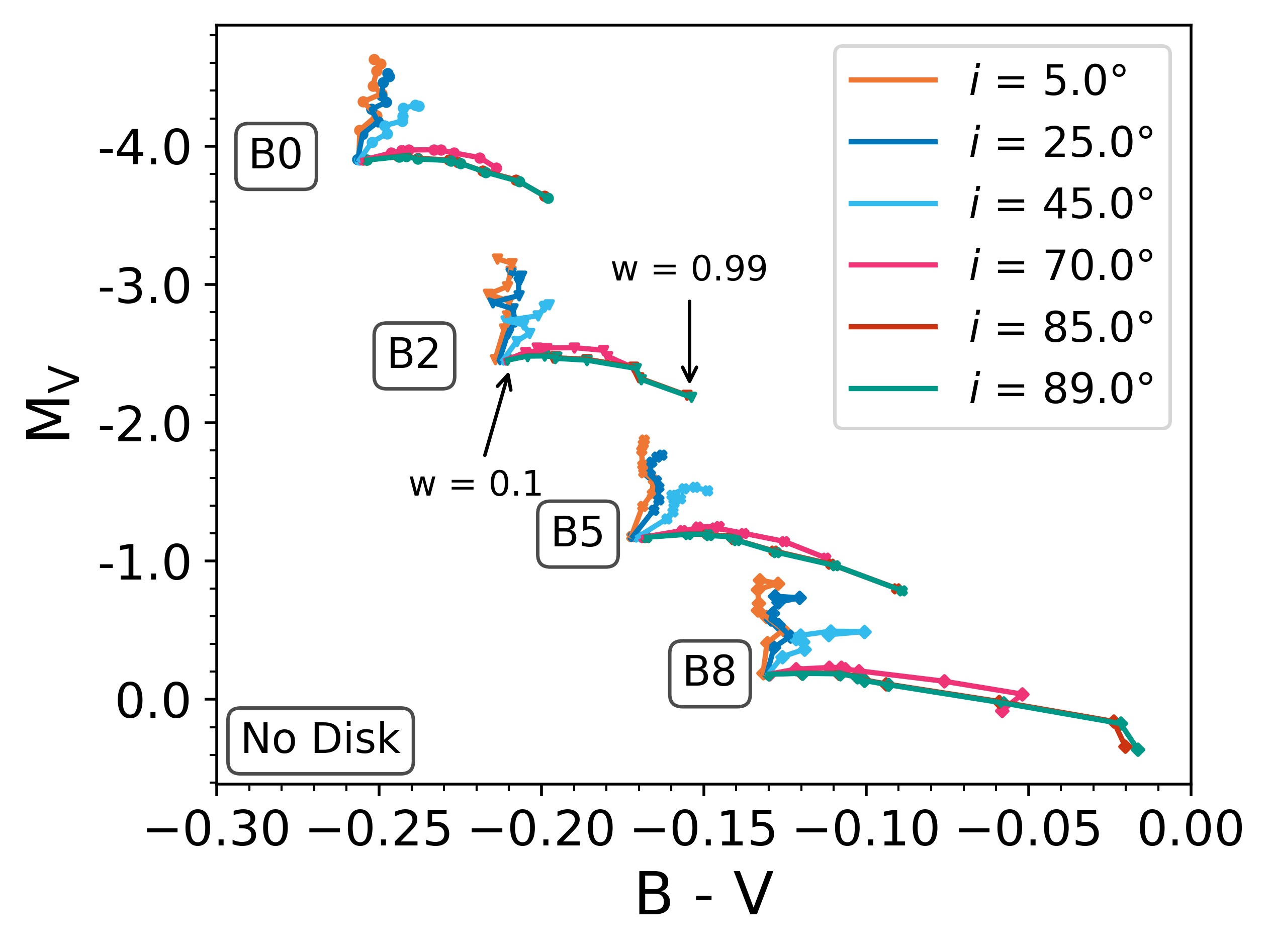}
    \end{subfigure}
    \begin{subfigure}{0.49\textwidth}
        \includegraphics[width=\textwidth]{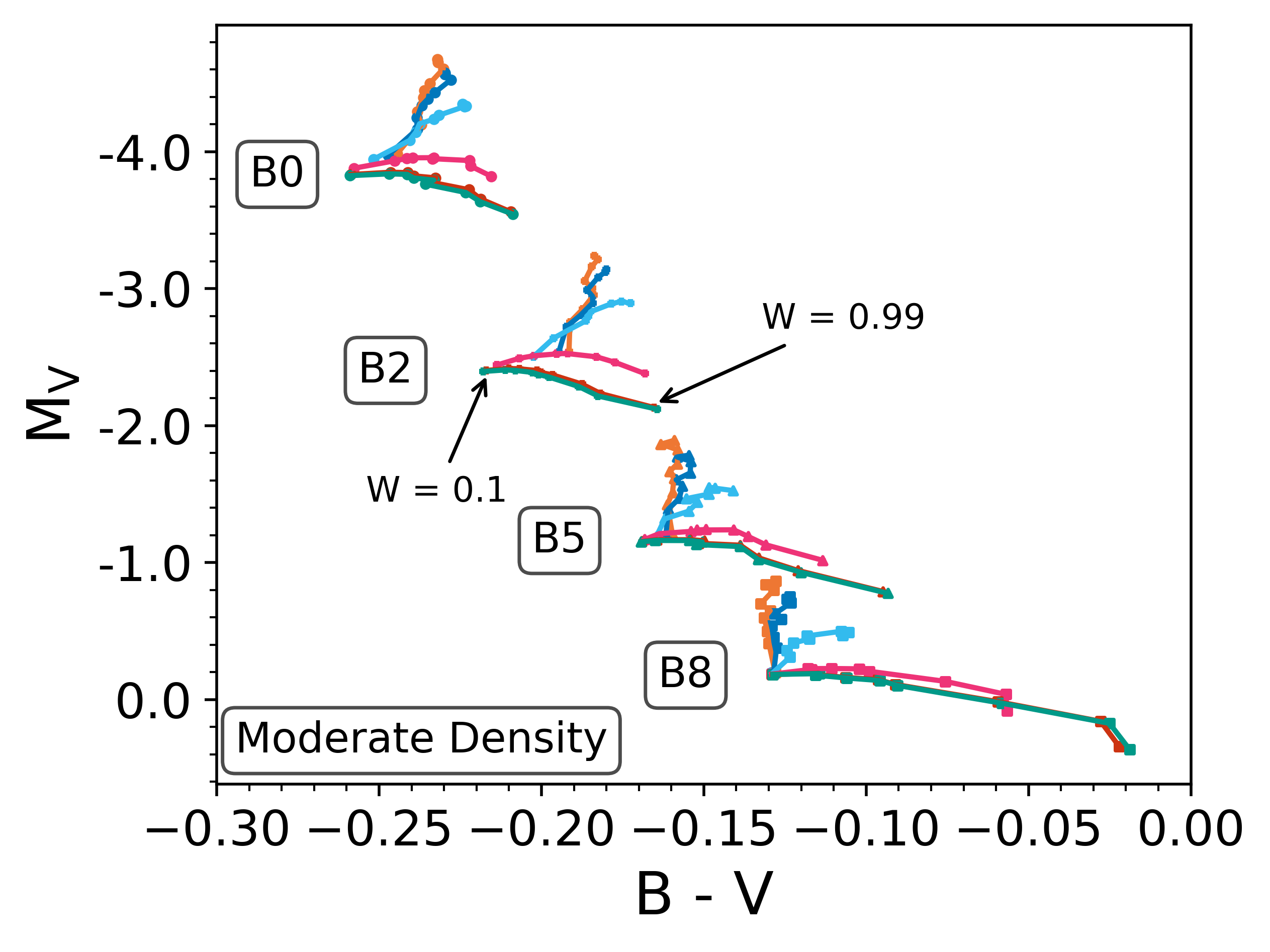}
    \end{subfigure}
    \begin{subfigure}{0.49\textwidth}
        \includegraphics[width=\textwidth]{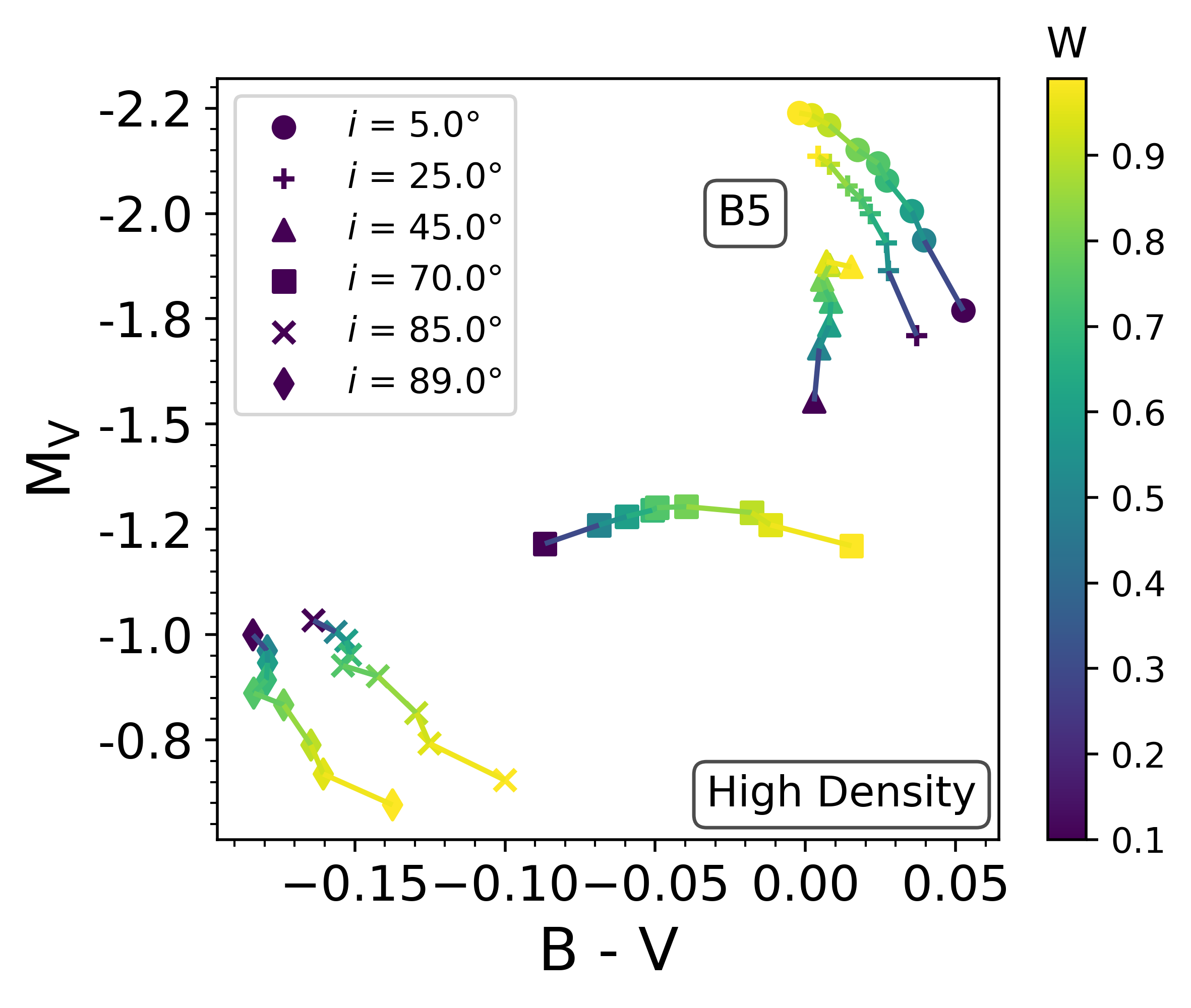}
    \end{subfigure}
    \begin{subfigure}{0.49\textwidth}
        \includegraphics[width=\textwidth]{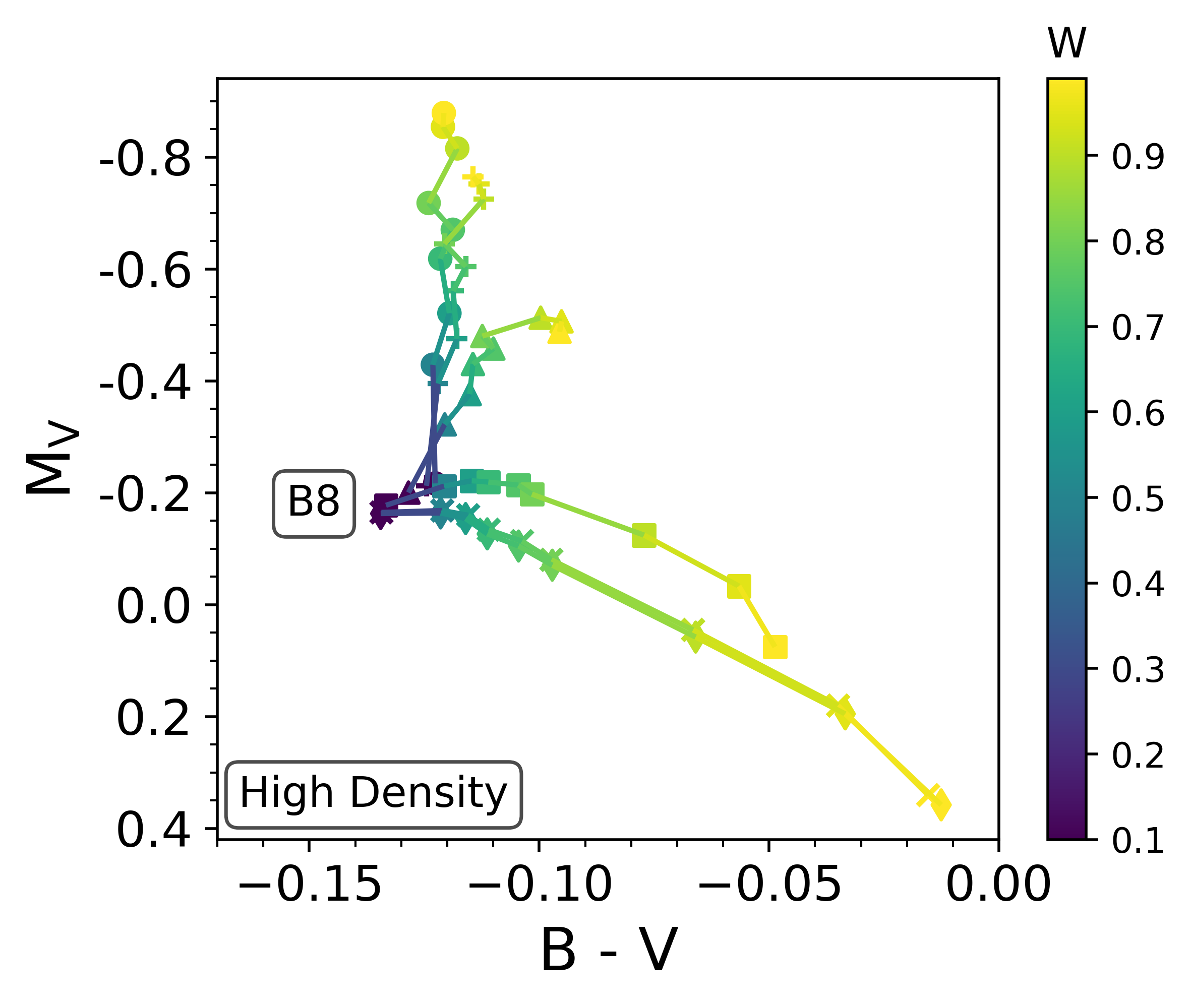}
    \end{subfigure}
\caption{Top: color-magnitude diagrams showing rotational displacement fans for the diskless models (left) in addition to the moderate density models (right). The legend in the left panel indicates the color for each tested inclination angle. Bottom: high density disks for B5 (left) and B8 (right) stars. Different inclination angles are denoted by different markers, shown in the legend in the left panel, and the rotation rate is indicated through the colorbar.}
\label{fig:colormag_fans}
\end{figure*}

Figure~\ref{fig:colormag_fans} shows $B-V$ color-magnitude diagrams for our models. We show the diskless models as well as moderate and high density disk models in separate panels. When there is no disk present, as seen in the top left panel, non-rotating stars will share the same color and magnitude values regardless of their inclination angle. As the rotation rate is increased, the brightness and color of the star change and the star moves across the diagram, following a ``track'' that depends on the inclination angle at which it is observed. Taken together, the tracks for different inclination angles form a rotational displacement ``fan" for a given spectral subtype. The paths followed by the stars viewed at low inclination angles are nearly vertical, exhibiting only small changes in color for significant changes in brightness. Conversely, the tracks for stars when viewed close to equator-on are nearly horizontal with large changes in color and only minor changes in brightness.

When there is a disk present, the track followed by the system with increasing rotation rate depends on the density of the disk. Low density disks have no significant impact on the track, and their color-magnitude diagram shows the same trends as the diskless models. For moderate density disks, shown in the top right panel of Figure~\ref{fig:colormag_fans}, the color and magnitude of the system become dependent on the inclination angle even for the lowest tested rotation rates. As a result, the migration tracks no longer radiate from the same common point, an effect that is especially prominent in the early types. For systems with the densest disks, the fan structure is entirely disrupted and, as we see in the bottom left panel of Figure~\ref{fig:colormag_fans}, the migrational tracks of the stars at different inclinations radiate from different regions of the diagram for all subtypes except for the B8 models, which are shown in the bottom right panel. The color and magnitude values exhibited by the earlier-type, high density systems are dominated by the disk and therefore highly dependent on both inclination angle and rotation rate. 

\subsection{H$\alpha$ equivalent width}\label{sec:ha_ew_results}

\begin{figure*}
    \centering
    \begin{minipage}[b]{0.45\textwidth} \centering\includegraphics[width=\textwidth]{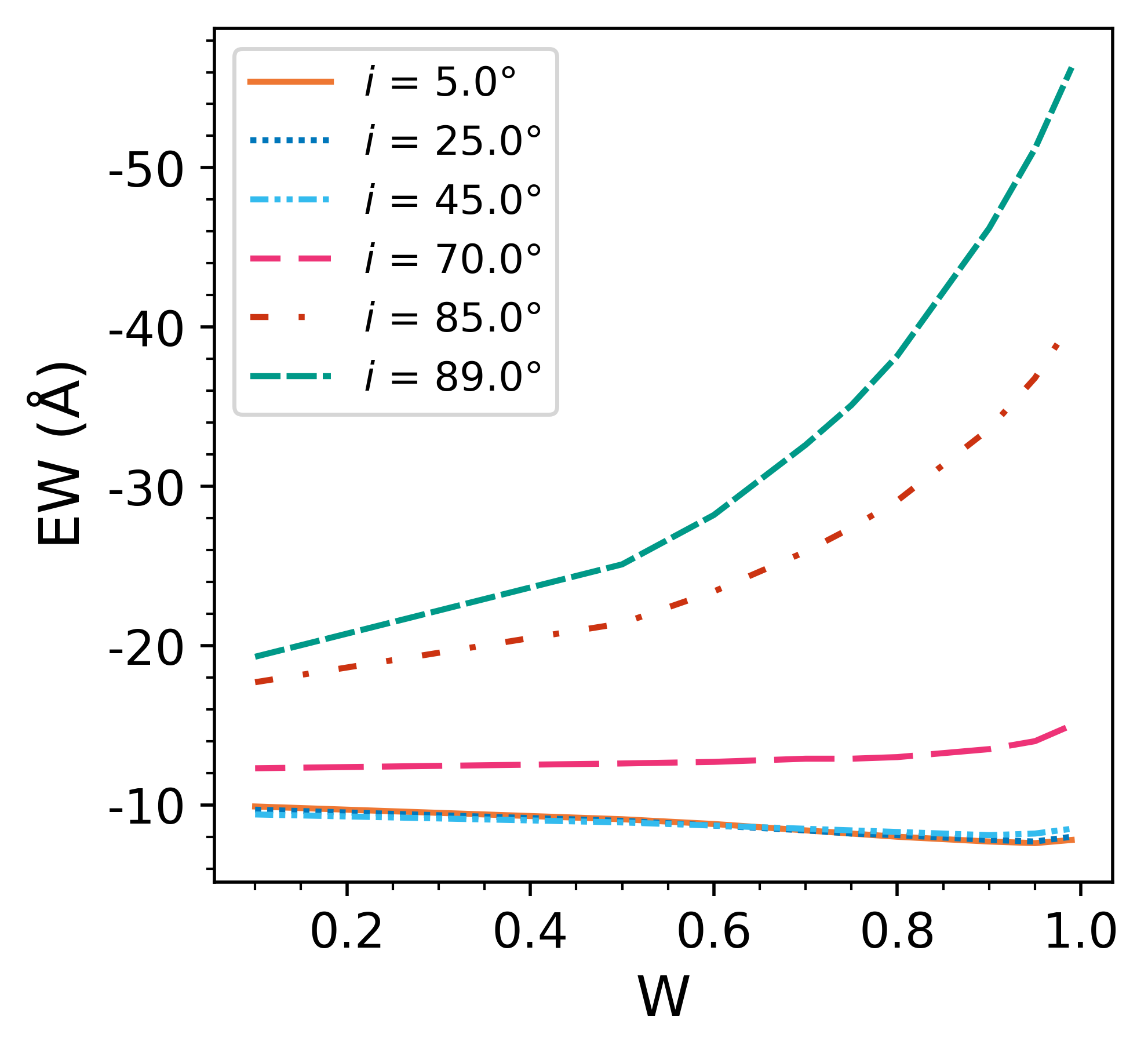}\par\vspace{0.5em}
    \includegraphics[width=\textwidth]{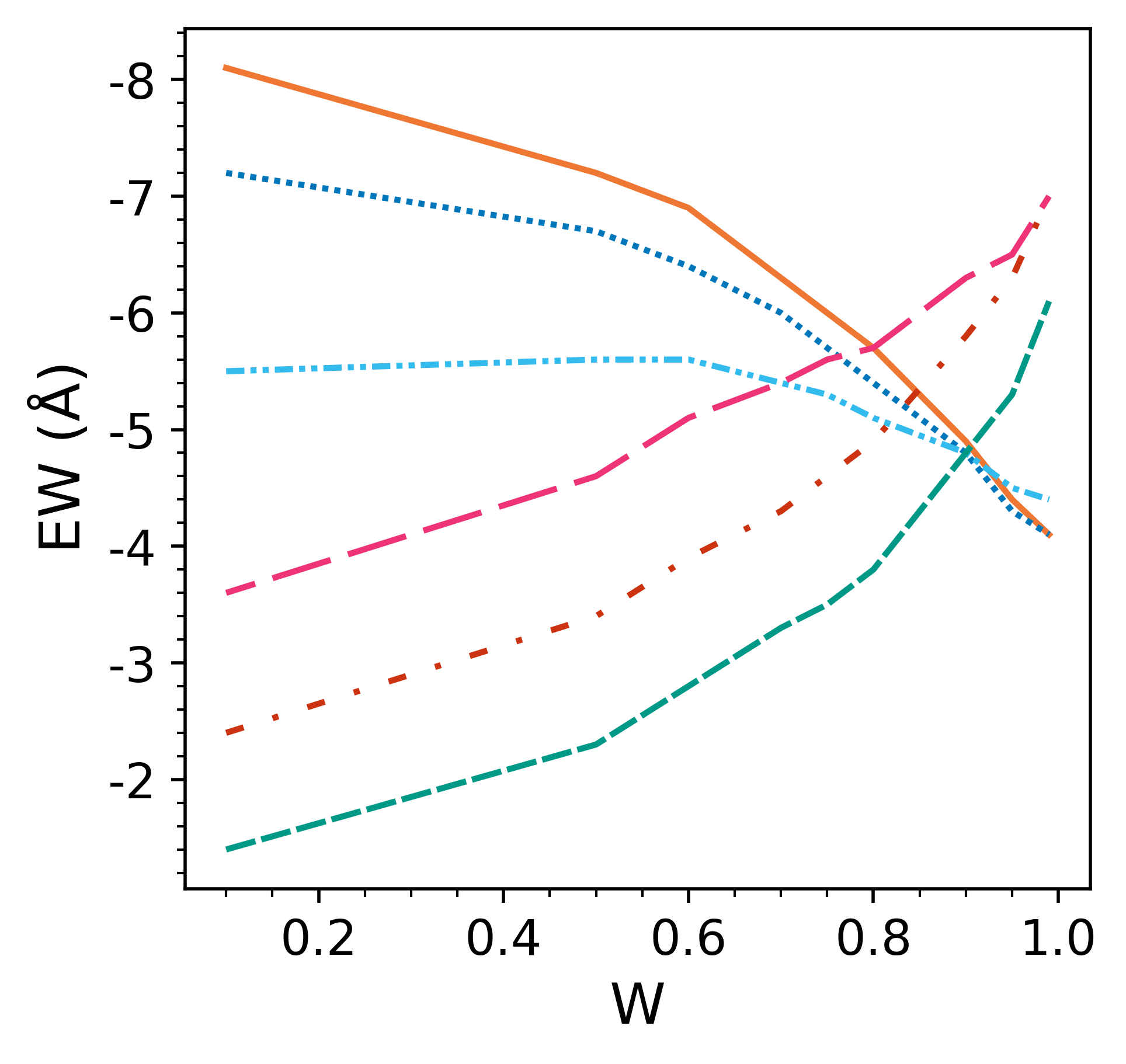}\par\vspace{0.5em}
    \includegraphics[width=\textwidth]{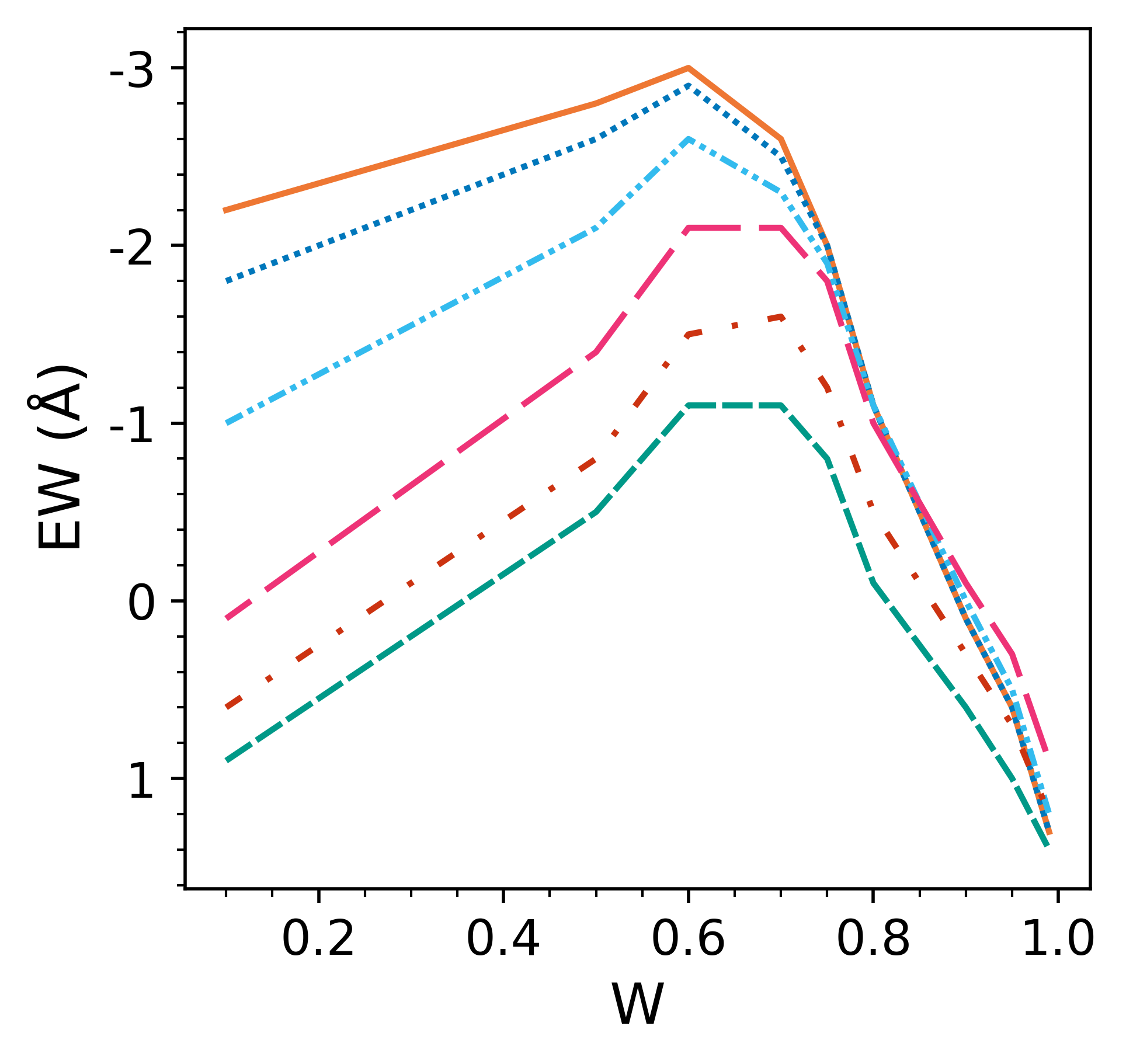}
    \end{minipage}%
    \hfill
    \begin{minipage}[b]{0.5\textwidth}  
        \centering
    \includegraphics[width=\textwidth]{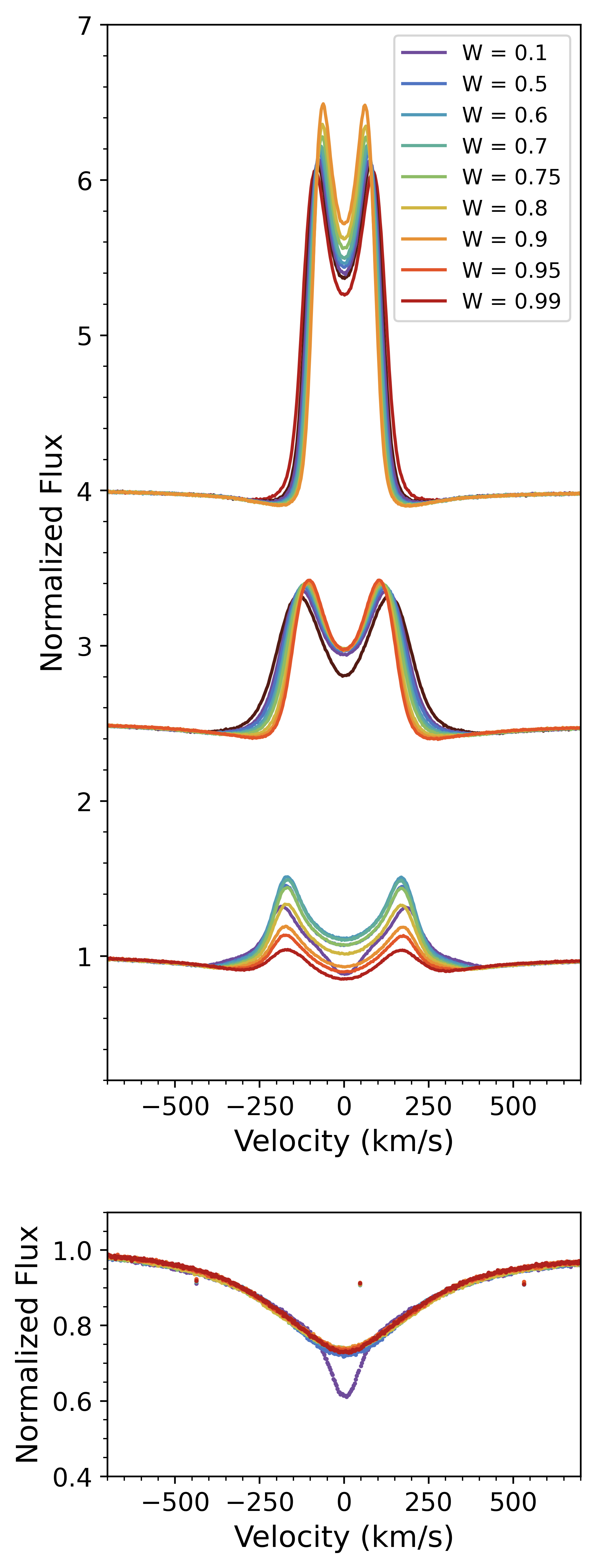}
    \end{minipage}
    \caption{Left: H$\alpha$ equivalent width trends with increasing $W$ for B2 models with high density disks (top left), moderate density disks (center left) and low density disks (bottom left), at various inclinations as indicated in the legend. Right: Normalized H$\alpha$ line profiles for B2 models at different rotation rates viewed at a constant inclination of 45$^{\circ}$. Profiles are shown in order of decreasing density, from high density at the top to no disk at the bottom. A vertical shift has been applied between densities to represent them on the same axes. The rotation rates are indicated in the legend.}
    \label{fig:B2_ew}
\end{figure*}

\begin{figure*}
    \centering
    \begin{minipage}[b]{\textwidth}  \centering\includegraphics[width=\textwidth]{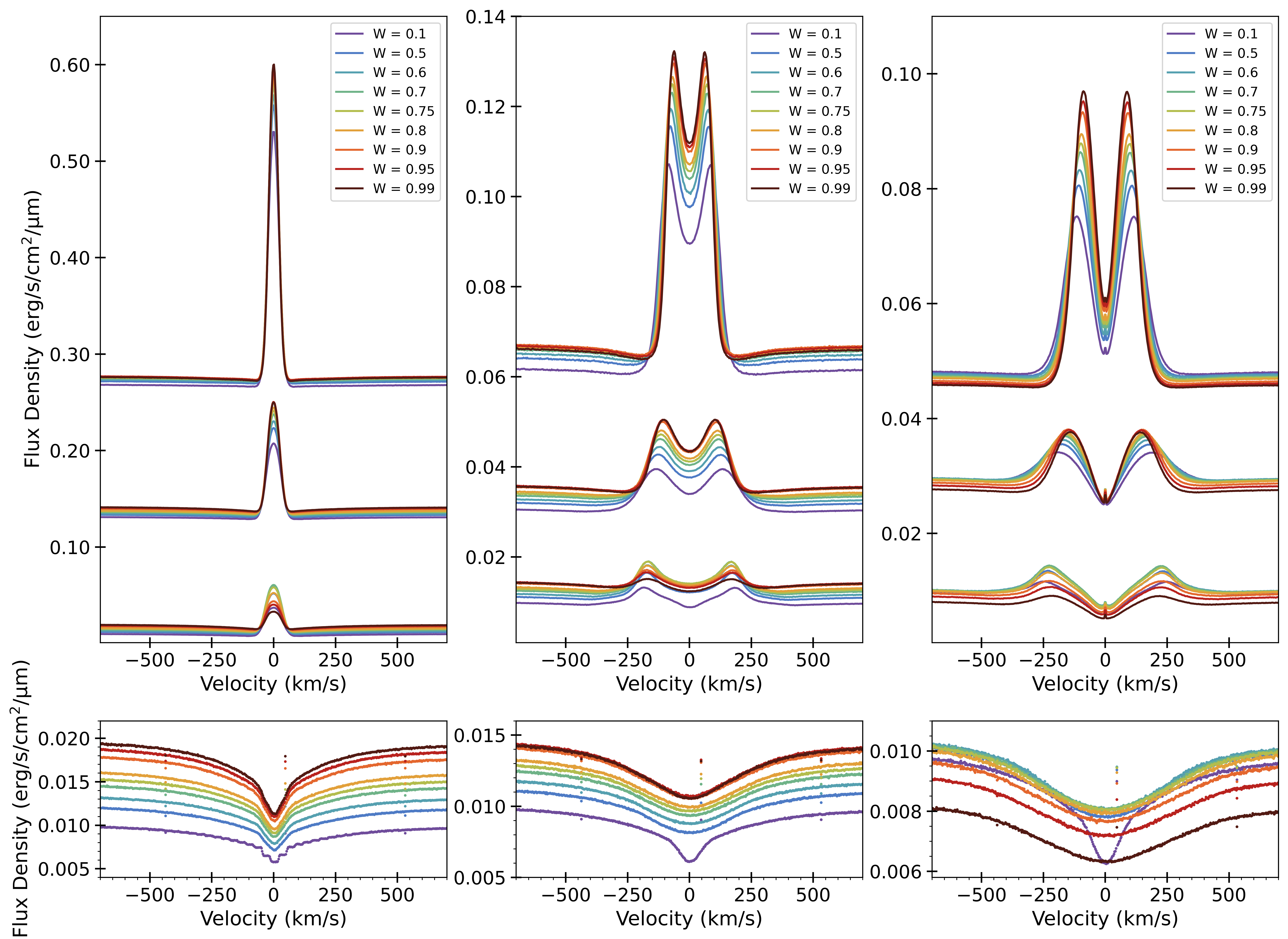}
    \end{minipage}%
    \hfill
    \caption{Unnormalized H$\alpha$ lines for the B2 models at an inclination of 5$^{\circ}$ (left), 45$^{\circ}$ (center), and 85$^{\circ}$ (right). In the top panels, we present models for high density (top), moderate density (center) and low density (bottom) disks, while the bottom panels show the diskless models. The rotation rates are indicated in the legend.}
    \label{fig:B2_unnormalized}
\end{figure*}

The strength of the H$\alpha$ emission, as measured by the EW, is strongly dependent on rotation rate for the early-type models. The left column of Figure~\ref{fig:B2_ew} shows the trends in H$\alpha$ EW with increasing rotation rate for all tested densities and inclination angles for the B2 simulations. We note that the B0 and B5 subtypes show qualitatively the same behavior in all cases except the lowest-density disks, and figures representing these models are included in Appendix~\ref{secA2}. The normalized H$\alpha$ lines of disks viewed at very low inclinations, near face-on, become less prominent for very large rotation rates ($W > 0.9$) when the brightening of the poles leads to the rise in the continuum levels. This is seen in the left and center panels of Figure~\ref{fig:B2_unnormalized}, which plots the unnormalized H$\alpha$ profiles for the B2 models at inclinations of 5$^{\circ}$ and 45$^{\circ}$. Meanwhile, stars viewed at moderate to high inclinations respond differently to increased rotation rates depending on the density of the disk. At low densities, moderate and high inclinations behave the same as low inclinations. However, at inclinations $\geq$45$^\circ$, the emission strength is enhanced for faster-rotating stars. For near-edge-on cases, the dependence on density is particularly strong, as seen in the right panel of Figure~\ref{fig:B2_unnormalized}.  As a result, the EW can become extremely large for high density disks, as seen in the top left panel of Figure~\ref{fig:B2_ew}.

The morphology of the H$\alpha$ line changes somewhat with rotation rate, as well. For a disk with a given density viewed at low inclinations, the peak intensity is reduced as the rotation rate increases, while the same disk viewed from high inclinations show the opposite effect and its line profile exhibits larger peak intensities for faster rotation rates. Holding the inclination angle constant and varying the disk density also produces different rotational effects. As an example, the right panel of Figure~\ref{fig:B2_ew} shows the H$\alpha$ profiles for the B2 models with different disk densities at an inclination of 45$^\circ$. We see that the height of the peaks when viewed at this inclination may drop with increasing rotation rate for low density disks, increase with rotation for moderate density disks, and stay roughly the same for high density disks. At the same time, the width of the line changes with rotation rate. This change in line width is likely due to the increasing disk radius with $W$, as described in Section~\ref{sec2}, rather than an effect of gravity darkening. Disks with more extended H$\alpha$ emitting regions will appear to have narrower lines. 

\begin{figure*}[!ht]
\centering
    \begin{subfigure}{0.49\textwidth}
    \includegraphics[width=\textwidth]{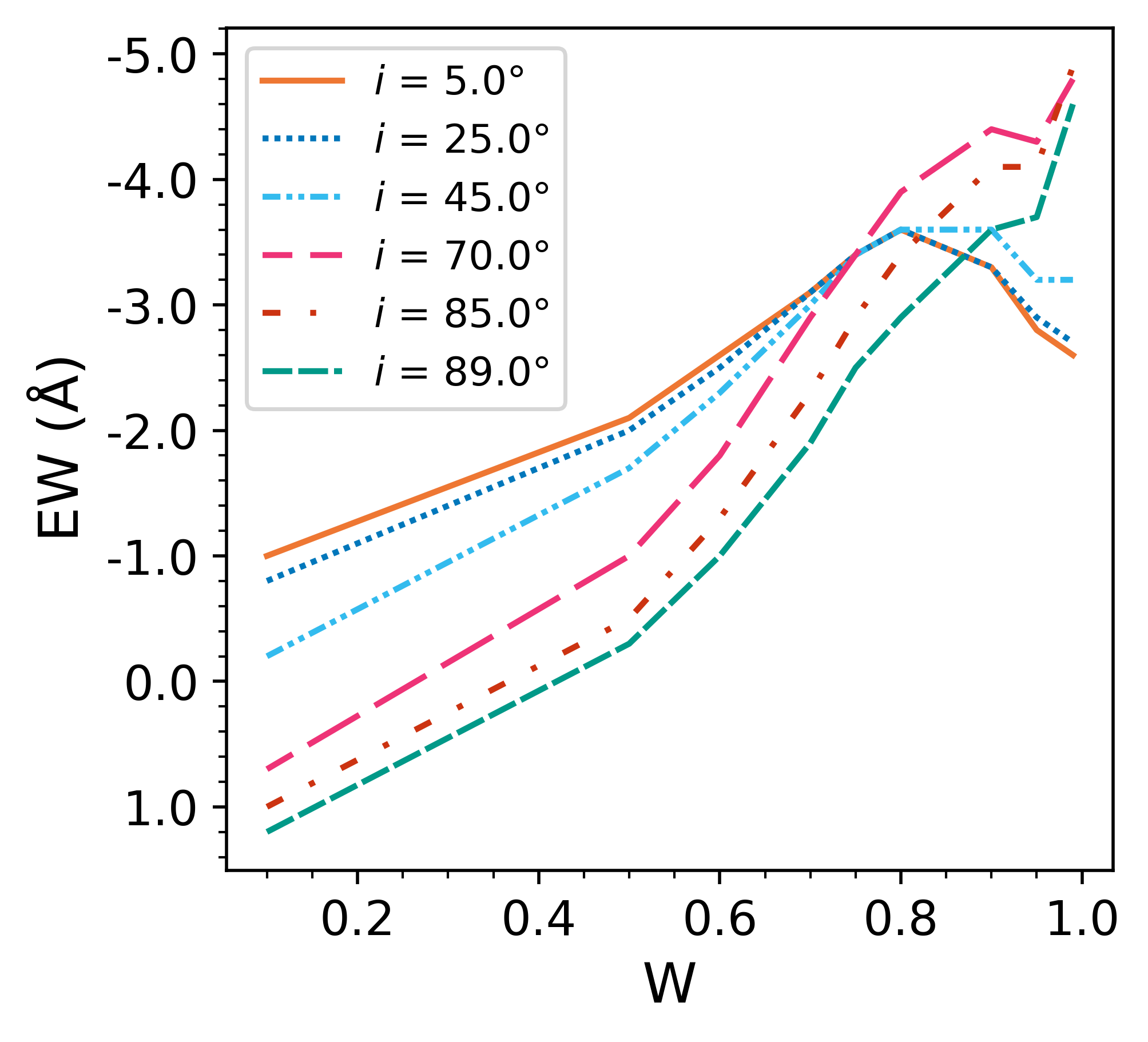}
    \end{subfigure}
    \begin{subfigure}{0.49\textwidth}
    \includegraphics[width=\textwidth]{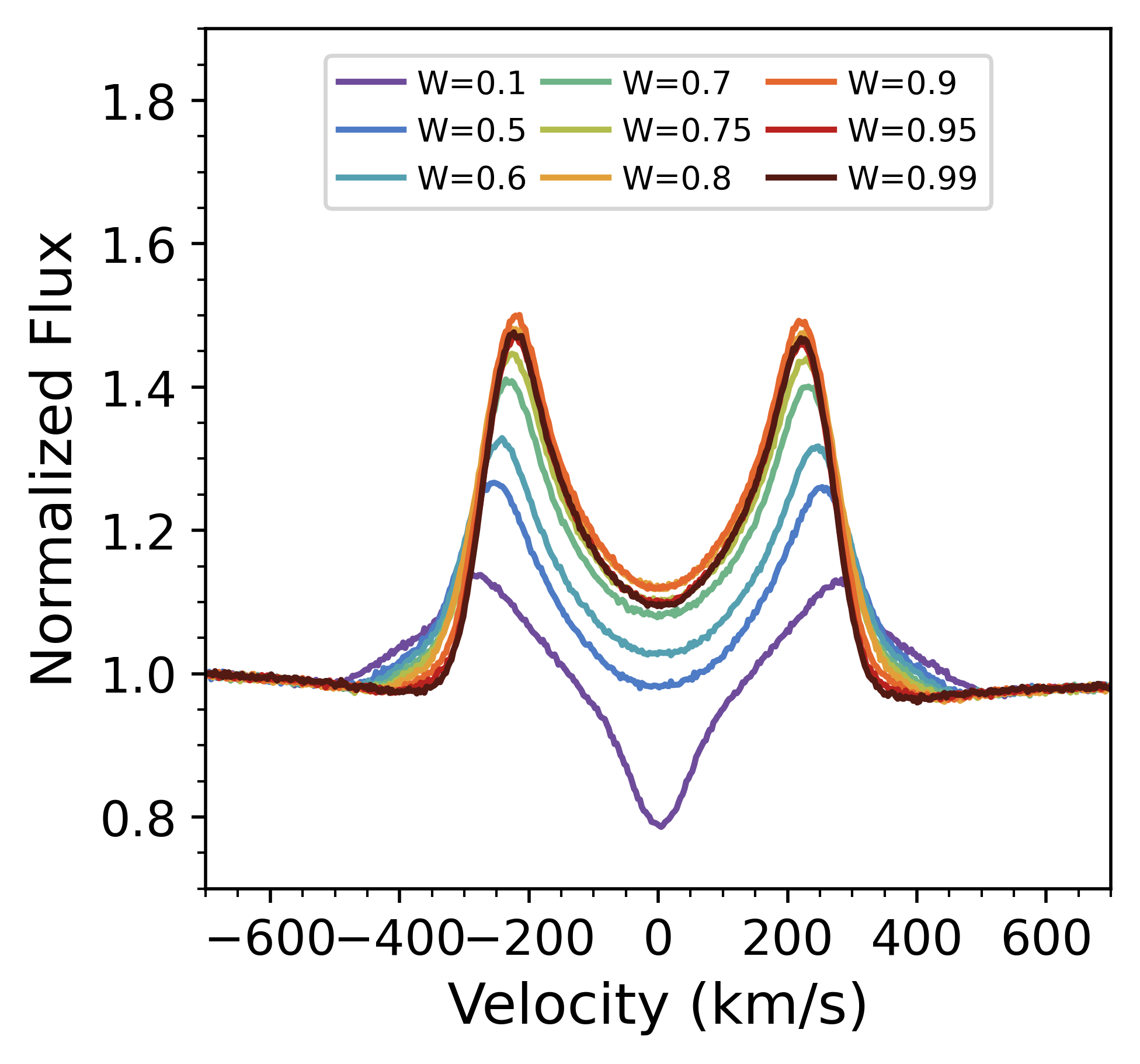}
    \end{subfigure}
        \begin{subfigure}{0.49\textwidth}
    \includegraphics[width=\textwidth]{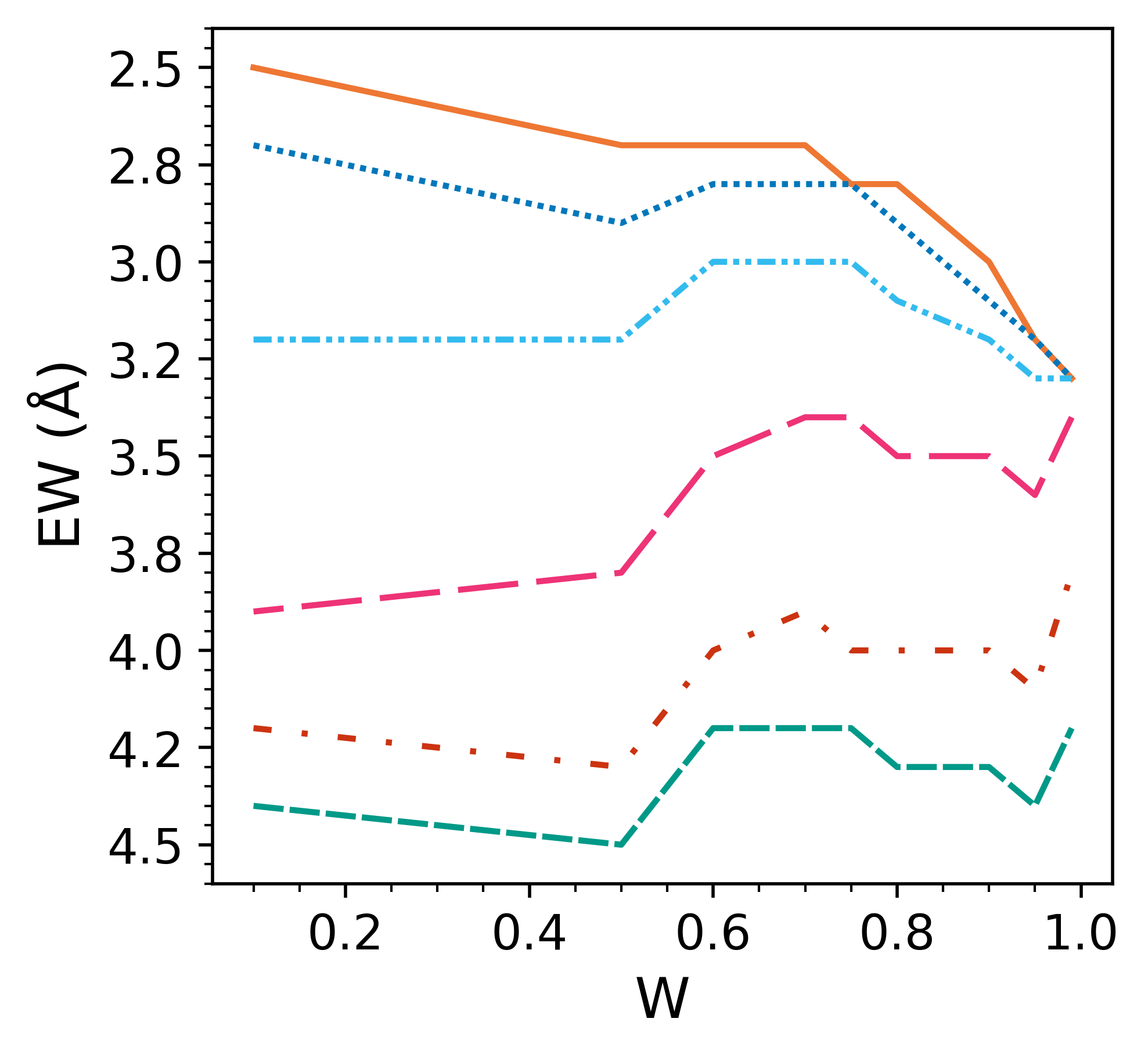}
    \end{subfigure}
    \begin{subfigure}{0.49\textwidth}
    \includegraphics[width=\textwidth]{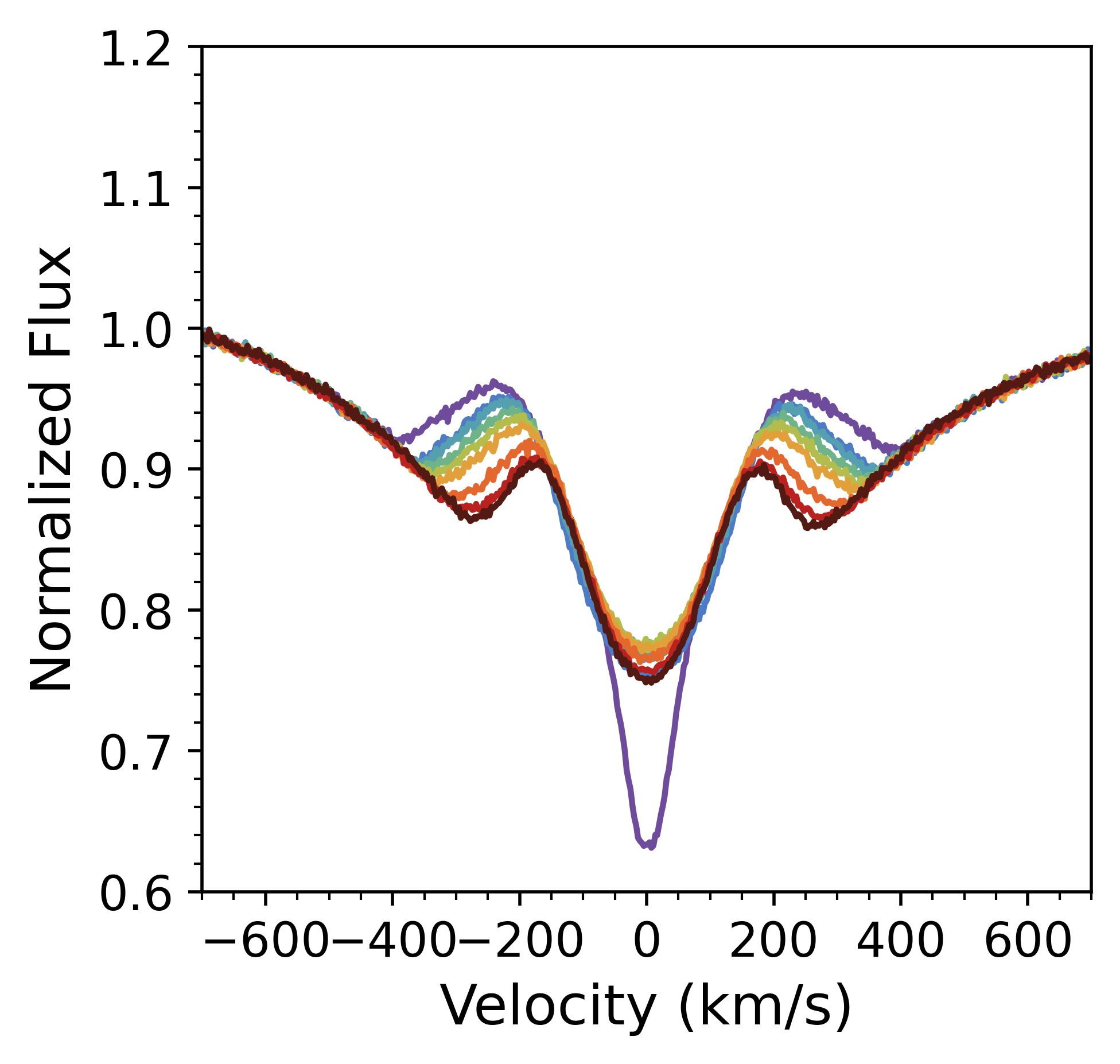}
    \end{subfigure}
\caption{H$\alpha$ EW trends as a function of rotation rate (left) and H$\alpha$ line profiles at an inclination of 45$^{\circ}$ (right) for B0 (top) and B5 (bottom) models at low disk densities.}
\label{fig:B0_B5_ew}
\end{figure*}

As mentioned previously, the B0 and B5 models show the same trends as the B2 for the moderate and high densities, but appear different at the lowest density. Figure~\ref{fig:B0_B5_ew} shows the H$\alpha$ EW trends for the low density disk models in both B0 and B5, along with their normalized H$\alpha$ lines. For the B0 spectral type, the effects of rotation near the critical value are dependent on inclination angle, with EWs at the lower inclinations dropping somewhat while those at the larger inclinations rise. The B0 normalized H$\alpha$ line at this density, represented by the profile at 45$^\circ$ in the top right panel of Figure~\ref{fig:B0_B5_ew}, becomes taller with increasing rotation rate.  Meanwhile, the EW shows little dependence on rotation for the B5 model.

\begin{figure*}
    \centering
    \begin{minipage}[b]{0.45\textwidth} \centering\includegraphics[width=\textwidth]{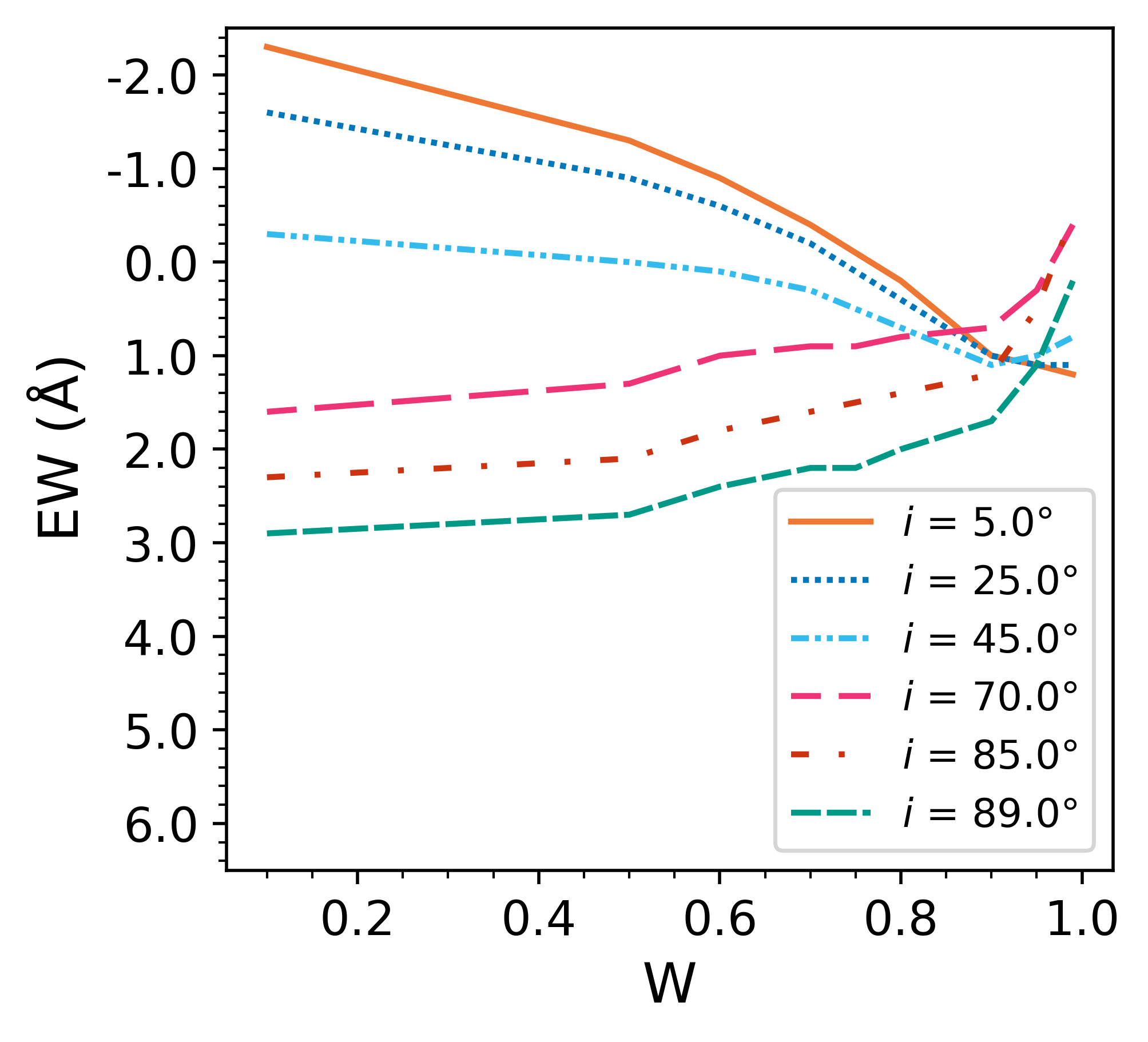}\par\vspace{0.5em}
    \includegraphics[width=\textwidth  ]{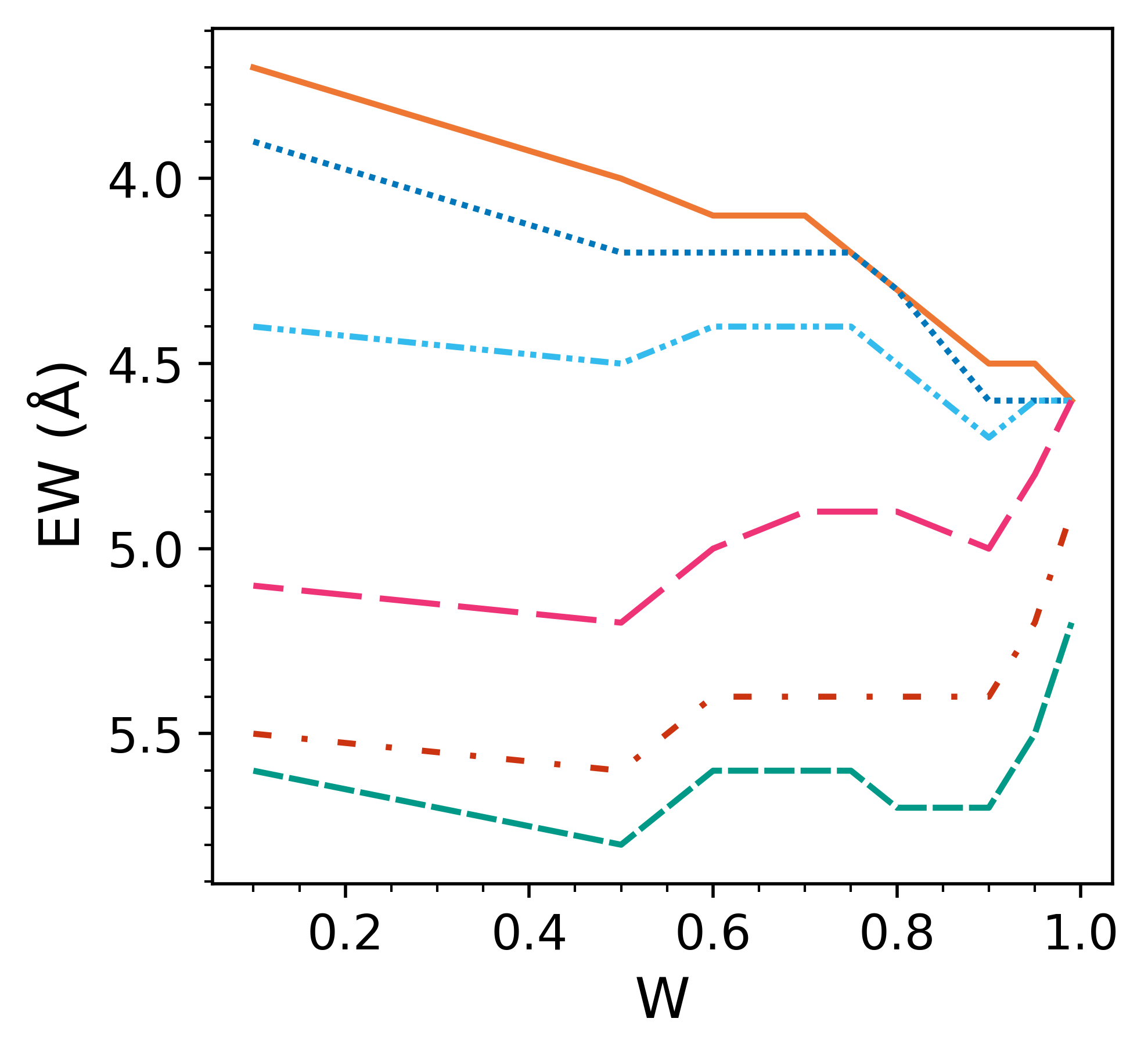}\par\vspace{0.5em}
    \includegraphics[width=\textwidth]{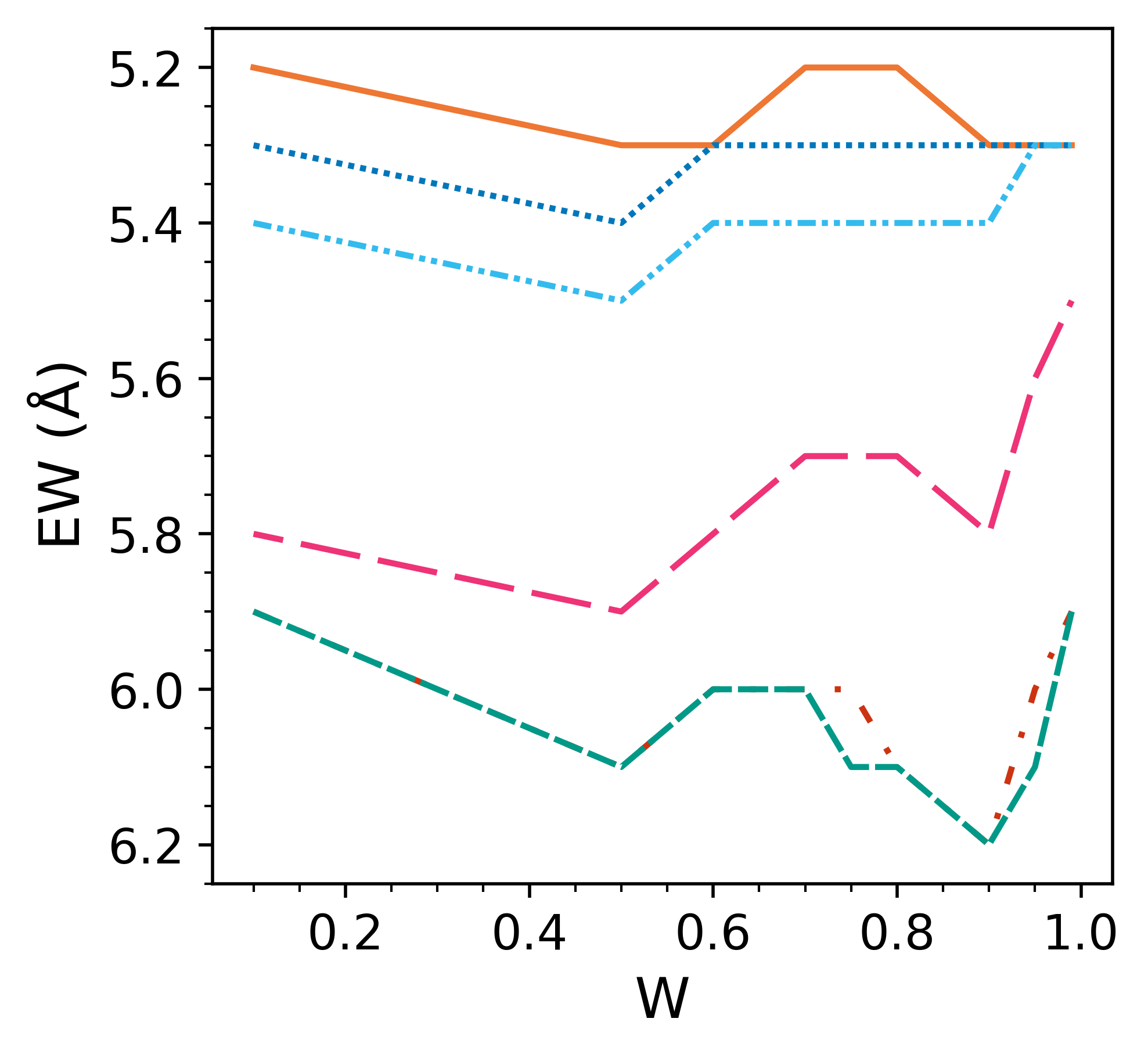}
    \end{minipage}%
    \hfill
    \begin{minipage}[b]{0.5\textwidth}  
        \centering
    \includegraphics[width=\textwidth]{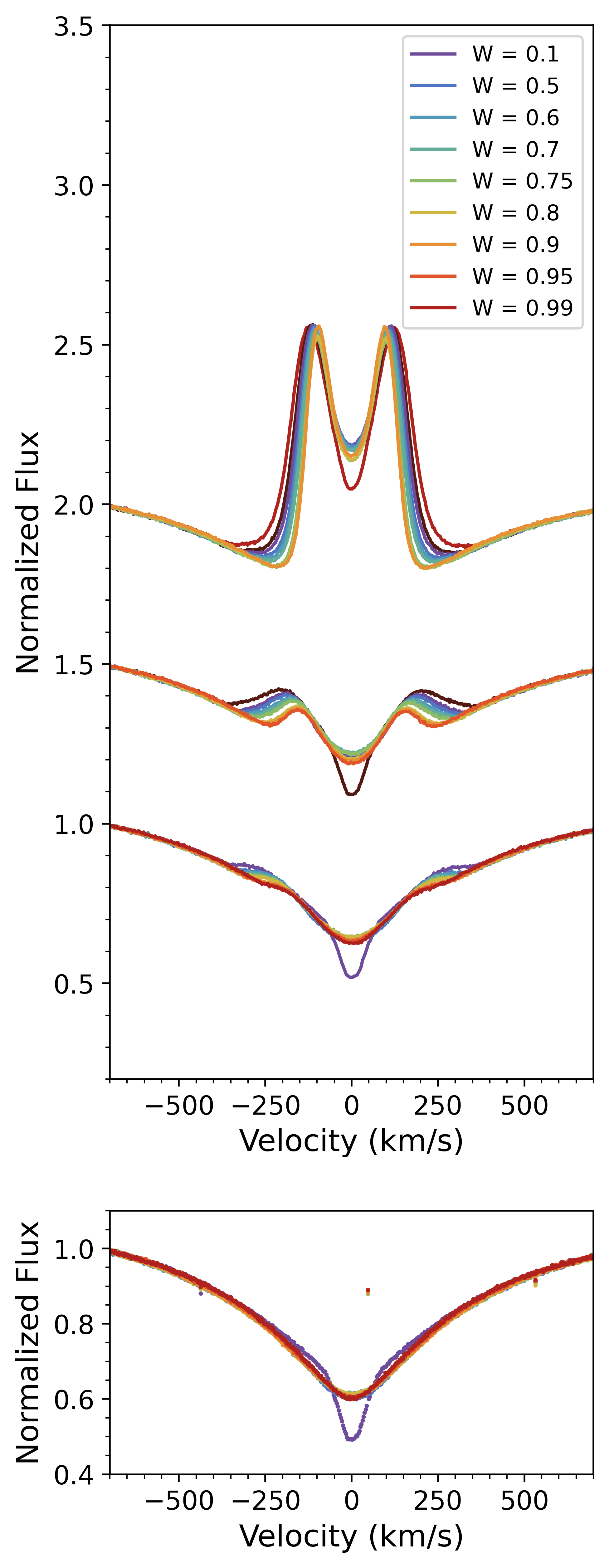}
    \end{minipage}
    \caption{Same as Figure~\ref{fig:B2_ew}, but for the B8 models.}
    \label{fig:B8_ew}
\end{figure*}

Our B8 models show trends that are distinct from those seen in the earlier types, exhibiting less extreme changes in EW with rotation rate. The left panels of Figure~\ref{fig:B8_ew} show that the low density and moderate density models have positive EW values, indicating a net absorption line rather than emission line. This is also apparent in the right panel of Figure~\ref{fig:B8_ew}, where the emission features are faint for the moderate density disks, and negligible in the low density models. In addition to being very faint, the H$\alpha$ EW for low and moderate density models varies only by a few tenths of an \AA, as seen in the center and bottom panels of Figure~\ref{fig:B8_ew}. The subset of B8 models with the densest disks, in the top left panel of Figure~\ref{fig:B8_ew}, are similar to the moderate density cases seen in the center panel of Figure~\ref{fig:B2_ew}.

\subsection{Intrinsic polarization}\label{sec:pol_results}

\begin{figure*}
    \centering
    \begin{minipage}[b]{0.47\textwidth} \centering\includegraphics[width=\textwidth]{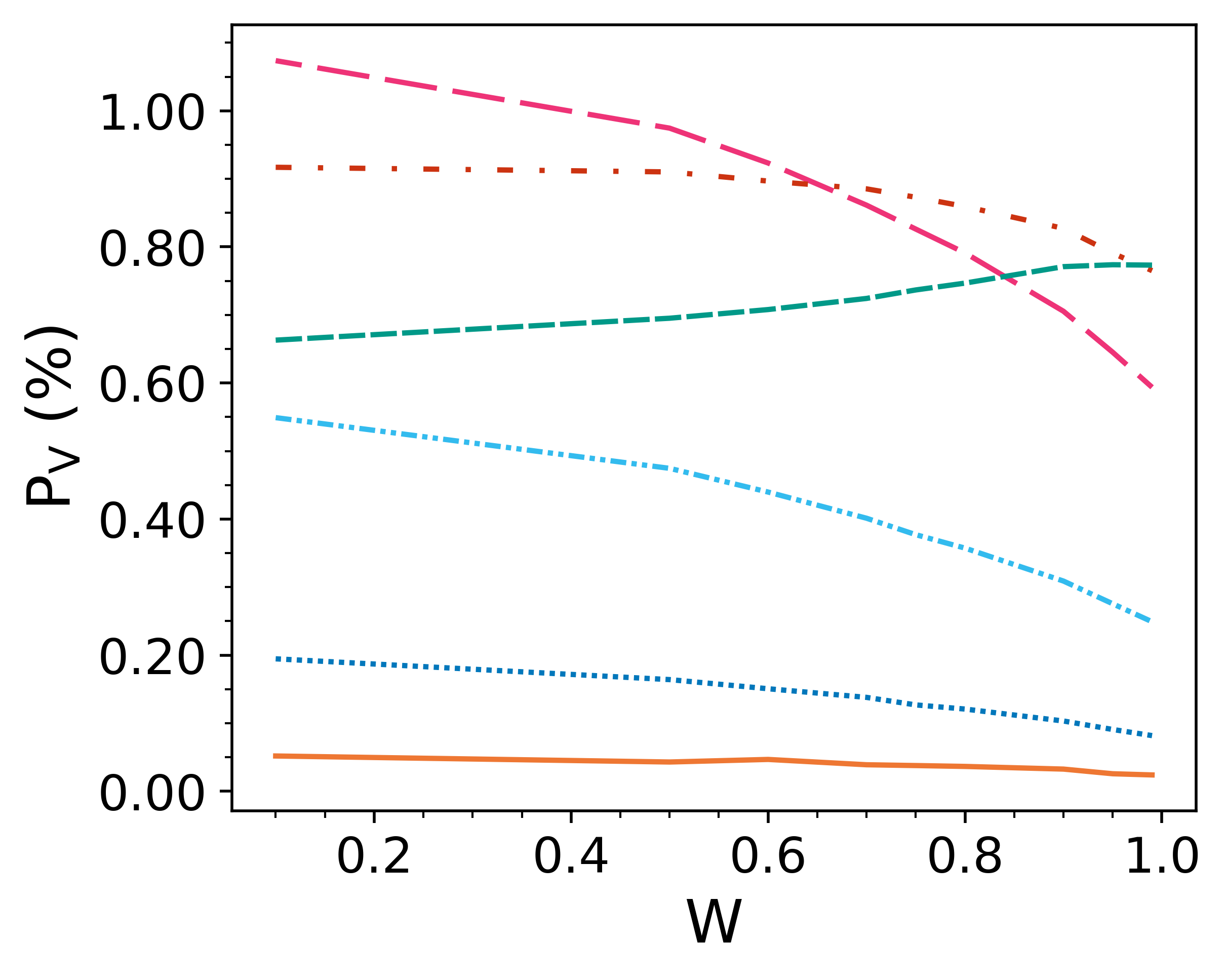}\par\vspace{0.5em}
    \includegraphics[width=\textwidth  ]{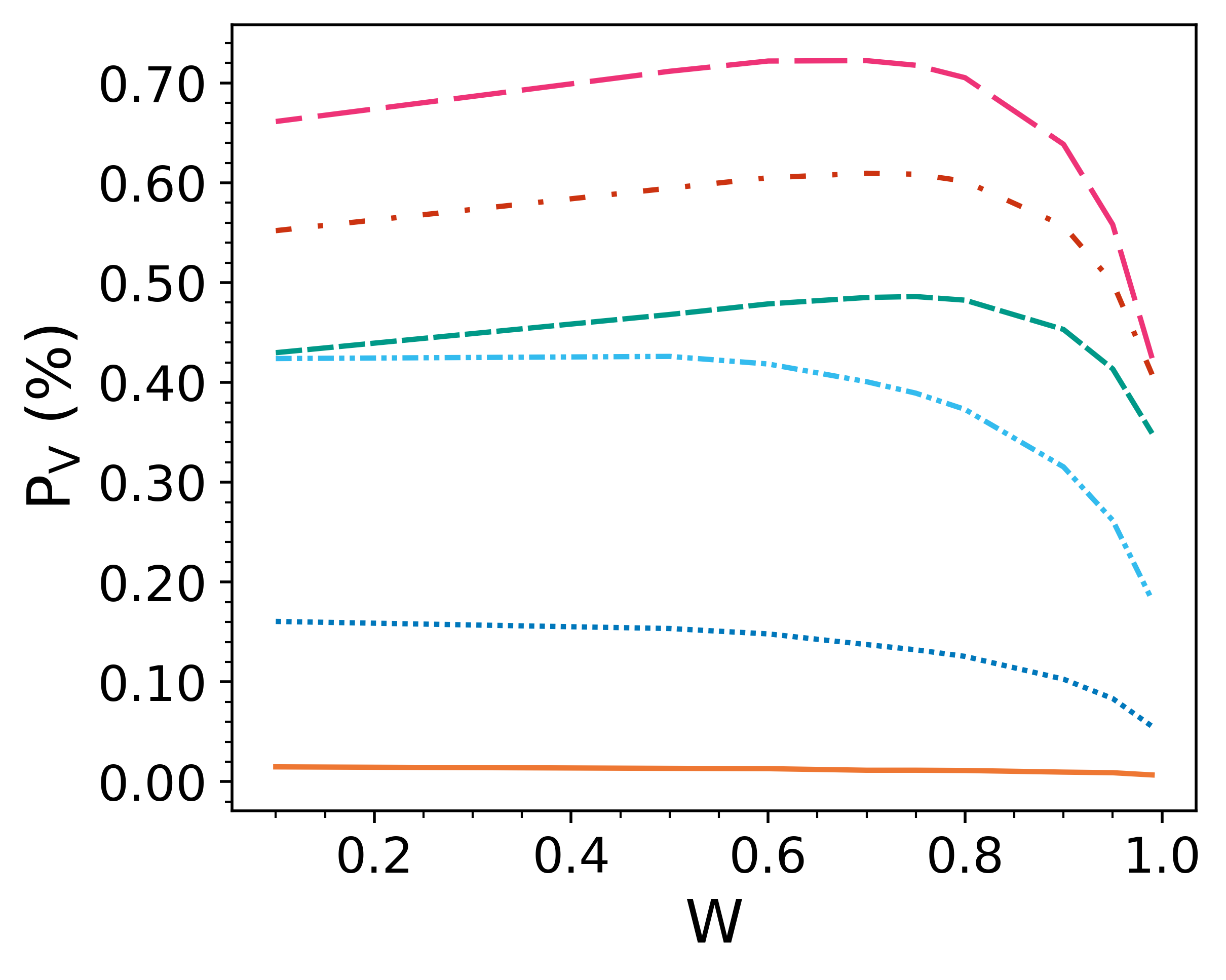}\par\vspace{0.5em}
    \includegraphics[width=\textwidth]{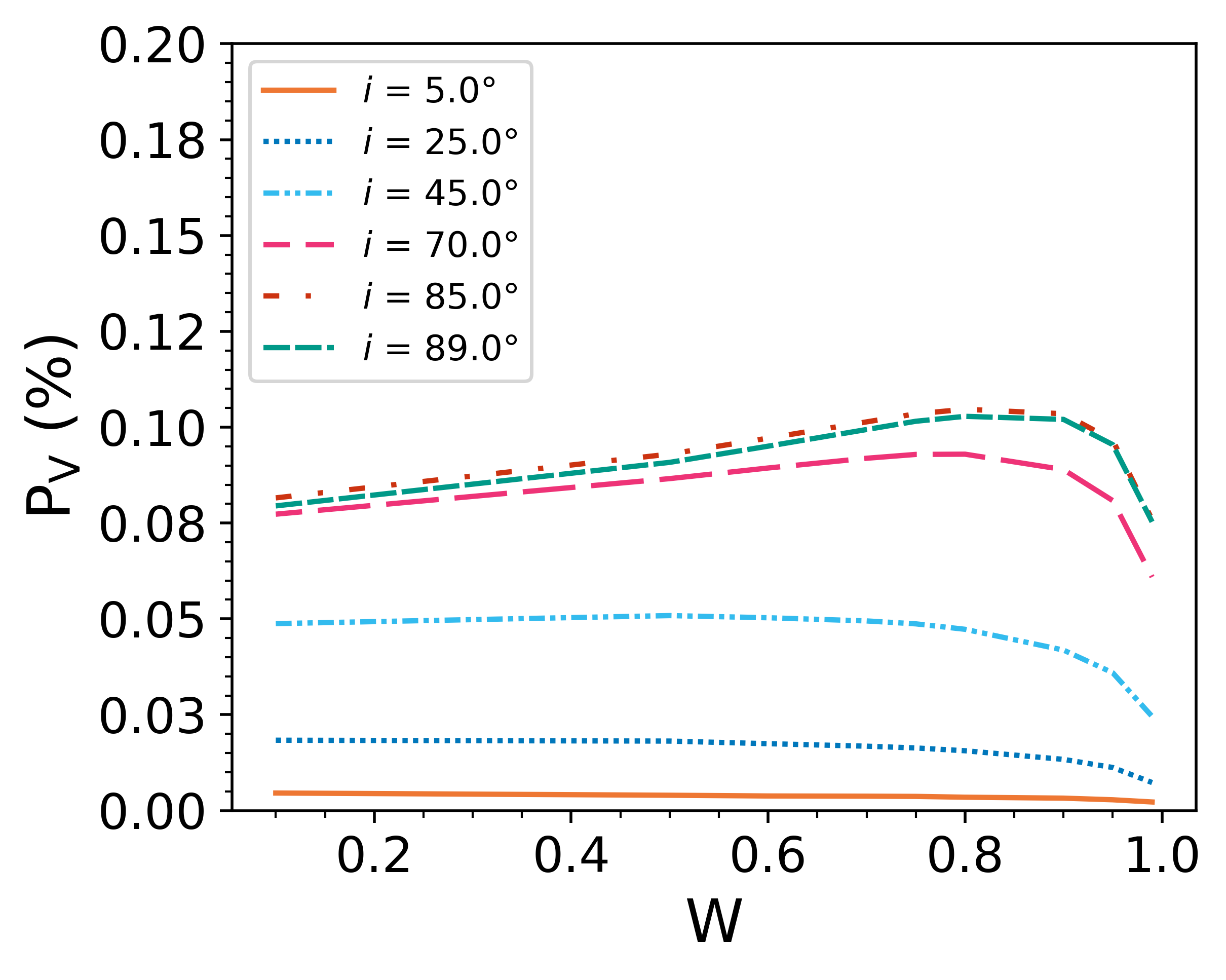}
    \end{minipage}%
    \hfill
    \begin{minipage}[b]{0.47\textwidth} \centering\includegraphics[width=\textwidth]{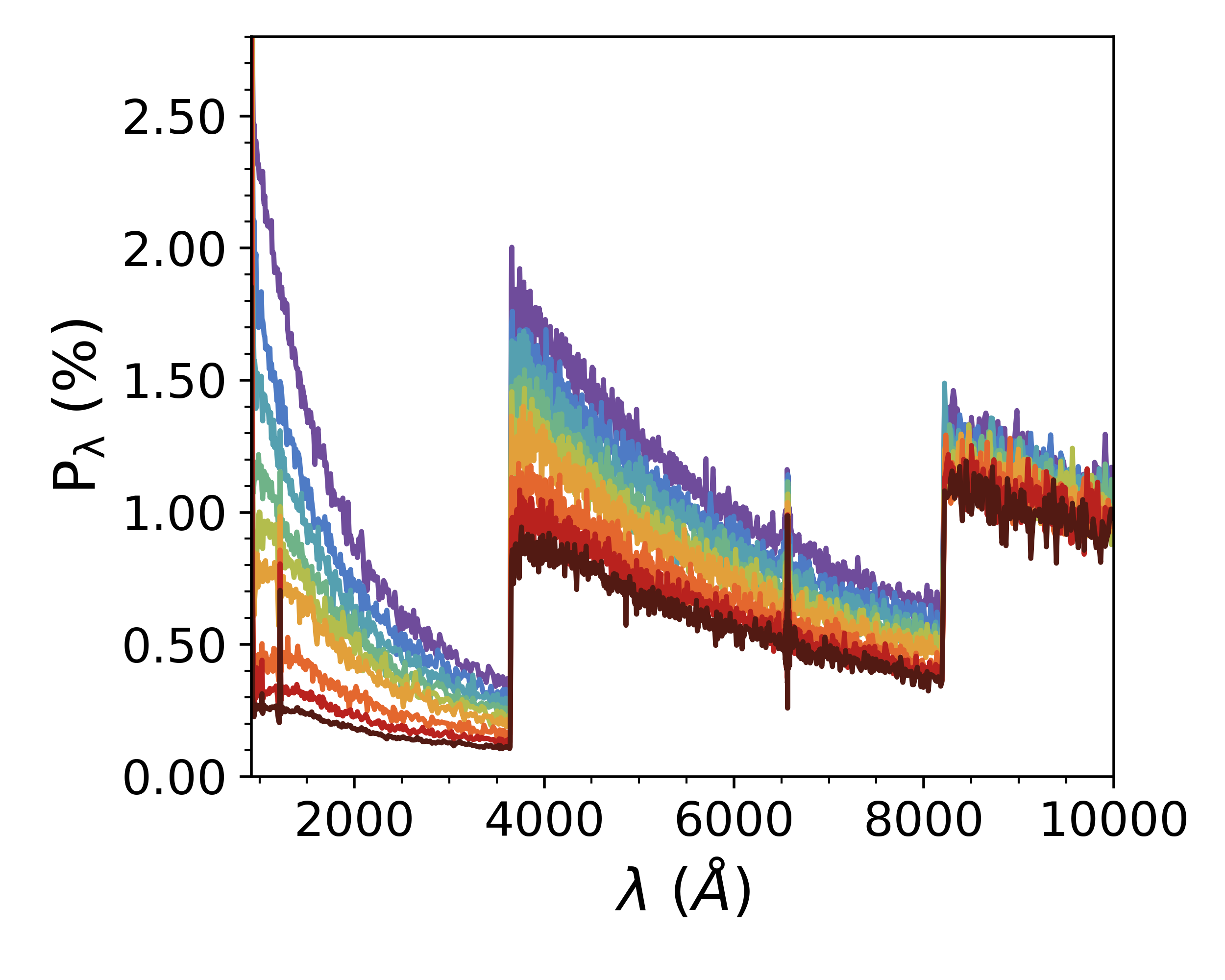}\par\vspace{0.5em}
    \includegraphics[width=\textwidth  ]{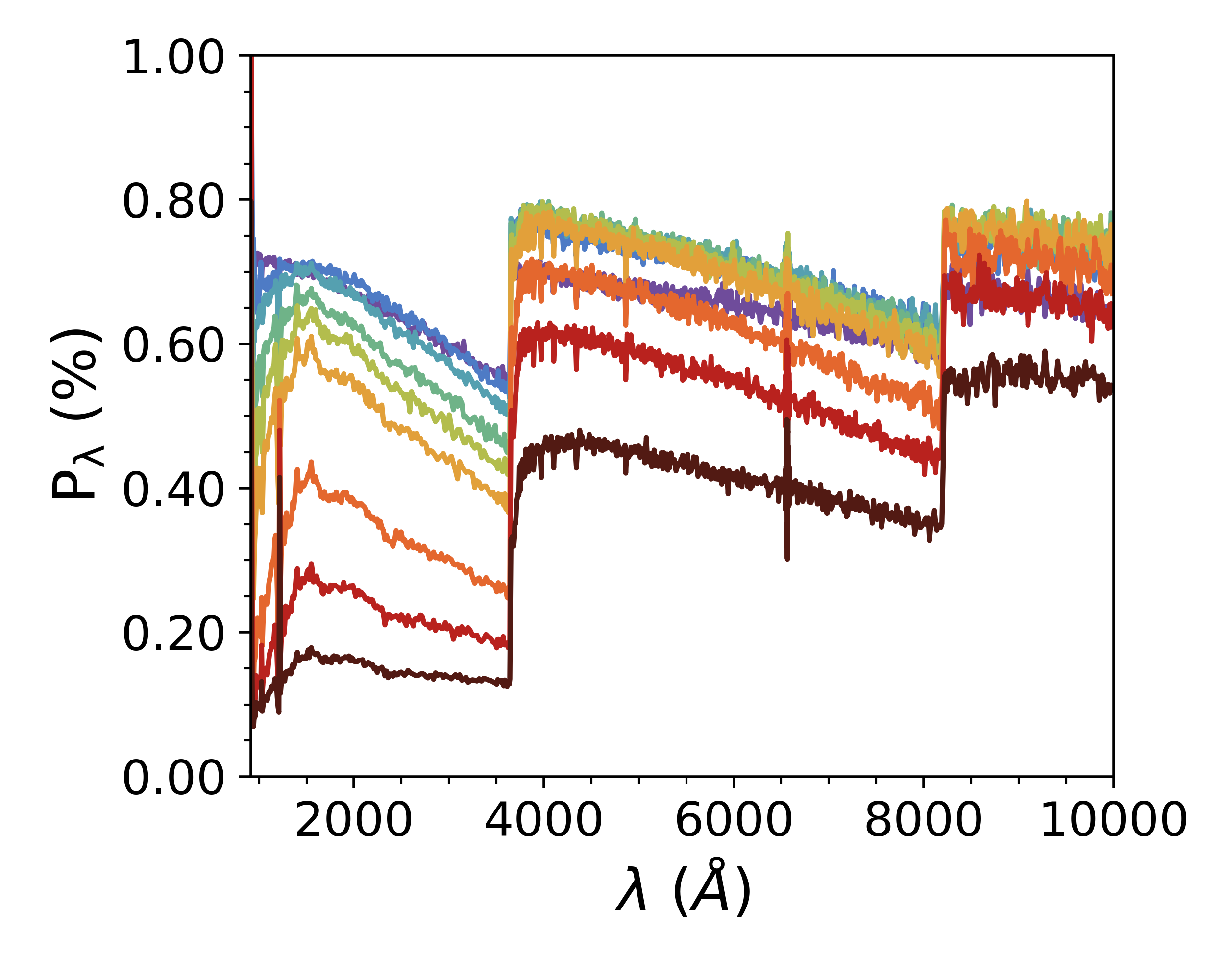}\par\vspace{0.5em}
    \includegraphics[width=\textwidth]{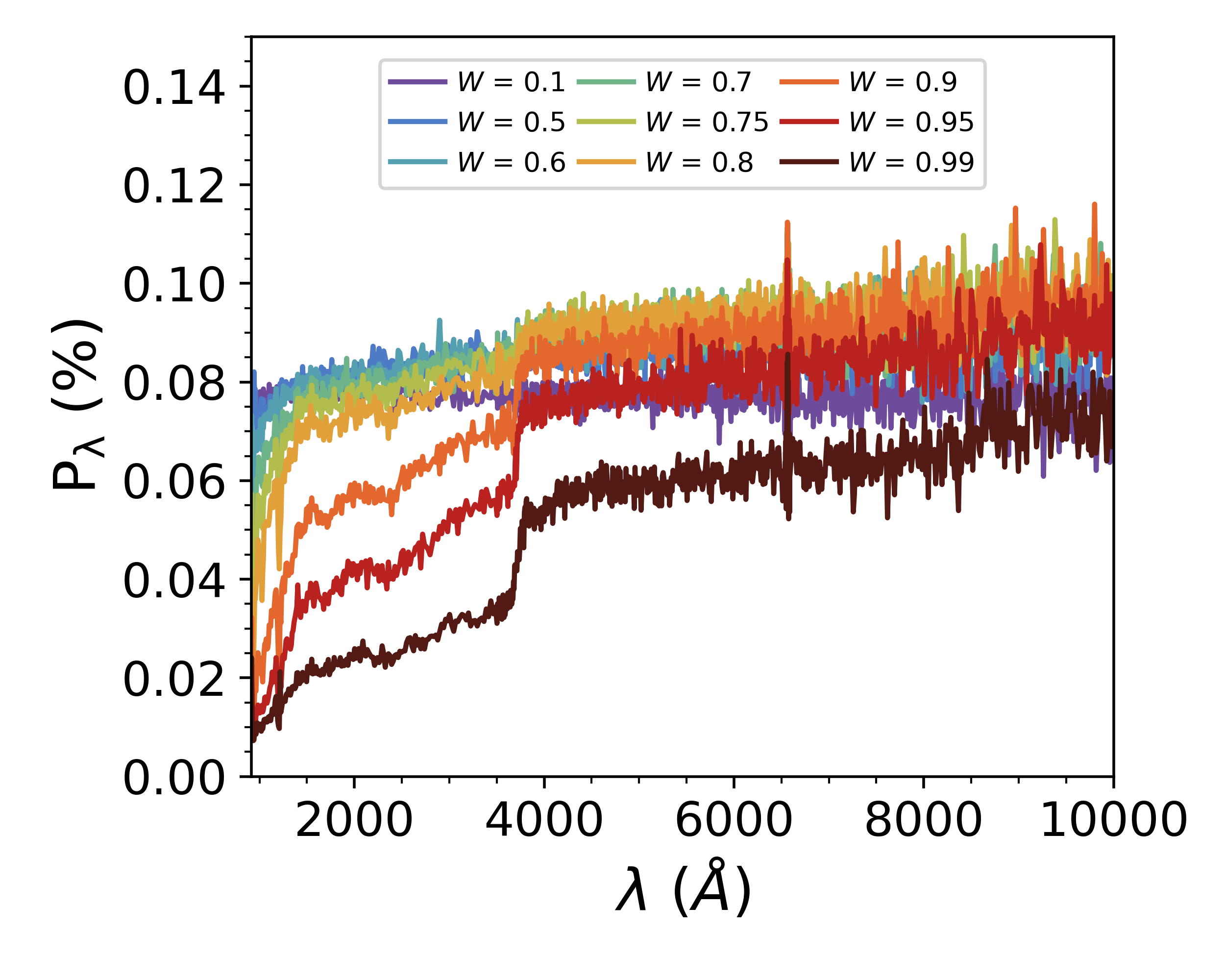}
    \end{minipage}%
    \hfill
    \caption{Left: $V$-band (averaged across 5000-6500 \AA) polarization degree with increasing $W$ for the B2 models, at different inclinations as indicated in the legend. Right: polarized spectra viewed at a constant inclination of 70$^\circ$ at different rotation rates, indicated in the legend, for the same B2 models. For both columns, profiles are shown in order of decreasing density, from high density at the top to low density disks at the bottom.}
    \label{fig:B2_pol}
\end{figure*}

\begin{figure*}
    \centering
    \begin{minipage}[b]{0.47\textwidth} \centering\includegraphics[width=\textwidth]{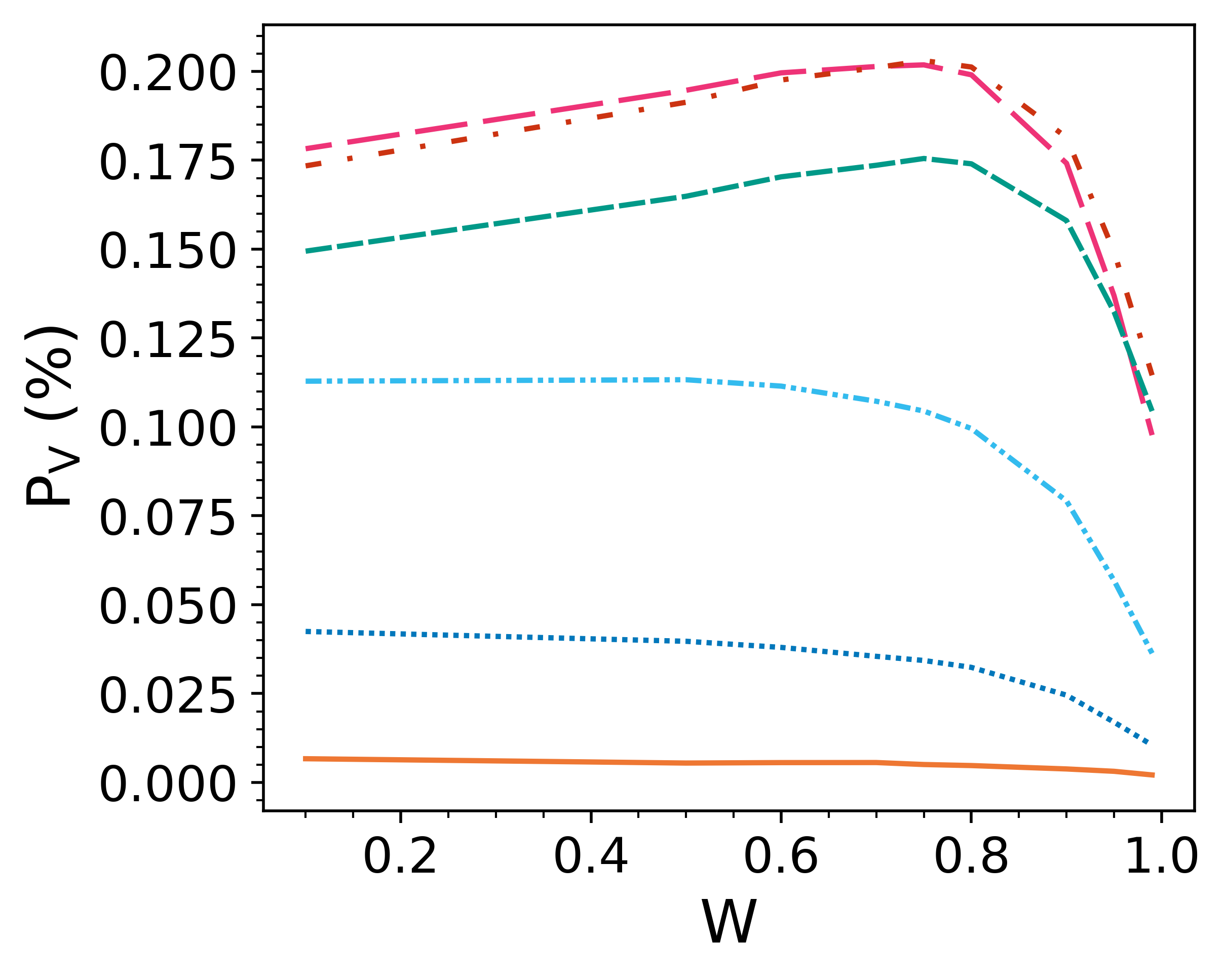}\par\vspace{0.5em}
    \includegraphics[width=\textwidth  ]{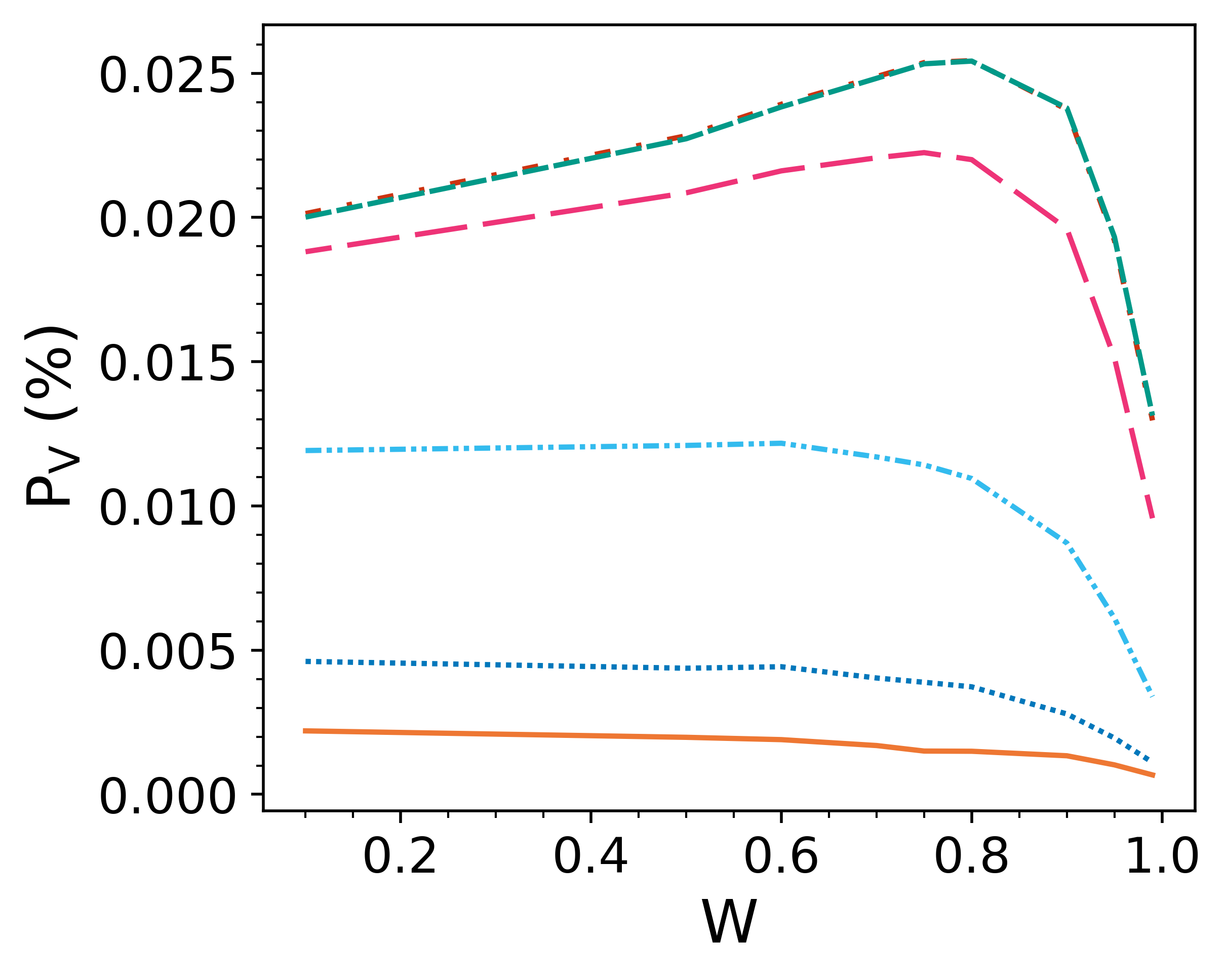}\par\vspace{0.5em}
    \includegraphics[width=\textwidth]{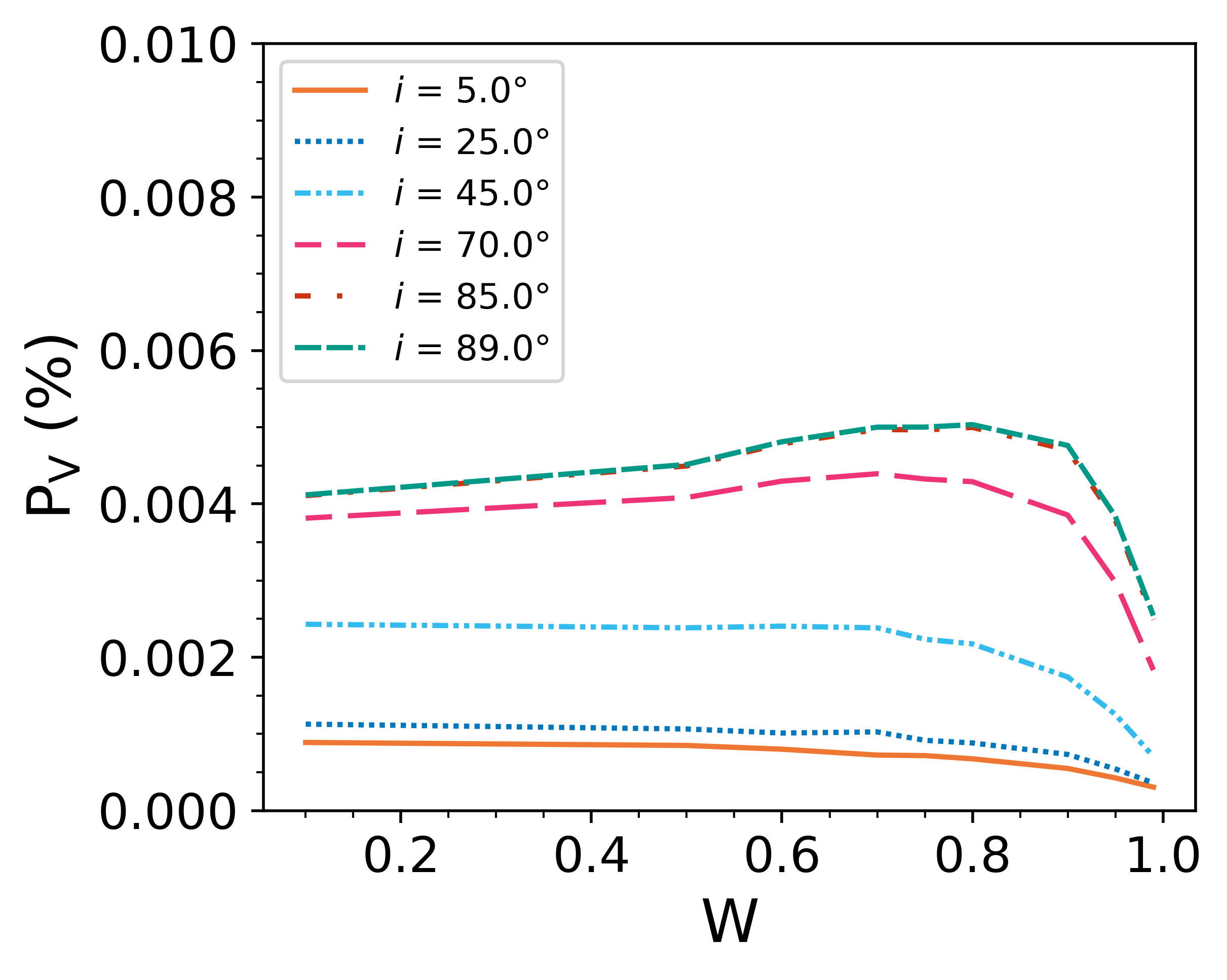}
    \end{minipage}%
    \hfill
    \begin{minipage}[b]{0.47\textwidth} \centering\includegraphics[width=\textwidth]{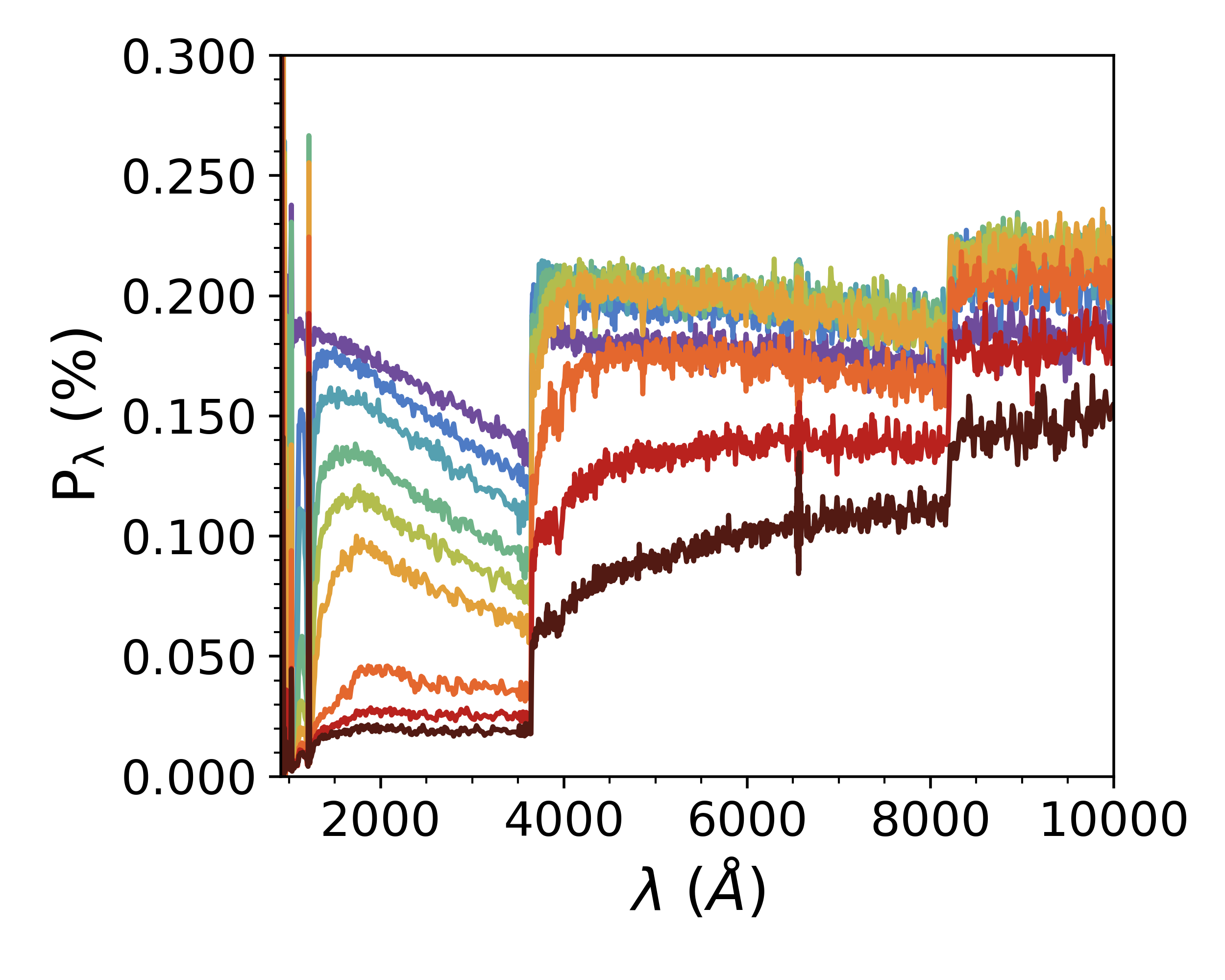}\par\vspace{0.5em}
    \includegraphics[width=\textwidth  ]{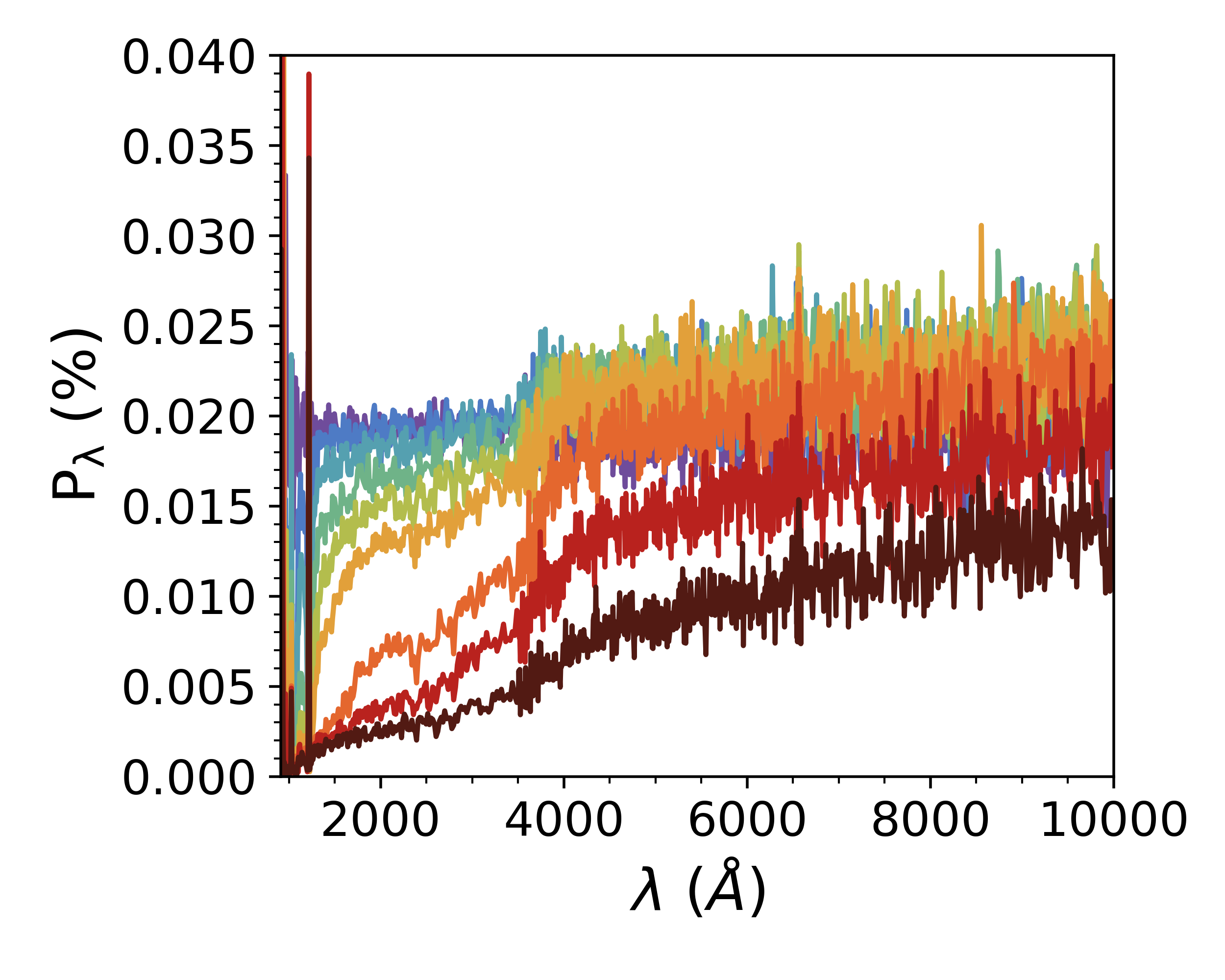}\par\vspace{0.5em}
    \includegraphics[width=\textwidth]{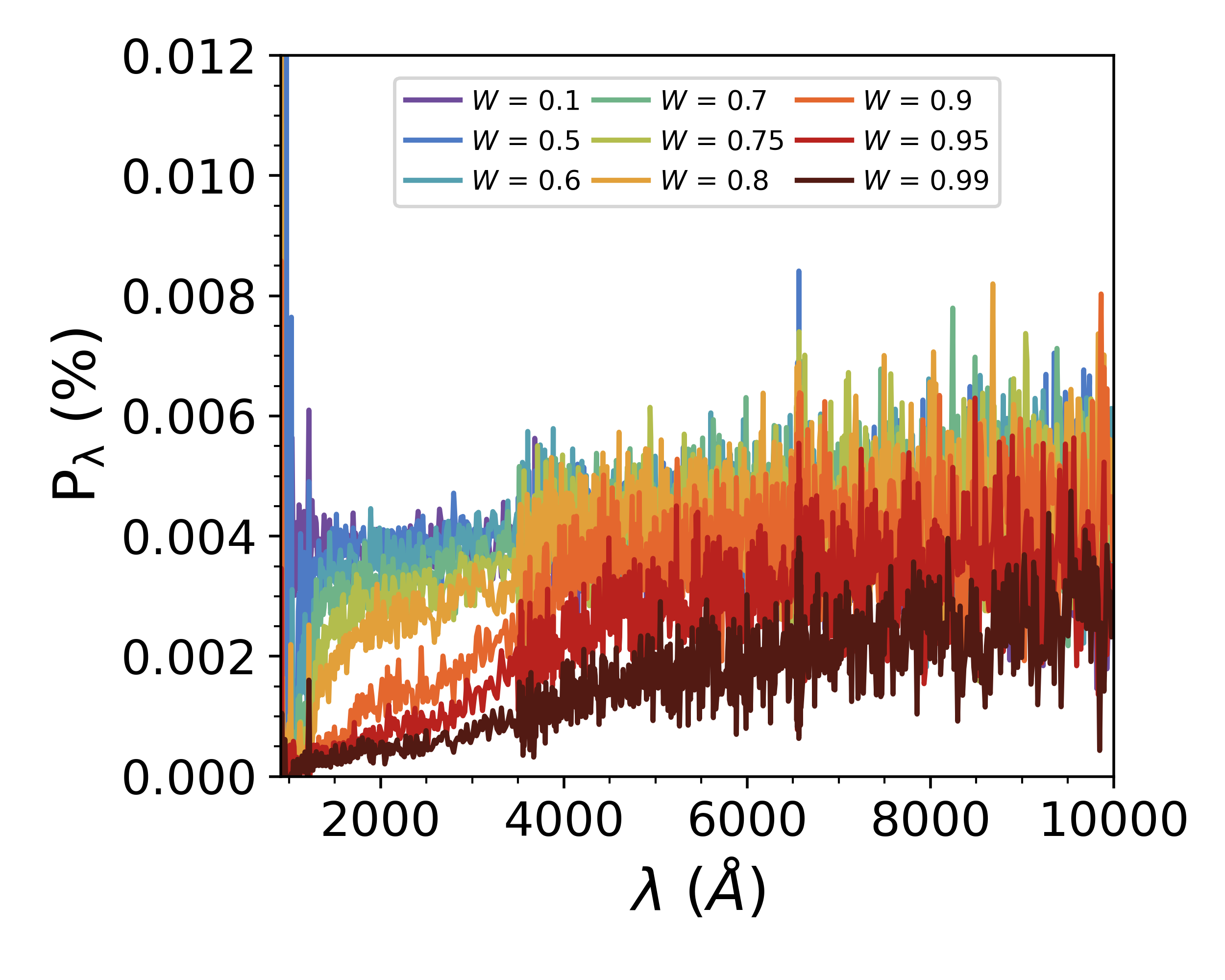}
    \end{minipage}%
    \hfill
    \caption{Same as Figure~\ref{fig:B2_pol}, but for the B8 models.}
    \label{fig:B8_pol}
\end{figure*}

The intrinsic polarization, or polarization degree, of a Be star is produced via Thomson scattering in the disk, and is a powerful tool for modeling disks of Be stars \citep{woo97, car07, hau14}. As with the H$\alpha$ EW, the polarization degree trends with changing rotation rate across the B0, B2 and B5 models are similar, while the B8 models are distinct. Representing the early-type models, Figure \ref{fig:B2_pol} shows the intrinsic polarization averaged across the $V$-band (5000-6500 \AA) as well as the polarized spectrum at an inclination angle of 70$^{\circ}$ (the inclination shown to generally produce the largest polarization degree, see e.g. \citealt{woo96a}) for the B2 case. The same figures for the B0 and B5 models are included in Appendix~\ref{secA3}. When the disk is densest, as we see in the top row of Figure~\ref{fig:B2_pol}, the polarization degree shows a slight to moderate decrease with increasing rotation rate for all inclinations except 89$^{\circ}$. In the top right panel of Figure~\ref{fig:B2_pol}, we see that the polarized continuum levels drop as the star rotates faster and the slope of the continuum, which is initially strongly negative, flattens. 

Moderate density disks surrounding early-type Be stars, shown in the center panels of Figure~\ref{fig:B2_pol}, show more complex behavior. The polarization degree at inclinations $\geq45^{\circ}$ rises with increasing rotation rate for slow or moderate rotators, then drops off for very high rotation rates. The rotation rate associated with the downturn varies from $W=0.8$ to $W=0.9$ depending on the spectral subtype and density. Comparing the polarized spectra of the moderate density disks in the center right panel of Figure~\ref{fig:B2_pol} reveals that the slope of the continuum is strongly responsive to the rotation rate. For low density disks, represented in the bottom row of Figure~\ref{fig:B2_pol}, the trend is the same as for moderate densities, but a higher rotation rate is reached before the polarization degree decreases, and the decrease is more rapid. For these models, the polarized continuum slope becomes positive for the largest rotation rates, instead of remaining negatively sloped or flat. This effect is especially noticeable for $W\geq0.8$. 

The B8 models, shown in Figure~\ref{fig:B8_pol}, show complex behavior that is similar to the moderate and low density models from the early types. The depolarization at high rotation rates is present for all densities but is most pronounced for the moderate and low density cases. For these densities, the right column of Figure~\ref{fig:B8_pol} shows that the changes in polarization level across H\,\textsc{i} ionization thresholds, particularly the Balmer discontinuity at 3650 \AA, are also highly sensitive to stellar rotation rate. Additionally, the slope of the polarized spectrum in the high density disk shown in the top right panel of Figure~\ref{fig:B8_pol} switches signs from negative to positive at the highest rotation rates. The slope for the moderate and low density disks, shown in the center and lower right panels of Figure~\ref{fig:B8_pol}, change from mostly flat to positive as well, for $W\geq0.8$. 

\begin{figure*}[!ht]
\centering
    \begin{subfigure}{0.49\linewidth}
        \includegraphics[width=\linewidth]{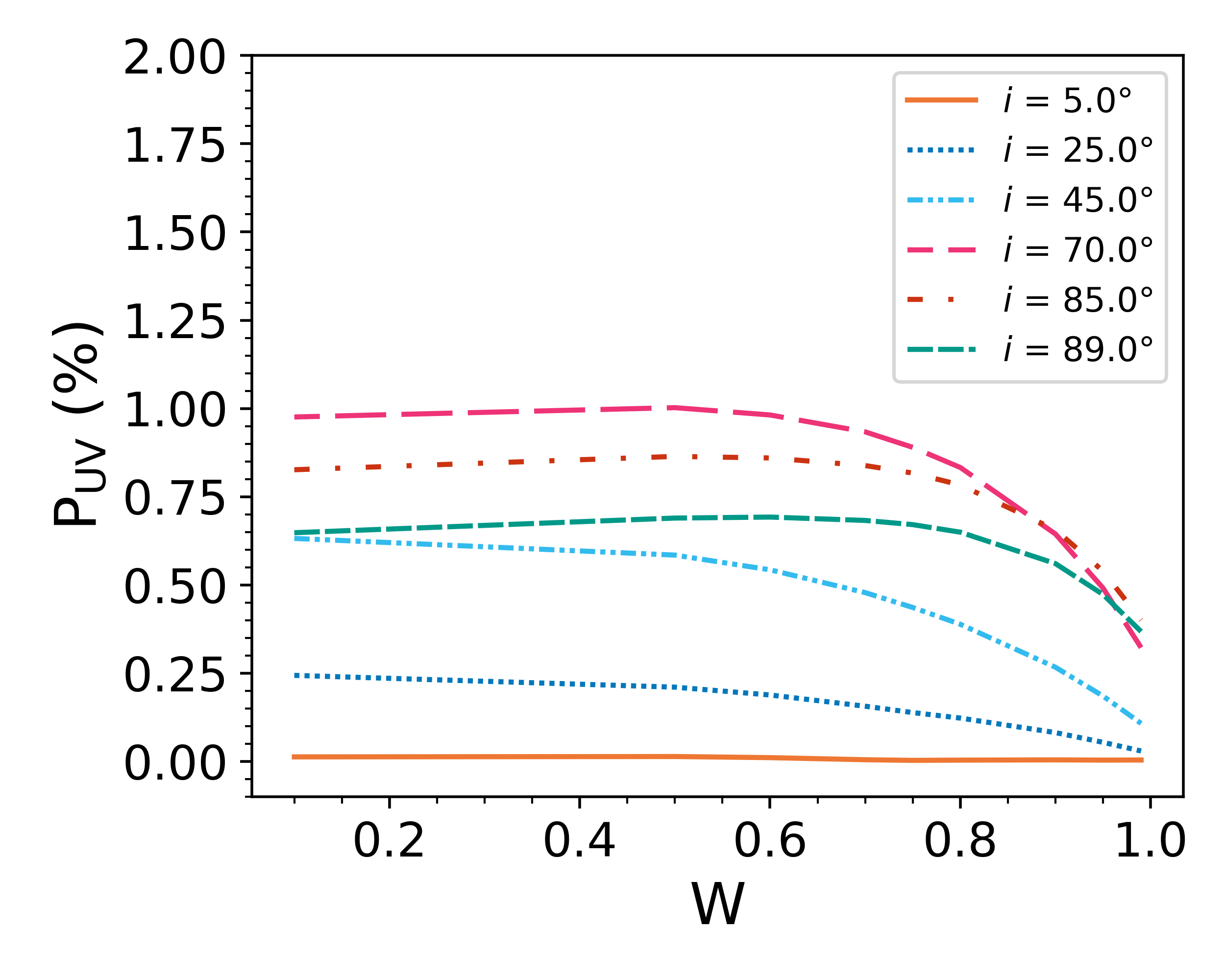}
    \end{subfigure}
    \begin{subfigure}{0.49\linewidth}
        \includegraphics[width=\linewidth]{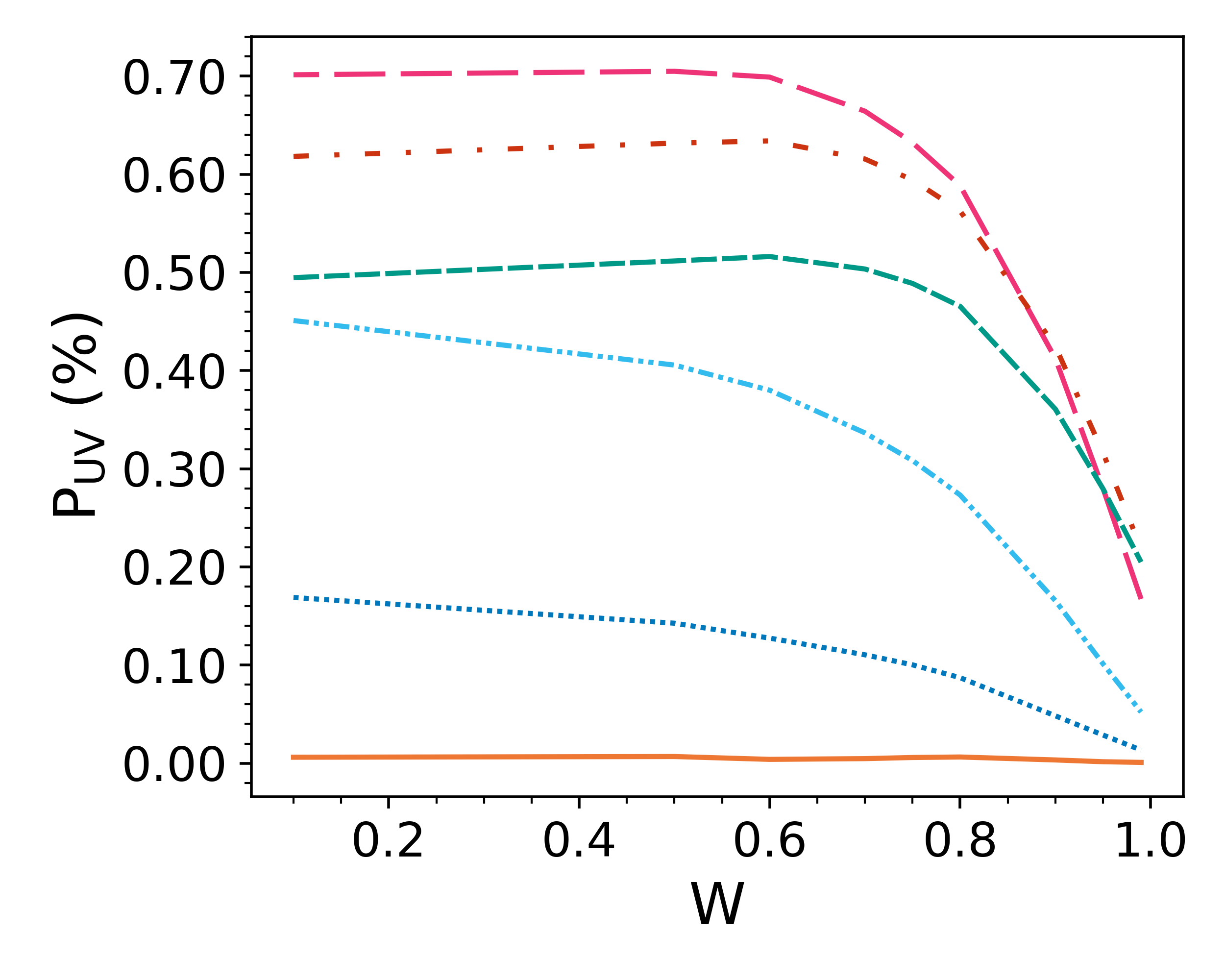}
    \end{subfigure}
    \begin{subfigure}{0.49\linewidth}
        \includegraphics[width=\linewidth]{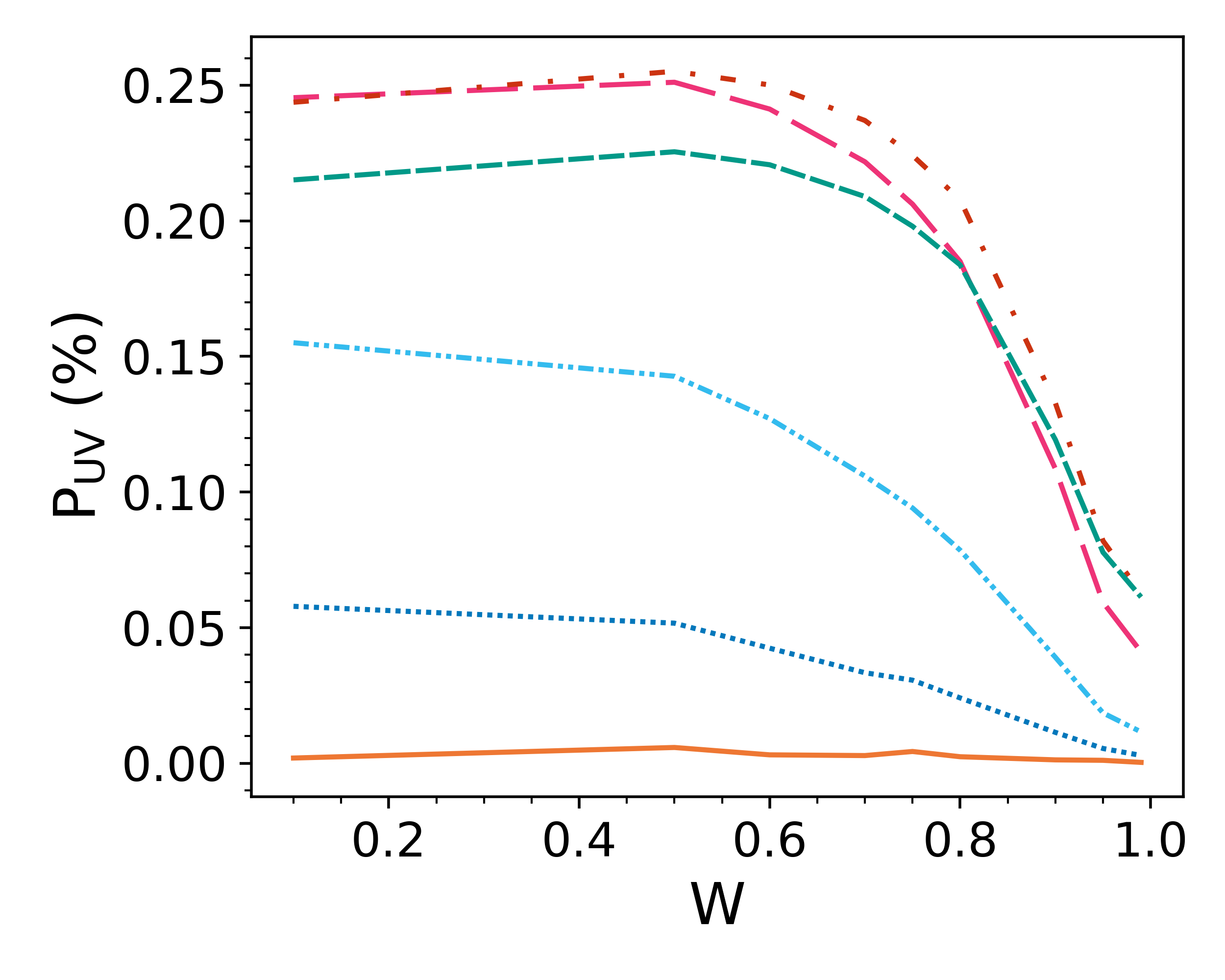}
    \end{subfigure}
    \begin{subfigure}{0.49\linewidth}
        \includegraphics[width=\linewidth]{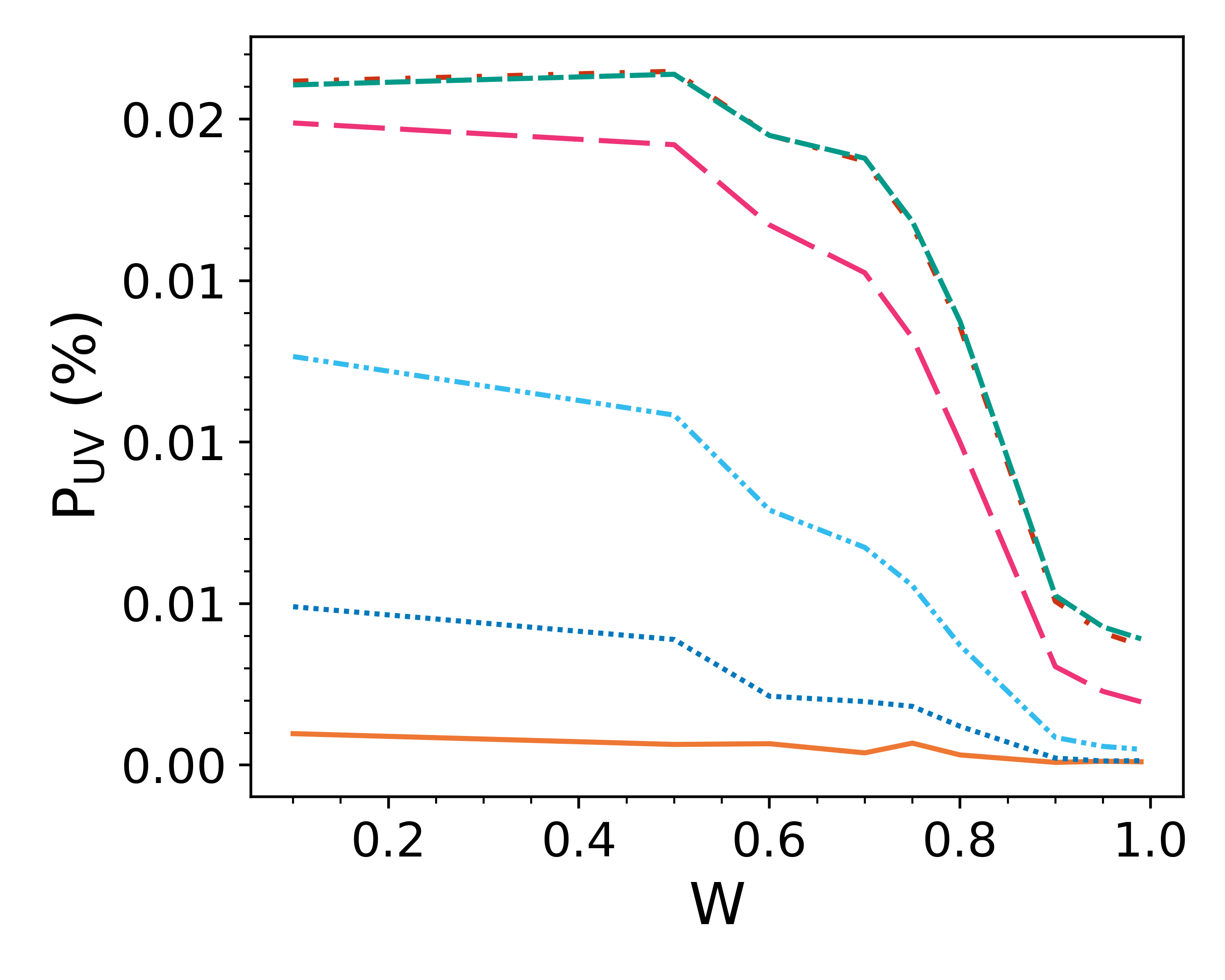}
    \end{subfigure}
\caption{UV polarization, calculated at an effective wavelength of 1500 \AA, for the B0 (top left), B2 (top right), B5 (bottom left), and B8 (bottom right) moderate density models.}
\label{fig:uv_pol}
\end{figure*}

Figure~\ref{fig:uv_pol} shows the UV polarization degree at an effective wavelength of 1500 \AA for the moderate density disks across all tested subtypes. We selected this wavelength because it is the fiduciary wavelength for the proposed Small Mission Explorer concept for UV spectropolarimetry mission {\em Polstar}. The polarization degree increases only minimally for moderate rotation rates, and then changes rapidly at high rotation rates, resulting in more than a 50\% decrease. 

\section{Discussion}
\label{sec4}

As before, we organize our discussion by observable type, considering photometry, emission line profiles, and polarization separately.

\subsection{Magnitude and color}

When no disk is present, classical gravity darkening effects are apparent when comparing the same star at different inclinations. High rotation rates have a dimming effect for stars viewed near edge-on and a brightening effect when the viewing angle is near pole-on. This is clear in the diskless models provided in the top left panel of Figure~\ref{fig:b2_vmag}, where we can see effects on the order of several tenths of a magnitude for rotation rates that exceed roughly $W=0.8$. The dimming effect at high inclination occurs because as $W$ increases, the size of the star's projected ellipse also increases, but this is counterbalanced by the decrease in equatorial temperature so the total brightness remains roughly constant at moderate rotation rates. However, for $W \gtrsim 0.8$, the temperature decrease due to the decrease in local effective gravity at the equator becomes the dominant factor. At the same time, the poles, which we assume to stay roughly constant in radius (e.g., \citealt{col66}), have higher temperatures and larger emitted flux in comparison to the equatorial regions, so stars viewed at low inclinations (near pole-on) will become brighter as they rotate faster. These effects are seen across all spectral subtypes.

The variation in $T_{\rm{eff}}$ with rotation rate at different stellar latitudes can result in measurable changes in a Be star’s color, even when there is no disk. When viewed from low inclinations, near pole-on, the increase in total observed area of the star at faster rotation rates is more significant than when the star is viewed equator-on. The added area at the equators, which is cooler, balances the increase in temperature at the poles and so no rotational effects on color are detectable for stars viewed near pole-on. However, when the star is viewed from high inclinations, the observed area of the star does not change as much with rotation and the equator, which becomes cooler for faster rotation rates, is in full view. This causes fast-rotating stars viewed at high inclinations to appear cooler and redder than one that is rotating more slowly. Together with the change in flux, this creates the tracks seen in Figure~\ref{fig:colormag_fans}. Stars viewed at 5$^{\circ}$ inclination will appear brighter as they rotate faster because the oblate shape causes more flux to be directed toward the poles. At the same time, the star’s color does not change significantly because the hotter temperature of the poles is canceled by the cooler equatorial temperature. This leads to the nearly vertical track seen in the top-left panel of Figure \ref{fig:colormag_fans}. On the other hand, stars viewed at 85$^{\circ}$ or 89$^{\circ}$ will redden and become simultaneously dimmer as they rotate more quickly, since their equatorial temperature drops and the surface is pointed away from the observer due to the star's oblate shape. 

Low density disks do not have a significant impact on the brightness or color of a Be star, regardless of the rotation rate. We see this in the nearly negligible differences between the top left and top right panels of Figure~\ref{fig:colormag_fans}. Since even the densest disks for B8 stars are comparatively diffuse, the photometric response of these systems to rotation rate is dominated by the star itself. However, a sufficiently dense disk contributes to the brightness of the system through bound-free and free-free emission \citep{geh74}. This contribution can begin to significantly affect the system's response to rotation rate at densities achieved by the densest B5 disks, where the maximum densities exceed \SI{e-12}{\gram \per \cubic \centi \meter}. The bulk of this continuum emission is produced in the optically thick region of the disk, which can be described as a ``pseudo-photosphere” \citep{car06a, hau12, vie15}. The size of this pseudo-photosphere increases as a function of wavelength, so emission at longer wavelengths is produced across increasingly larger areas of the disk. The visible excess is produced in the innermost few stellar radii of the disk, depending on disk density \citep{hau12}. The brightness of the disk in this region is dependent on the combined effects of the absorption of stellar radiation, and emission from the gas itself \citep{hau12}. Therefore, we expect to see significant effects from both inclination angle and disk density.

The effects of inclination can be explained through changes in the observed area of the emitting region. When the disk is viewed near face-on, the entire emitting region is observed and since the disks are optically thick along their planes, radiation is preferably toward the poles. Figure~\ref{fig:b2_vmag} shows that at these small inclinations, the addition of a disk causes the entire system to be brighter than when no disk is present. With increasing inclination, an observer can see smaller areas of the disk, and the stellar surface becomes obstructed by the disk which absorbs radiation rather than emitting it. Again, Figure~\ref{fig:b2_vmag} shows that for high inclinations, the star-disk system is dimmer than a diskless star. 

The effects of disk density vary somewhat with spectral subtype, since different spectral subtypes are associated with different base disk densities \citep{vie17} and ionization levels are higher for the hotter, early-type stars. The denser yet more ionized disks around the B0 and B2 stars are able to contribute more continuum emission than in late-type stars. This can mask gravity darkening effects (e.g., the system does not brighten as significantly with increasing W when viewed pole-on). 

The increase in the disk mass for our models, discussed in Section~\ref{sec2}, impacts the color of the systems with moderate and high disk densities, particularly at low inclinations. When the star-disk system is viewed near face-on, we would expect minimal changes in the color as the rotation rate increases. However, as seen in the upper right panel of Figure~\ref{fig:colormag_fans}, the systems become redder as the stars rotate faster. This is likely due to an increase in the mass of the disk which reddens the total observed flux. 

\subsection{H$\alpha$ emission line}

The effects of rotation rate on the H$\alpha$ line are primarily due to three factors: the disk density, the amount of ionizing radiation traveling through the disk, and the inclination angle at which the star-disk system is observed. The first two factors are interdependent, since disk density varies with the subtype of the Be star. The amount of ionizing radiation emitted by the Be star depends on its effective temperature, which in turn affects the ionization levels of the disk and therefore the emission strength. As $W$ becomes larger, the $T_{\rm{eff}}$ at the equator, which dominates the surface area, drops and the mean intensity becomes redder, so the amount of ionizing radiation traveling through the disk diminishes. This results in lower ionization levels in the disk, and therefore a higher H\,\textsc{i} bound-bound opacity. This can have important effects on the H$\alpha$ line, if the disk is sufficiently dense. The inclination angle of the system determines the relative strength of the continuum and disk emission through changing the region of the stellar surface (e.g. pole or equator), and surface area of the disk, that is visible to the observer. 

At low disk densities, the rotational effects on the H$\alpha$ EW differ by spectral subtype. The B0 stars generally show stronger emission for faster rotation, except when viewed at low inclinations (see the top left panel of Figure~\ref{fig:B0_B5_ew}). However, the emission seen in the B2 stars becomes weaker with increasing rotation rate for all inclinations as demonstrated by the bottom left panel of Figure~\ref{fig:B2_ew}. Conversely, as we see in the bottom left panels of Figure~\ref{fig:B0_B5_ew}~and~\ref{fig:B8_ew}, the EW values for the B5 and B8 models alike are largely unaffected by changes in the rotation rate. For these cool stars, the emission features are barely present, and any increase in disk emission is counteracted by the broadening of the H$\alpha$ absorption contribution from the photosphere.

\begin{figure*}[!ht]
\centering
    \begin{subfigure}{0.49\textwidth}
    \includegraphics[width=\textwidth]{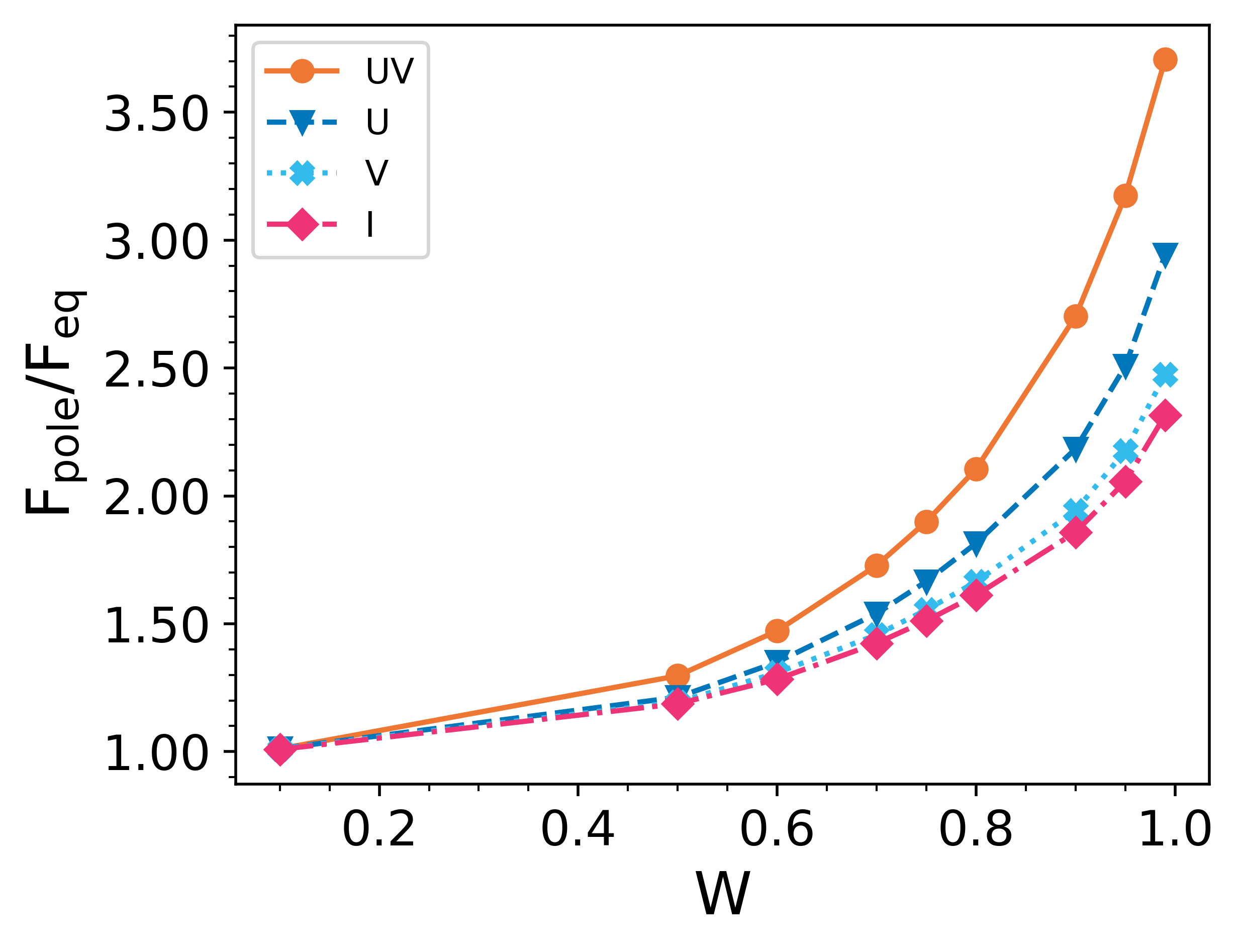}
    \end{subfigure}
    \begin{subfigure}{0.49\textwidth}
    \includegraphics[width=\textwidth]{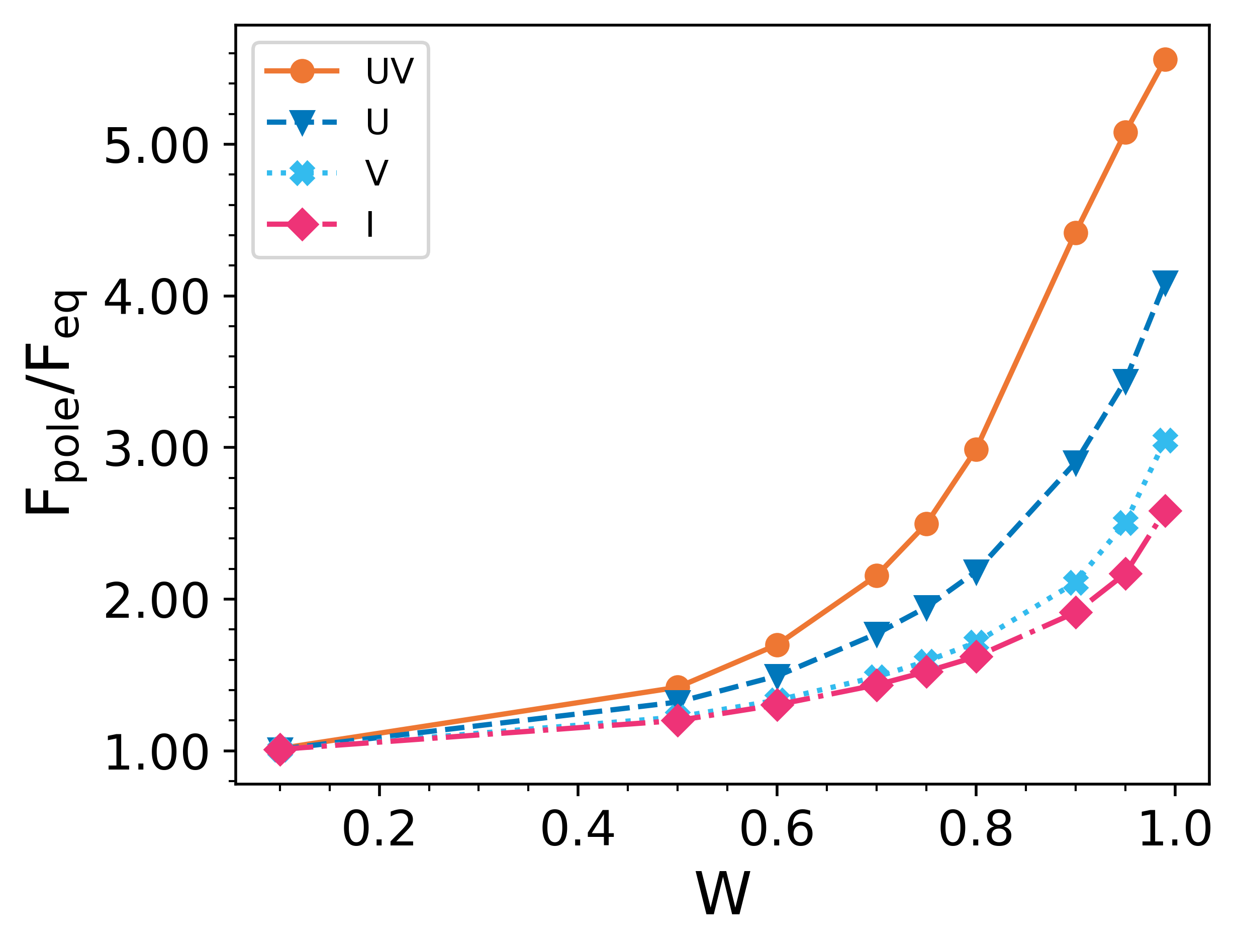}
    \end{subfigure}
    \begin{subfigure}{0.49\textwidth}
    \includegraphics[width=\textwidth]{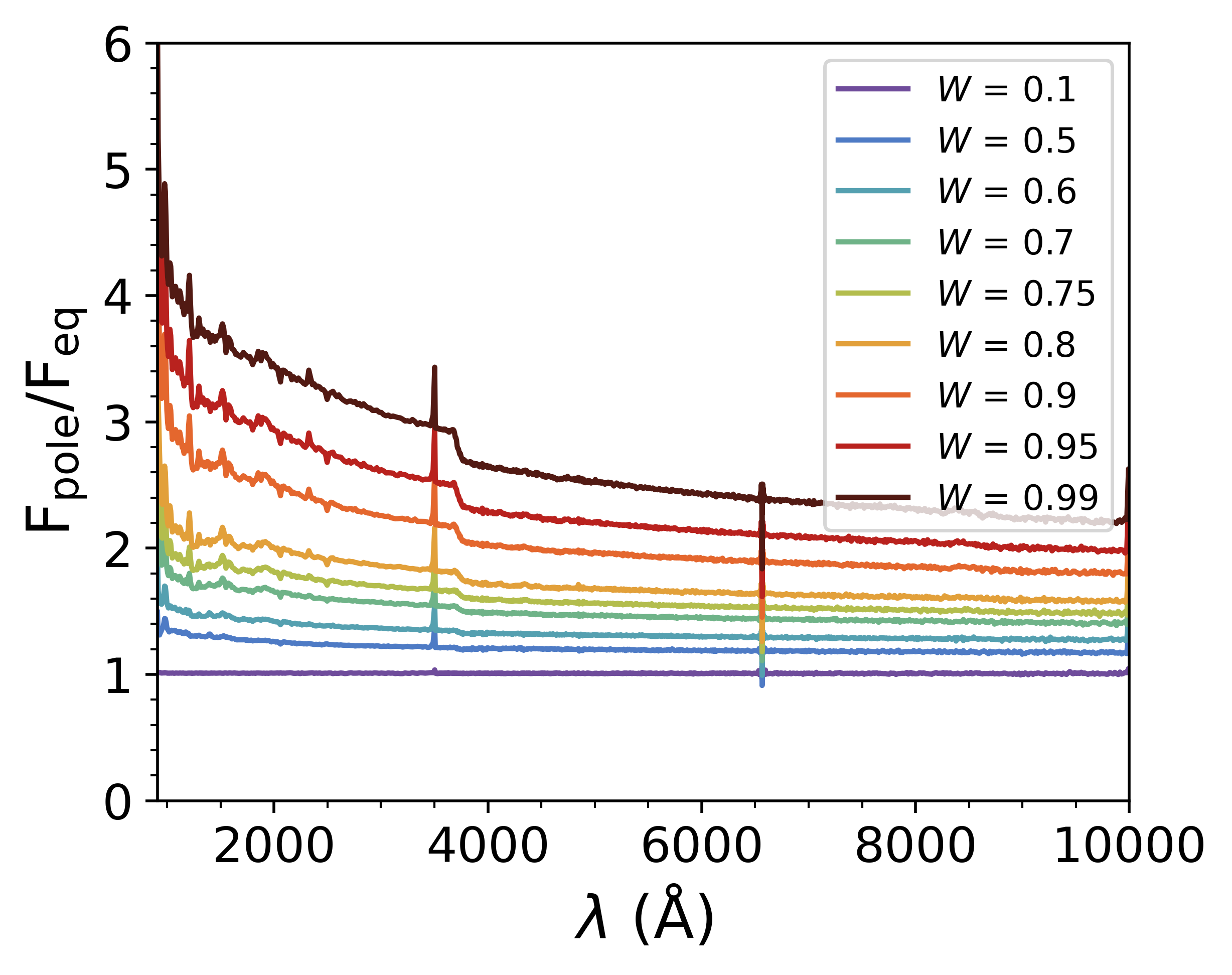}
    \end{subfigure}
    \begin{subfigure}{0.49\textwidth}
    \includegraphics[width=\textwidth]{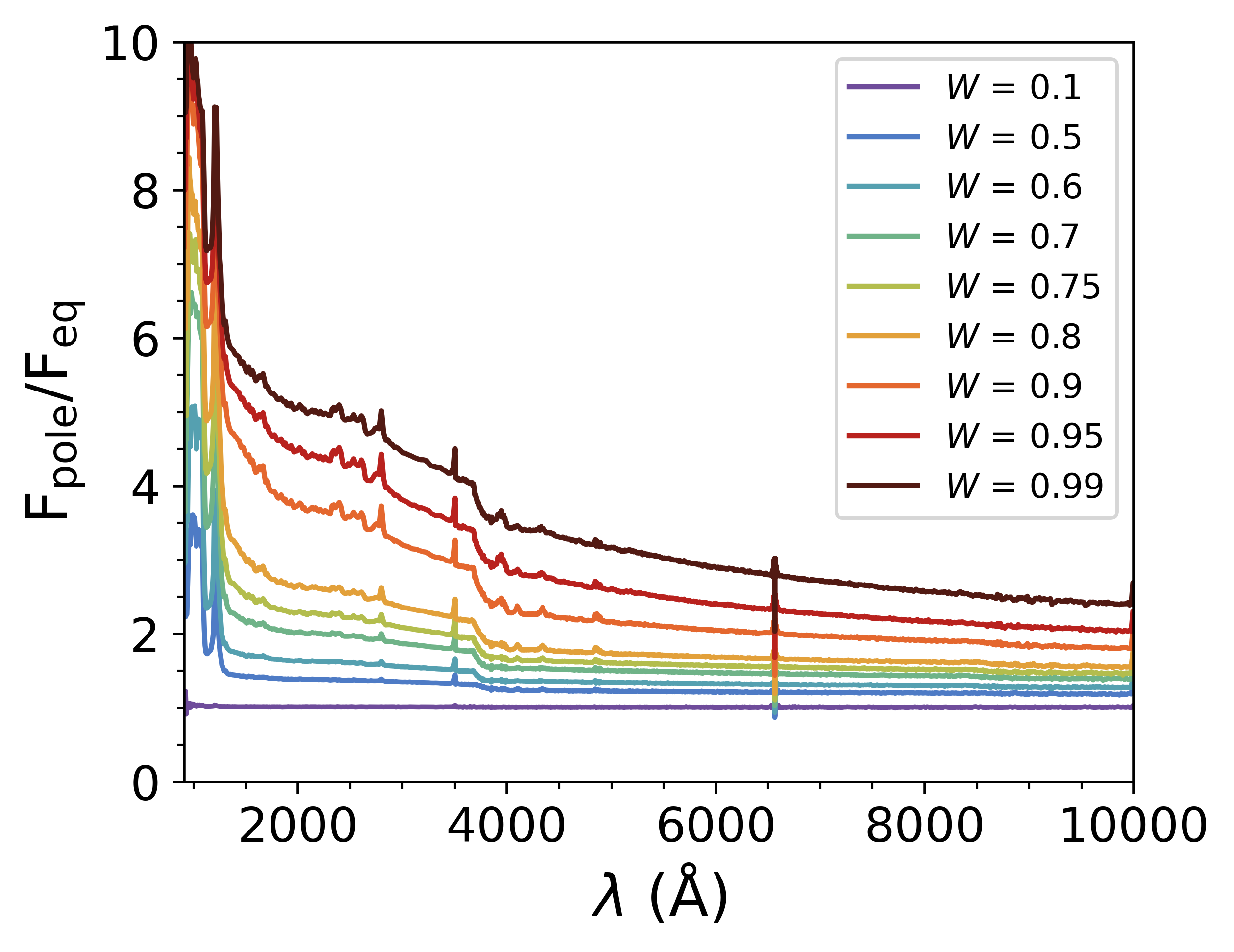}
    \end{subfigure}
\caption{Top row: Ratio of the polar flux, as measured at 85$^\circ$, to the equatorial flux, as measured at 5$^\circ$, for models with no disks and plotted as a function of $W$. We show the B0 (left), and B8 (right) models, which show the least and greatest changes in flux ratio in response to rotation rate, respectively. The flux ratio is calculated at the effective wavelengths for the UV (1500 \AA) as well as the $U$- (3600 \AA), $V$- (5450 \AA) and $I$- (7900 \AA) bands, as indicated in the legend. Bottom row: Ratio of the polar flux to the equatorial flux, as a function of wavelength for the B0 (left) and B8 (right) diskless models.}
    \label{fig:fpol_feq}
    \end{figure*}
  
For disks with moderate densities, all spectral subtypes except B8 show the same rotational effects: faster-rotating stars produce diminished emission at small inclinations, and enhanced emission at large inclinations. This can be attributed to changes in the continuum flux at different viewing angles. At small inclinations, the star is directed near pole-on toward the observer and the disk is nearly face-on. For larger rotation rates, the poles become hotter and contribute to an increased continuum flux. This is illustrated in Figure~\ref{fig:fpol_feq}, which shows that the polar flux (F$_{\rm{pole}}$) can exceed the equatorial flux (F$_{\rm{eq}}$) by a factor of two to three at the effective wavelength for the $V$-band, or three to four at the effective wavelength for the $U$-band and even stronger at 1500 \AA. The elevated continuum flux for near-critical rotation at small inclinations, then, results in a smaller normalized H$\alpha$ line and therefore smaller EW. Conversely, at high inclinations when the star-disk system is viewed edge-on, the flux from the hotter, brighter poles is directed away from the observer, while the flux from the cooler equatorial regions becomes lower for increasing rotation rates. As a result, the continuum levels are inversely proportional to the rotation rate, as we see in the right column of Figure~\ref{fig:B2_unnormalized}. At the same time, the dropping ionization levels in the disk at faster rotation rates cause the bound-bound opacity to increase. Together, these effects result in the enhanced H$\alpha$ emission for moderate density disks at high inclinations. 

The effect of inclination on emission strength is still more prominent among high density disks. Here, as before, the early-type models show the most striking differences between slowly- and fast-rotating stars. The disk's H$\alpha$ emission, which originates in the optically thick region of the disk, scales with the square of the density. As a result, the densest disks produce the most prominent H$\alpha$ lines. The early-type models have the densest disks, which explains why they show the most significant changes with rotation rate. The effects of rapid rotation are particularly striking on the normalized emission profiles of such star-disk systems when viewed at high inclinations because the ratio of the line flux to the continuum flux becomes very large and the effects on opacity, causing the EW to become very strongly negative. Meanwhile, the EWs of H$\alpha$ lines seen at inclinations $\leq$70$^\circ$ behave the same as their moderate density counterparts. 

The unique behavior of the B8 models can be explained in terms of their disk densities. Since disk density scales with spectral subtype, becoming weaker for cooler stars, the B8 systems were modeled with lower density as seen in Table~\ref{tab:parameter_summary}. As a result, only the highest-density disks among them have sufficient density to produce significant emission, as seen in the right panel of Figure \ref{fig:B8_ew}. Moderate and low density disks have EW values in absorption and show only weak emission features on their line profiles. When the system is viewed at small inclinations, corresponding to large disk surface areas, the absorption is weaker, or the emission is stronger, than when viewed from moderate to high inclinations, which have lower disk surface areas. Since the ratio between the continuum and the line profile fluxes remains small, even for large inclinations, the EW value for a slowly-rotating star and fast-rotating star at the same inclination is roughly the same. Small changes in line morphology, most significantly the narrowing of the emission features, are still noticeable.

\subsection{Intrinsic polarization}

Changes in the stellar rotation rate impact the predicted polarization in two key ways: altering the overall polarization levels, and modifying the slope of the polarized spectrum. Both of these quantities are quite responsive to variations in the rotation rate. While the models show a diverse range of phenomena, each can be explained through a careful analysis of the system geometry, gravity darkening, and the effects of disk density. 

\begin{figure}
    \centering
    \includegraphics[width=0.7\linewidth]{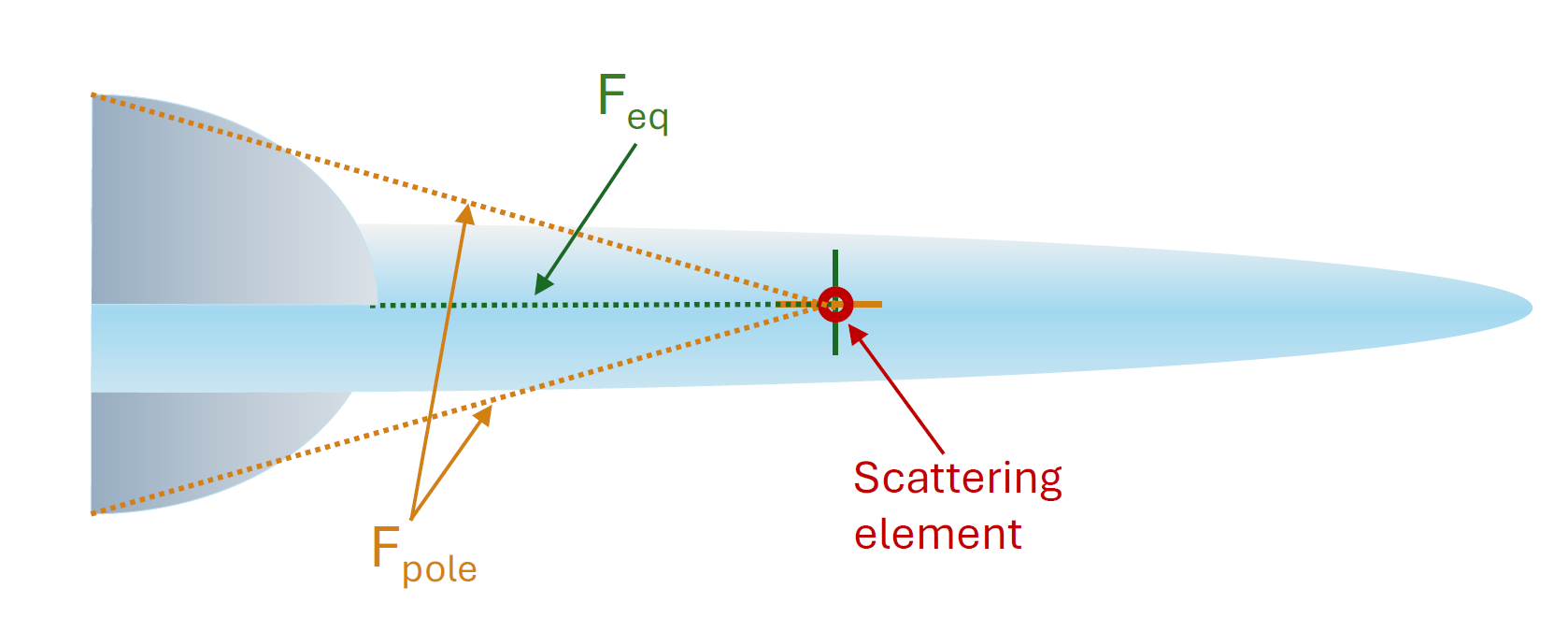}
    \caption{Cartoon showing the directions of the polar and equatorial flux, as well as their scattering planes as they intersect at a scattering element in the disk.}
    \label{fig:pol_visual}
\end{figure}

Two effects contribute to the downturn in polarization at very high rotation rates seen at both visual wavelengths, as seen in Figures~\ref{fig:B2_pol} and \ref{fig:B8_pol}, and UV wavelengths as demonstrated in Figure~\ref{fig:uv_pol}. The first is related to the amount of radiation the disk receives from the stellar poles versus the equator at different stellar rotation rates. The amount of radiation heating a scattering element in the disk, as seen in Figure~\ref{fig:pol_visual}, depends on the ratio F$_{\rm{pole}}$/F$_{\rm{eq}}$. As demonstrated by the bottom row of Figure~\ref{fig:fpol_feq}, this ratio increases with $W$ and is especially strong at short wavelengths. If the majority of the scattered radiation is supplied by the stellar equator, the simultaneous decrease in scattered equatorial flux and increase in direct, unpolarized polar flux results in a lower polarization level, particularly in the UV. Such an effect was suggested by \citet{bjo91}. The second effect contributing to the decline in polarization at high rotation rates can be understood through considering the multiple scattering planes producing the total observed polarization degree, which are at different orientations. Flux from the stellar equator has a scattering plane perpendicular to the disk, while the flux from the poles is scattered in a plane parallel to the disk. As $W$ increases and the polar flux dominates the equatorial flux, the total detected polarization (which is a vector sum of the two polarization vectors) will be partially canceled. The effects of the flux ratio and the different scattering planes work together to produce the same outcome. In short, gravity darkening depolarizes the light; the faster the star rotates, the stronger the depolarization effect. This has been noted at UV wavelengths by \citet{bjo94}, and here we confirm that it can also occur in the $V$-band polarization. 

The increase in disk mass in our models as $W$ increases results in a slight enhancement in the polarization levels, however the overall depolarization trends are not affected. We note that the \textsc{hdust} models do not include the effects of metal line blanketing, which are more significant at UV wavelengths. Therefore, the UV polarization values seen in Figure~\ref{fig:uv_pol} likely represent the upper limit for the disk densities we present. 

In addition to the polarization levels, the slope of the polarized spectrum can yield useful information about the rotation rate of Be stars. This slope has been referred to as the ``polarization color" by \citet{hau14}, and here we adopt the same terminology. As mentioned previously, we find that the densest disks in our grid of models always have a negative polarization color (see the top right panel of Figure~\ref{fig:B2_pol}), while color in more tenuous disks can become positive at high rotation rates in both the Balmer (912-3647 \AA) and Paschen (3647-8207 \AA) continua (see the bottom right panel of Figure~\ref{fig:B2_pol}, and all B8 models shown in the right column of Figure~\ref{fig:B8_pol}). These predicted observations can be explained through the interplay of two effects. The first is the wavelength-dependent H\,\textsc{i} bound-free opacity, which becomes larger with increasing rotation rate as the ionization levels in the disk drop, and which contributes a negative polarization color. The second is the ratio between F$_{\rm{pole}}$ and F$_{\rm{eq}}$. The bottom row of Figure~\ref{fig:fpol_feq} shows a negative slope when the ratio is plotted as a function of wavelength, so it causes a positive slope in the polarization spectrum since the depolarizing effect is strongest where the flux ratio is largest. Therefore, this gravity darkening effect works to produce a positive polarization color. The complex behavior in all our models can be explained through this duo scheme, where the effects of bound-free opacity and gravity darkening are competing, and the factor that dominates depends on the disk density. Because density drops in later spectral types, we can refer to our early-type models to represent the high density cases, and the late-type models to represent the low density extremes. 

When the disk is very dense, the polarized spectrum is dominated by the bound-free opacity and shows strong discontinuities at the H\,\textsc{i} ionization energies. In this case, represented by the top right panel in Figure~\ref{fig:B2_pol}, increasing $W$ results in an increased bound-free opacity and therefore a net negative polarization color. The slope also flattens somewhat due to the effects of gravity darkening, but the disk density is high enough for the hydrogen opacity to overcome these effects.

However, in the moderate and low density regimes (see e.g. the entire right column of Figure~\ref{fig:B8_pol}), the wavelength independent electron scattering becomes the dominant opacity. As discussed in \citet{hau14}, we might then expect the polarized spectrum to be flat, with only small discontinuities at the wavelengths corresponding to hydrogen ionization energies. As seen in the center and lower panels of Figure~\ref{fig:B8_pol}, the lower-density models are indeed relatively flat for low stellar rotation rates. Since the bound-free opacity is limited for these models, the effects of gravity darkening are able to produce positive polarization colors. This behavior was noted by \citet{kle15} when modeling the disk surrounding the late-type Be star $\beta$ Cmi. Some models, such as the highest-density B8 model shown in the top row of Figure~\ref{fig:B8_pol}, are dominated by the hydrogen opacity and have negative polarization color for lower rotation rates, but the polarization color in the Paschen continuum completely changes its sign to become negative and dominated by gravity darkening at very high stellar rotation rates. 

\begin{figure*}[!ht]
\centering
    \begin{subfigure}{0.49\textwidth}
    \includegraphics[width=\linewidth]{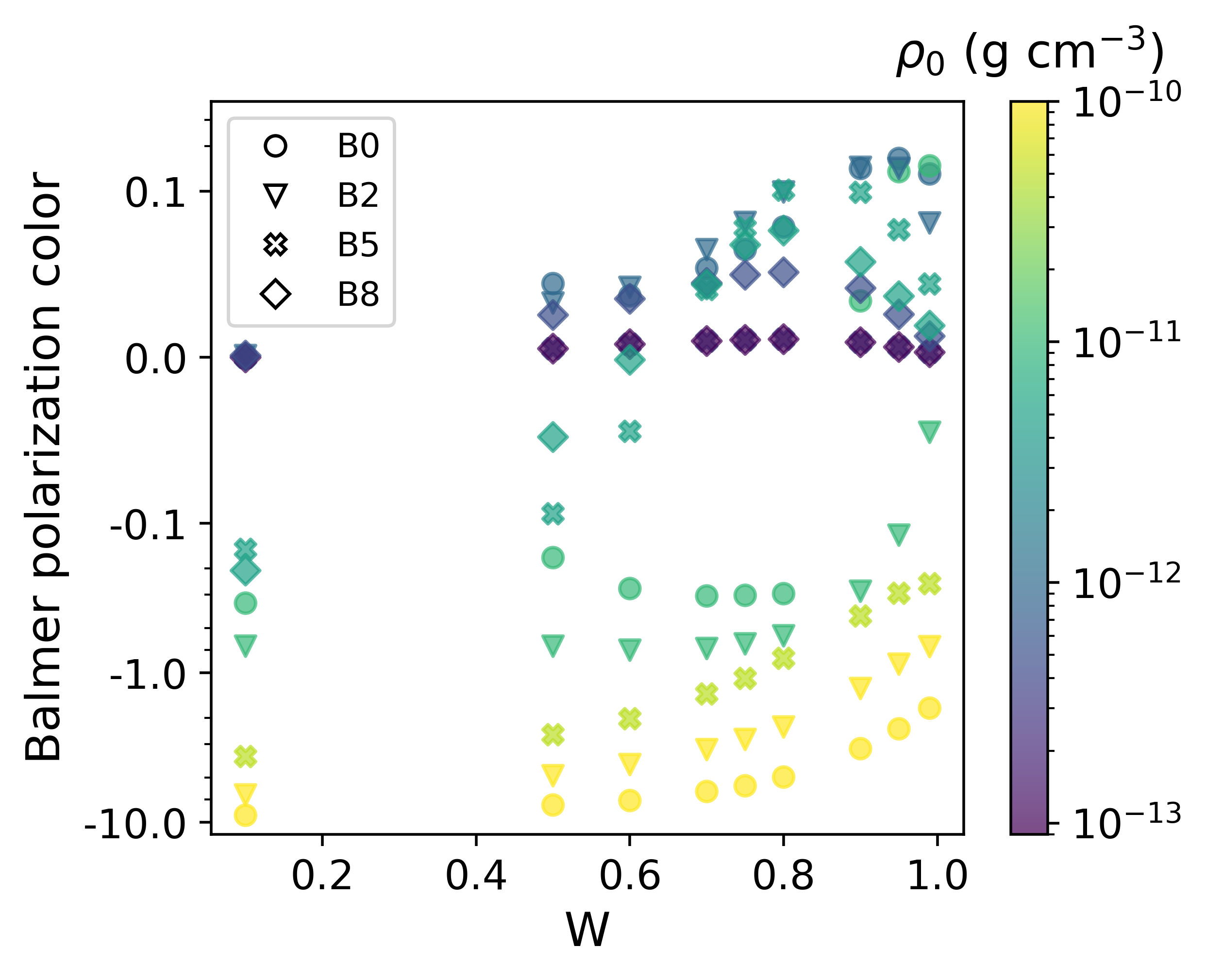}
    \end{subfigure}
    \begin{subfigure}{0.49\textwidth}
    \includegraphics[width=\linewidth]{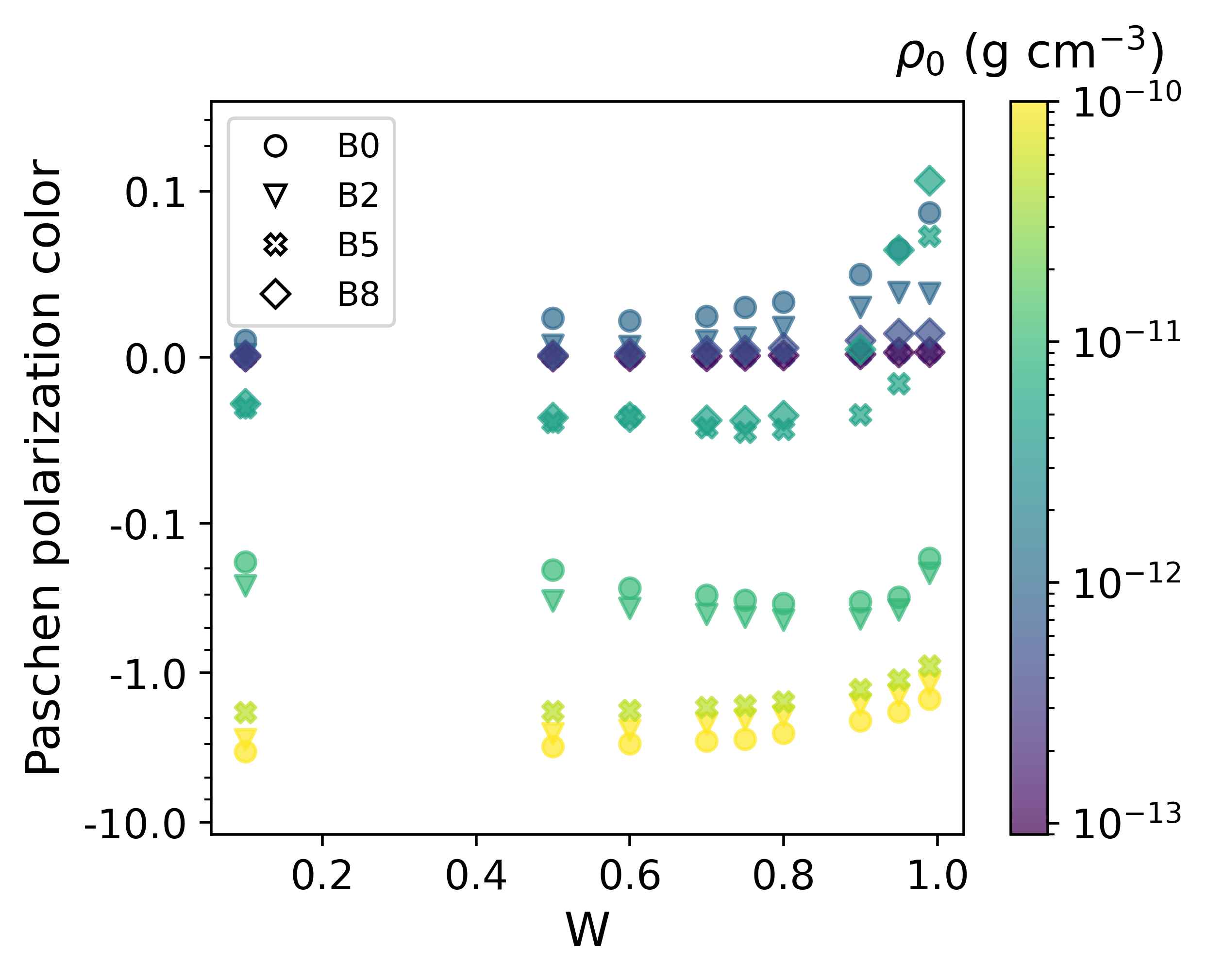}
    \end{subfigure}
\caption{Polarization color in the Balmer continuum (left) and Paschen continuum (right), as a function of $W$ at an inclination of 70$^{\circ}$. We include trends for all tested spectral subtypes, and the density is indicated in the colorbar. Note the change in y-axis scale, from linear between $\pm$ 0.1, and logarithmic for values outside this range.}
    \label{fig:pol_color}
    \end{figure*}

The utility of the polarization color as a diagnostic tool for stellar rotation rate is illustrated in Figure~\ref{fig:pol_color}. In the Balmer continuum, shown in the left panel, positive polarization color peaks between $W=0.7$ and $W=0.8$ in the gravity darkening dominated, low density regime. The moderate density regime is sometimes dominated by opacity, and other times dominated by gravity darkening, depending on the spectral type of the Be star. The high density region is dominated by opacity, but still shows some gravity darkening effects. The polarization color in the Paschen continuum, shown in the right panel of Figure~\ref{fig:pol_color}, is most responsive to rotation rate at $W\geq0.8$. Together, the polarization color in the Paschen and Balmer continua can be used to estimate the rotation rate of a Be star, if the spectral type of the Be star is known.

Our predicted polarization values at different inclinations show good agreement with the expected inclination dependence based on post-scattering absorption. This phenomenon has been discussed in detail by \citet{woo96a}, and was also observed in protoplanetary disks by \citet{whi92} as well as in several other studies on Be star disks including \citet{wat92}, \citet{hal13}, and \citet{hau14}. When the disk is viewed at very large inclinations, it is oriented so that the observed scattered flux has passed through the dense disk. The majority of the scattered photons will escape through the polar regions, which are less optically thick, so fewer photons escape from the directions near the equatorial plane. These photons must pass through more of the disk and a larger fraction of them are reabsorbed, decreasing the overall polarization level. The maximum polarization, then, occurs near an inclination angle of 70$^\circ$ rather than 90$^\circ$. While not an effect of gravity darkening, this observed effect in our models is a good confirmation of known theory. 

\section{Conclusions}\label{sec5}

As rapidly rotating stars, understanding the effects of gravity darkening on the observables produced by Be stars is key in interpreting their physical properties as well untangling the role of rotation in the Be phenomenon itself. Using radiative transfer models, we investigate the rotational effects on observables produced by the star-disk system across a wide range of spectral subtypes and disk densities. We find that the magnitude, color, H$\alpha$ EW, and polarization produced by the system can all be significantly impacted by rotation rate, especially near the critical value, and these effects must be considered when interpreting observations of Be stars. 

Gravity darkening effects on Be star brightness are most significant for models with low or moderate disk densities. Since late-type Be stars tend to have lower-density disks, we expect these stars to be most affected by the effects of rapid rotation on magnitude. For these systems, the disk does not contribute as significantly to the overall brightness of the system, and instead the brightness of the system is controlled by the relative flux received from the equator and the poles. Since this ratio is wavelength dependent, rotational effects are more significant in the $U$ and $V$ bands compared to longer wavelengths. The effects on magnitude are also inclination-dependent, and stars viewed at low inclinations may appear dimmer, while stars viewed at large inclinations may appear brighter, than their slowly-rotating counterparts. Rapid rotation can also cause reddening and the appearance of being more evolved when the star-disk system is viewed near edge-on, while rotational effects on color are insignificant at low inclinations. For systems with high-density disks, the disk contributes a higher percentage of the brightness and introduces a stronger inclination dependence at low rotation rates. This can slightly conceal the effects of rapid rotation close to critical rates. Since early-type Be stars tend to have denser disks, we expect the impacts on color and magnitude to be less significant for them as a whole. 

By contrast, the rotational effects on H$\alpha$ EW are most significant for Be stars with disks near the upper limit of observed densities. We therefore expect these impacts to be stronger on early-type Be stars, which tend to have denser disks. When these star-disk systems are viewed near face-on, the H$\alpha$ emission strength diminishes with increasing rotation rate due to a rise in the continuum level from the brightening of the poles. Based on the EW of the normalized H$\alpha$ line alone, this may cause a disk of a Be star rotating near the critical limit to appear less dense than it truly is. When dense disks are viewed near edge-on, the H$\alpha$ emission strength can be significantly enhanced. In this case, the H$\alpha$ EW will give the appearance of a much denser disk.

Rapid rotation has an attenuating effect on the polarization produced by the disk. This effect is especially significant at UV wavelengths, but is also seen in the $V$-band polarization. The depolarization at high rotation rates is primarily due to the combined effects of the changing flux ratio from the equator and poles with rotation rate, and the geometry of the scattering planes of the flux from different stellar latitudes. The slope of the polarized continuum (the polarization color) is also sensitive to the rotation rate through the competing effects of gravity darkening and changes in the H\,\textsc{i} opacity in the disk. Stars rotating near critical rates exhibit diminished polarization levels, particularly at UV wavelengths. The slope of polarized continuum flattens or becomes negative with increasing rotation rate, particularly across the Balmer continuum. As a result, we find that the polarization color can be used as a diagnostic tool to determine the rotation rate of a Be star.

Due to the wavelength-dependent effects of rotation, future observations of Be stars at short wavelengths will be important in investigating these stars as rapid rotators. Further monitoring of Be stars at visual and UV wavelengths could utilize polarization color to better constrain the observed stellar rotation rates. This could be achieved through new mission concept designs such as {\em Polstar}, a UV spectropolarimeter designed to study rapid rotation in massive stars \citep{sco25}, {\em Pollux}, a high-resolution spectropolarimeter capable of observing UV, visible and infrared wavelengths for the Habitable Worlds Observatory (HWO, \citealt{nas21, mus24}), or {\em Arago} \citep{mus23}, a UV and visible spectropolarimeter proposed to ESA. 

\backmatter


\bmhead{Acknowledgements}

We thank the anonymous referee for their comments which have improved this manuscript. CEJ and RGR acknowledge support from the Natural Sciences and Engineering Research Council of Canada. This work was made possible through the use of the Shared Hierarchical Academic Research Computing Network (SHARCNET). AuD acknowledges support from NASA through Chandra Award number TM4-25001A issued by the Chandra X-ray Observatory 27 Center, which is operated by the Smithsonian Astrophysical Observatory for and on behalf of NASA under contract NAS8-03060. ACC acknowledges support from the Conselho Nacional de Desenvolvimento Cient\'ifico e Tecnol\'ogico (CNPq, grant 314545/2023-9) and the S\~ao Paulo Research Foundation (FAPESP, grants 2018/04055-8 and 2019/13354-1). This research has made use of the SIMBAD database operated at CDS, Strasbourg (France), and of NASA’s Astrophysics Data System (ADS). 

\section*{ORCID iDs}
\scriptsize
RGR \url{https://orcid.org/0009-0007-9595-2133} \\
CEJ \url{https://orcid.org/0000-0001-9900-1000} \\
MWS \url{https://orcid.org/0000-0003-0696-2983} \\
JLB \url{https://orcid.org/0000-0002-2919-6786} \\
AuD \url{https://orcid.org/0000-0001-7721-6713} \\
ACC \url{https://orcid.org/0000-0002-9369-574X} \\
PQ \url{https://orcid.org/0000-0001-5611-9340} \\
CN \url{https://orcid.org/0000-0003-1978-9809} \\
JD \url{https://orcid.org/0000-0002-0210-2276}

\begin{appendices}

\section{Supplementary $V$ magnitude figures}\label{secA1}

This section contains supporting figures for the spectral types not included in the main results shown in Section~\ref{sec:Vband_results}. 

\begin{figure}[!ht]
\centering
   \begin{subfigure}{0.49\columnwidth}
        \includegraphics[width=\columnwidth]{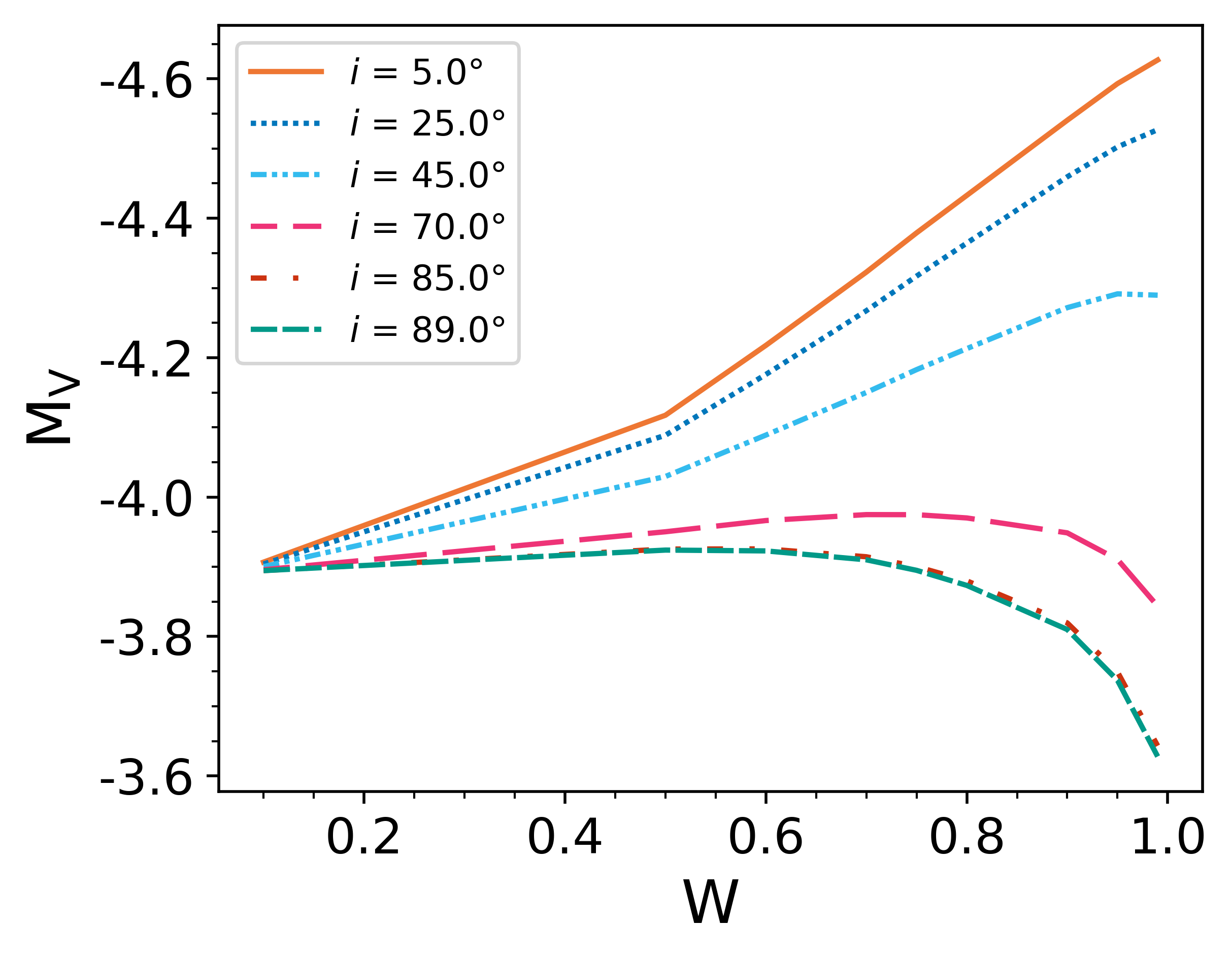}
    \end{subfigure}
    \begin{subfigure}{0.49\columnwidth}
        \includegraphics[width=\columnwidth]{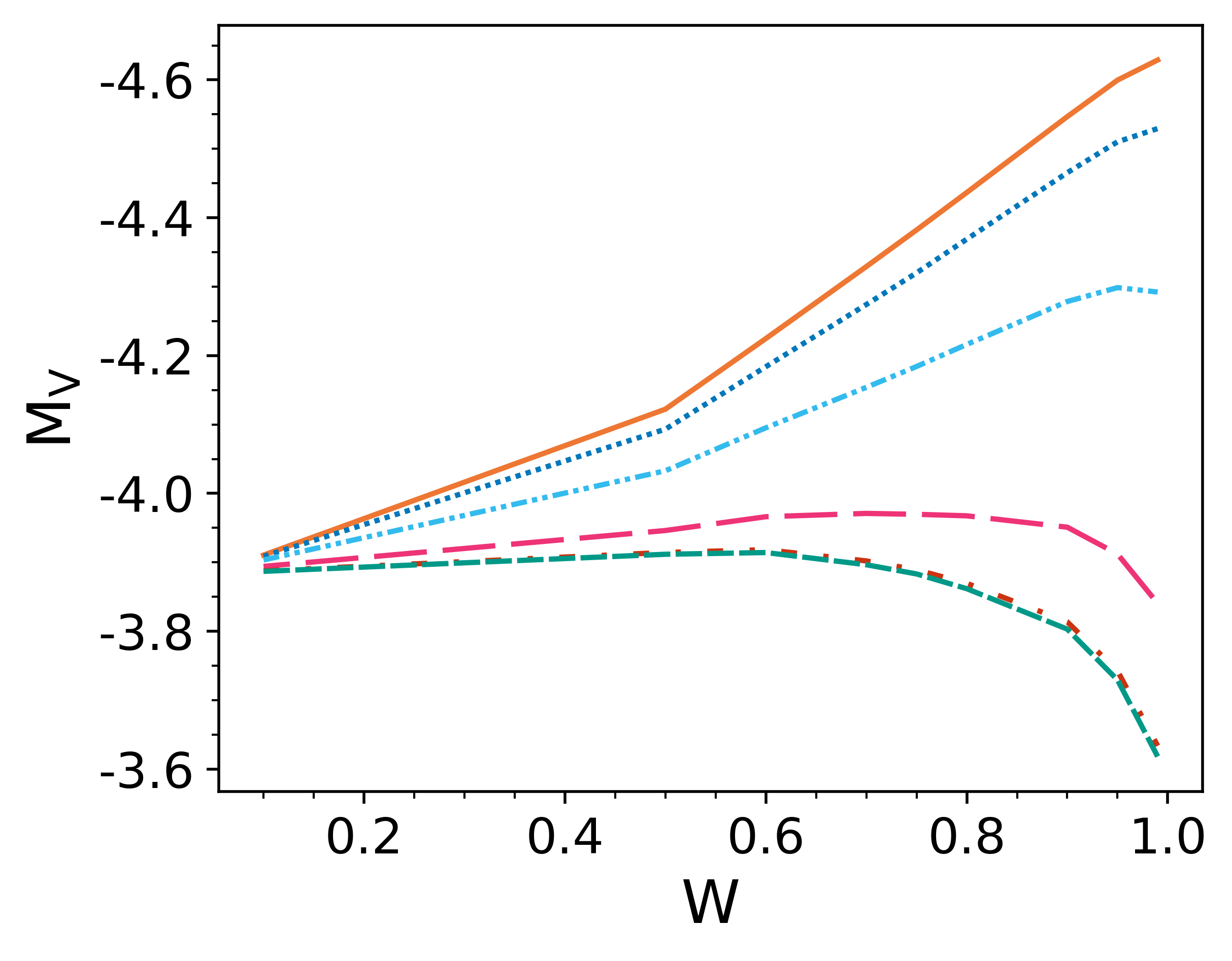}
    \end{subfigure}
    \begin{subfigure}{0.49\columnwidth}
        \includegraphics[width=\columnwidth]{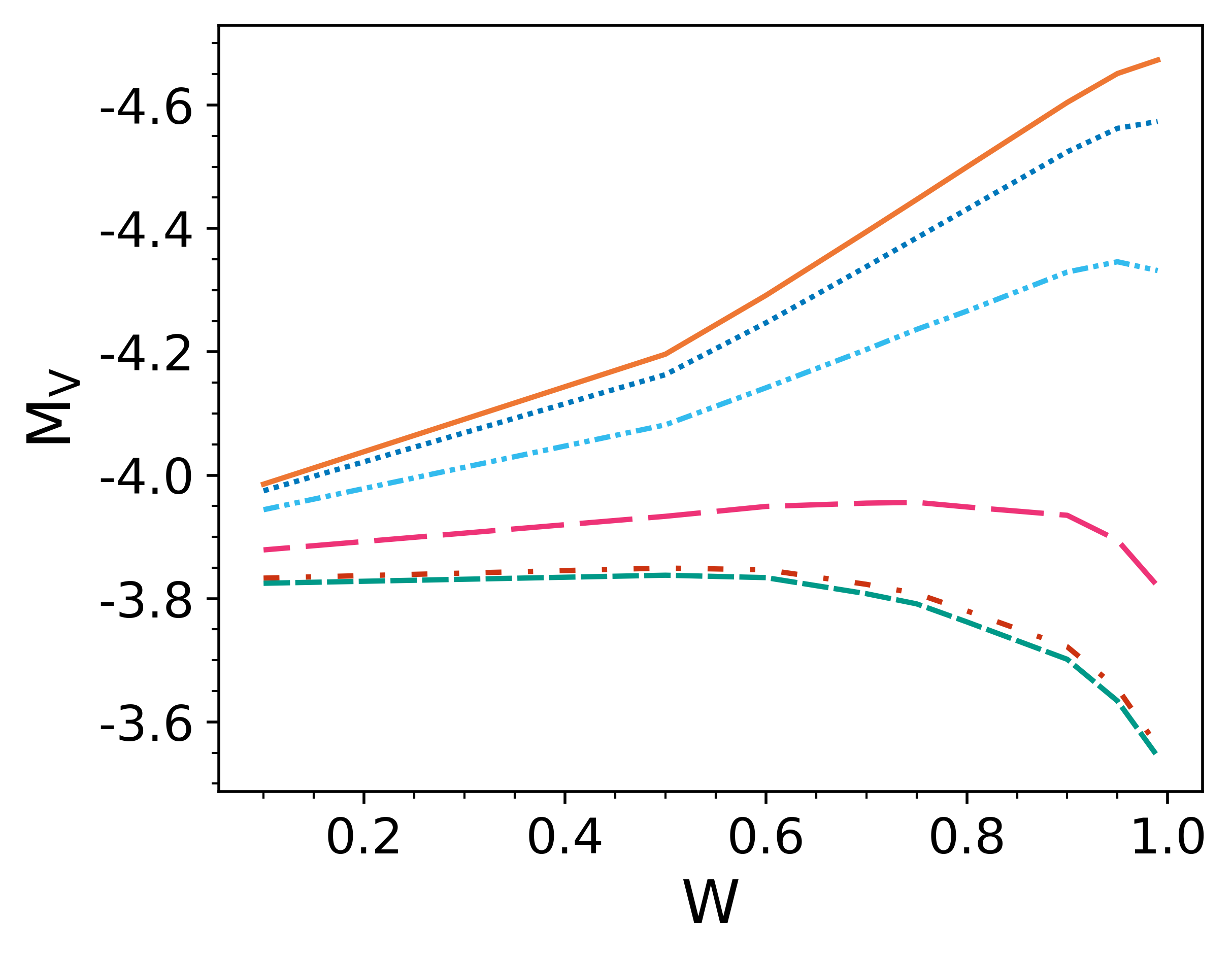}
    \end{subfigure}
    \begin{subfigure}{0.49\columnwidth}
        \includegraphics[width=\columnwidth]{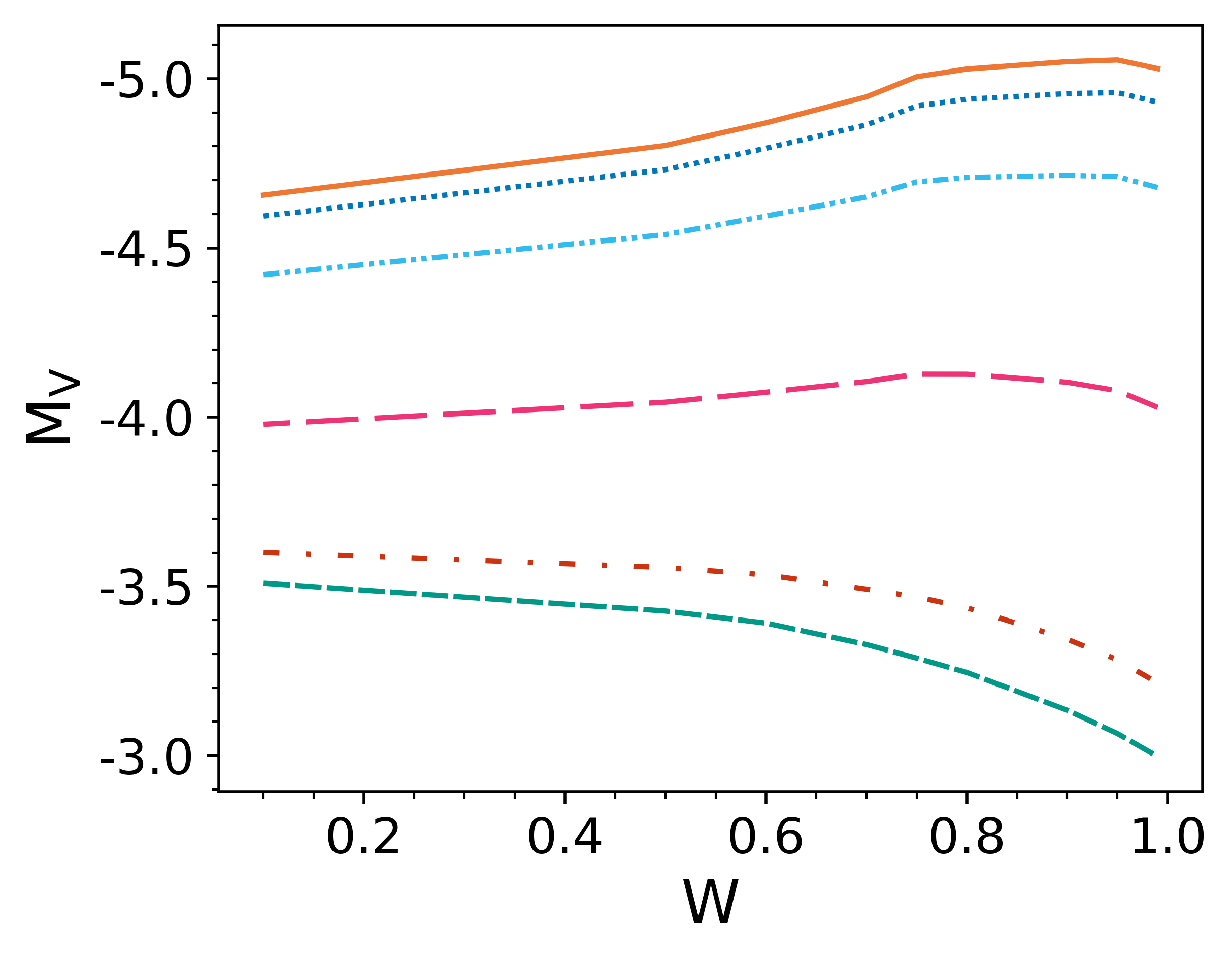}
    \end{subfigure}
\caption{Same as Figure~\ref{fig:b2_vmag}, but for the B0 models.}
    \label{fig:b0_vmag}
    \end{figure}

\begin{figure}[!ht]
\centering
   \begin{subfigure}{0.49\columnwidth}
        \includegraphics[width=\columnwidth]{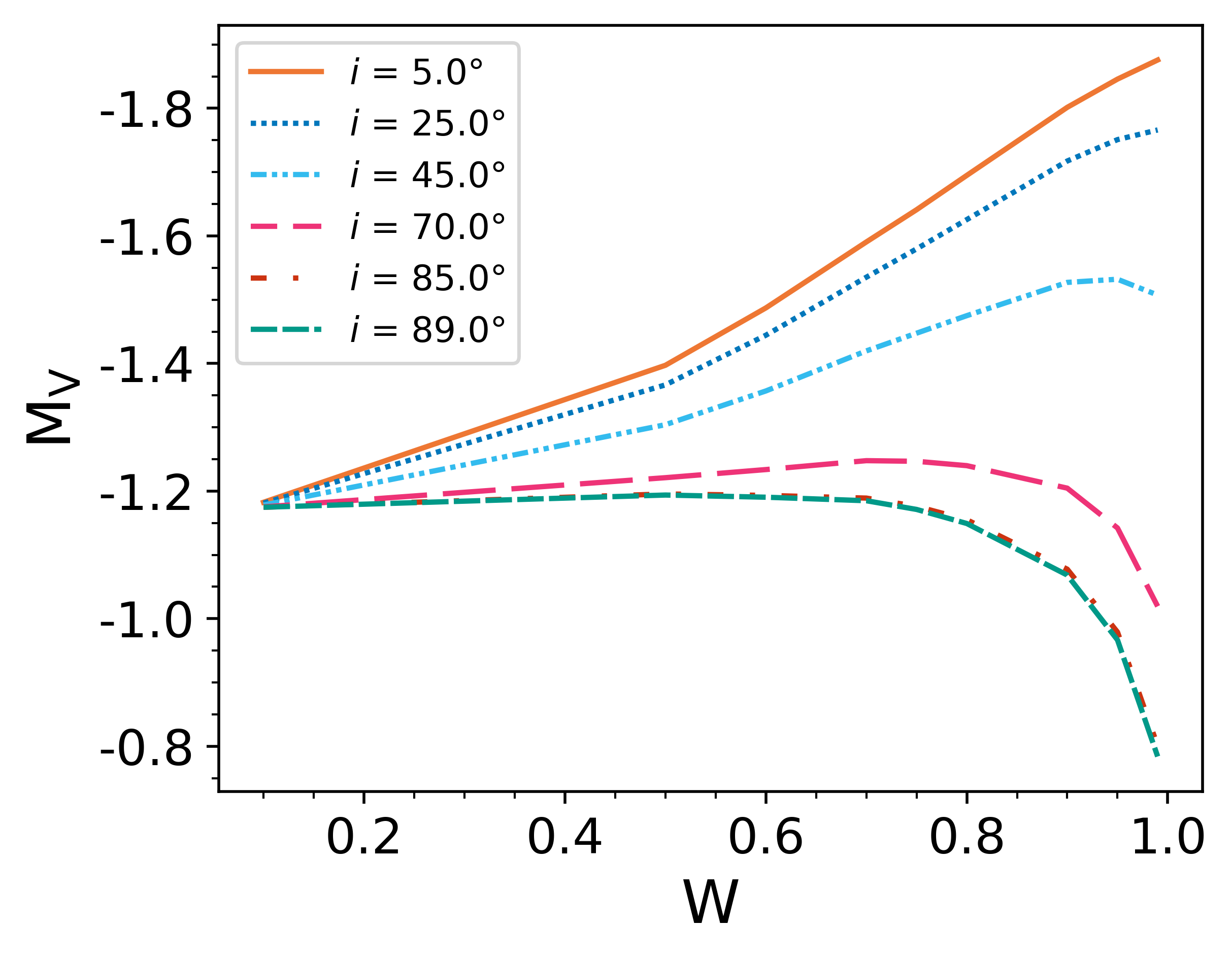}
    \end{subfigure}
    \begin{subfigure}{0.49\columnwidth}
        \includegraphics[width=\columnwidth]{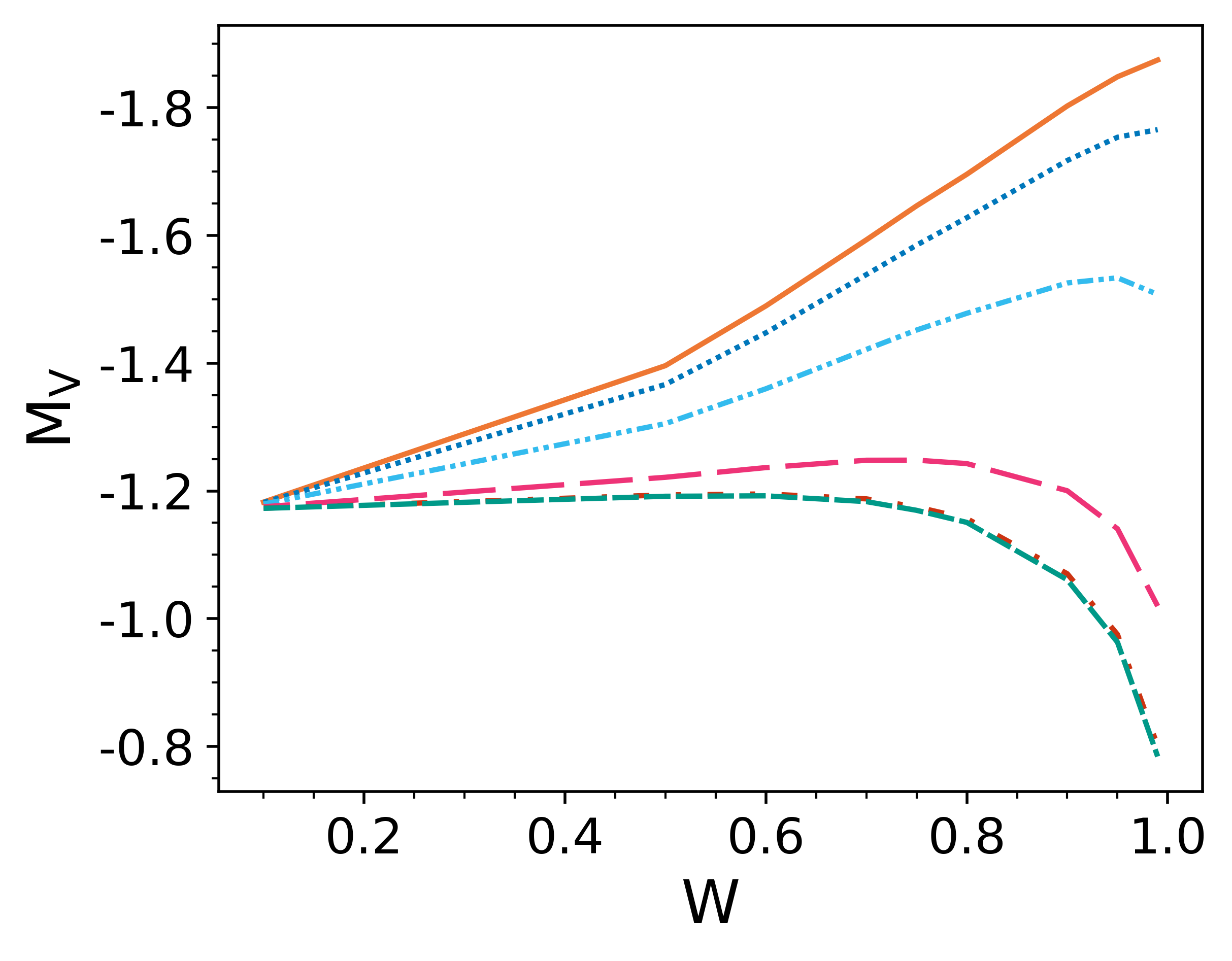}
    \end{subfigure}
    \begin{subfigure}{0.49\columnwidth}
        \includegraphics[width=\columnwidth]{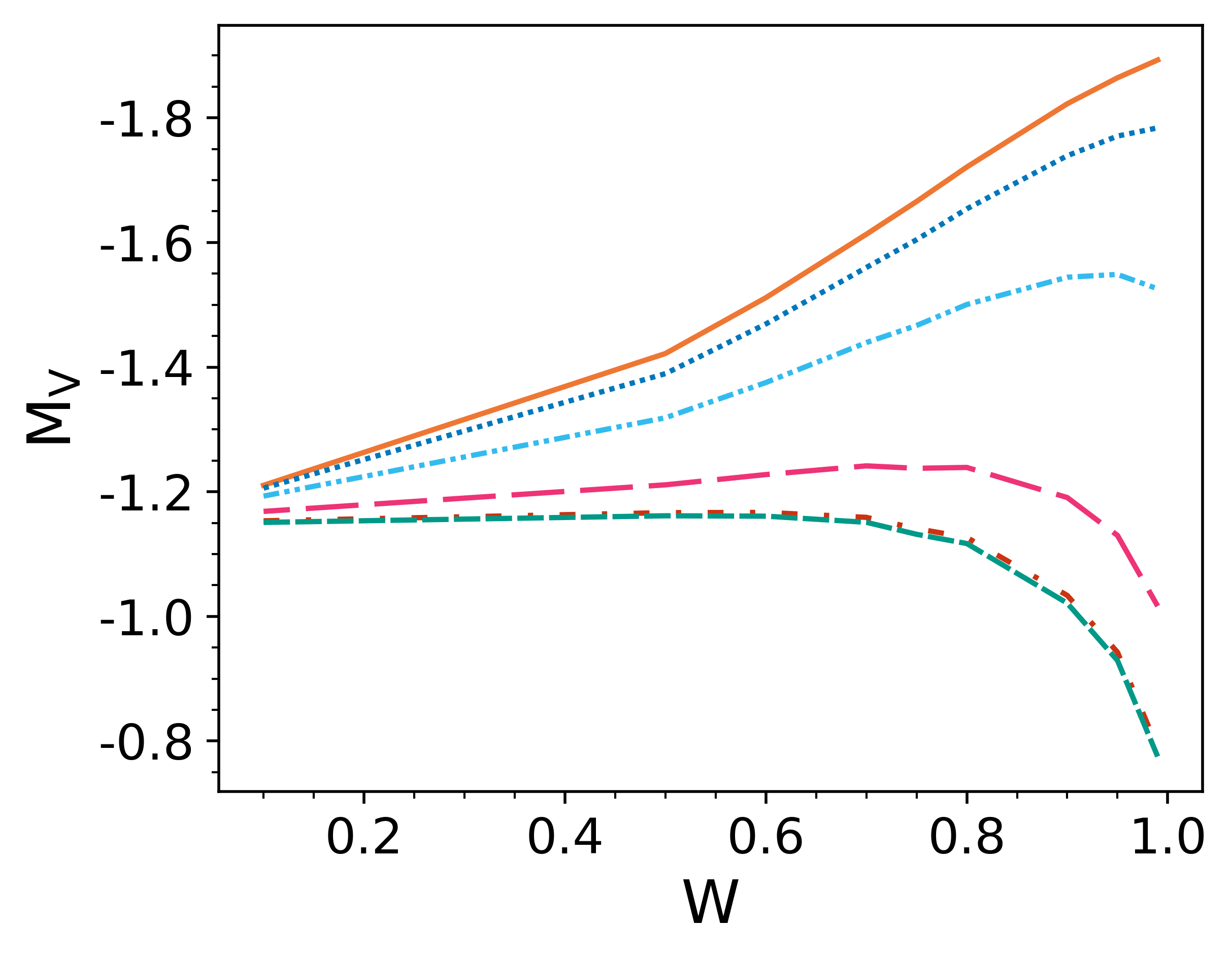}
    \end{subfigure}
    \begin{subfigure}{0.49\columnwidth}
        \includegraphics[width=\columnwidth]{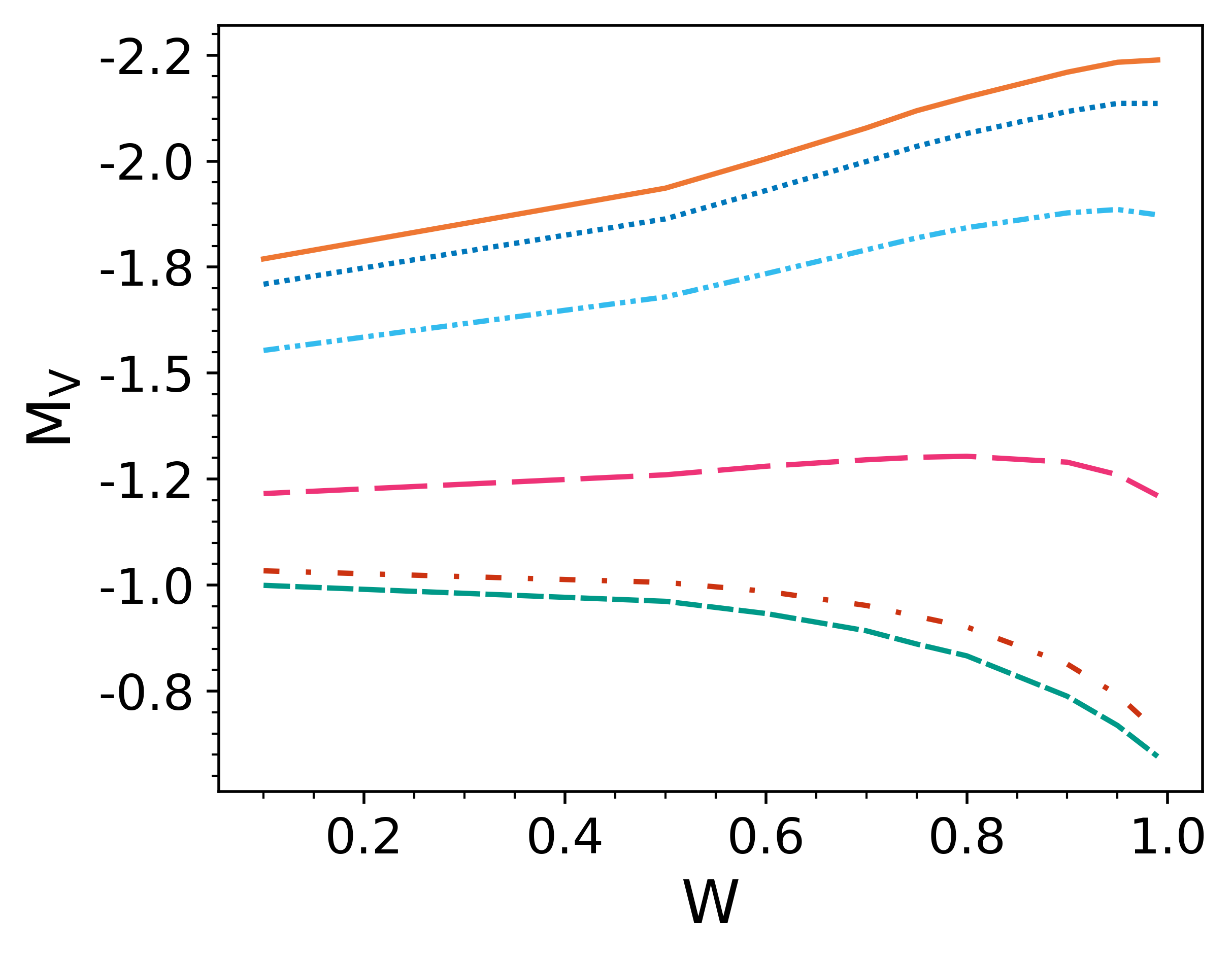}
    \end{subfigure}
\caption{Same as Figure~\ref{fig:b2_vmag}, but for the B5 models.}
    \label{fig:b5_vmag}
    \end{figure}

\begin{figure}[!ht]
\centering
   \begin{subfigure}{0.49\columnwidth}
        \includegraphics[width=\columnwidth]{b8/b8_Mv_comp_phot.png}
    \end{subfigure}
    \begin{subfigure}{0.49\columnwidth}
        \includegraphics[width=\columnwidth]{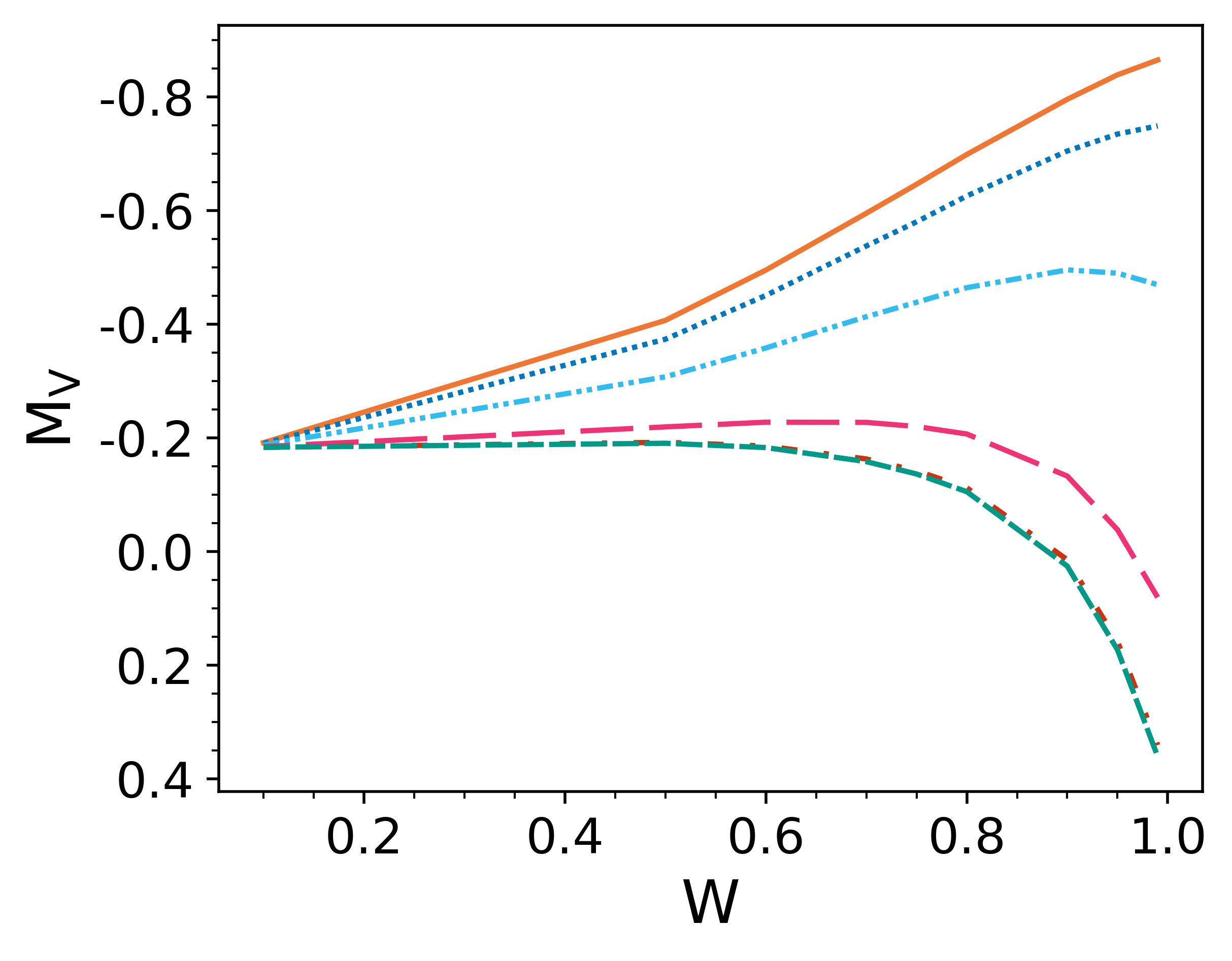}
    \end{subfigure}
    \begin{subfigure}{0.49\columnwidth}
        \includegraphics[width=\columnwidth]{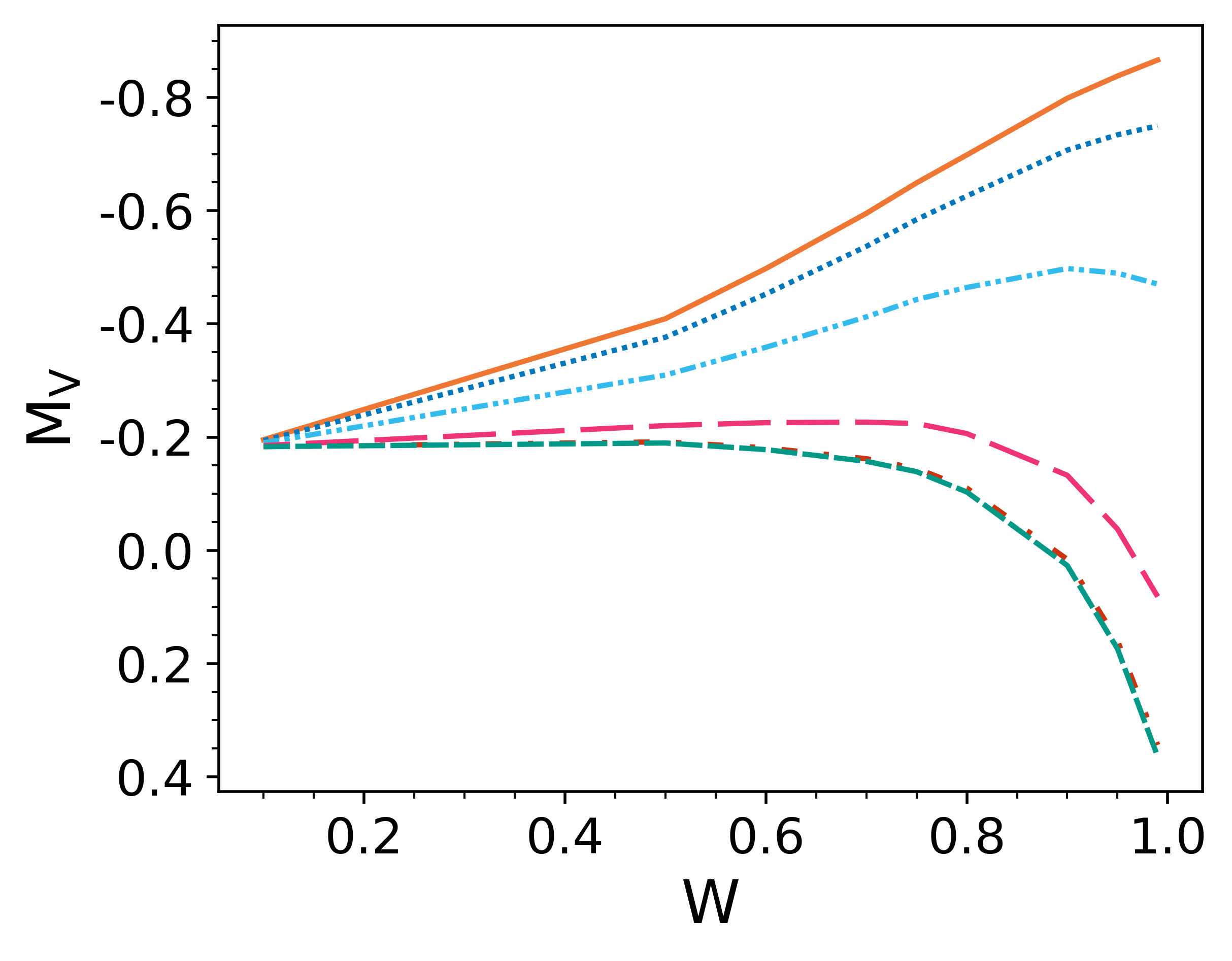}
    \end{subfigure}
    \begin{subfigure}{0.49\columnwidth}
        \includegraphics[width=\columnwidth]{b8/b8_rho0_5e-12_Mv_comp.png}
    \end{subfigure}
\caption{Same as Figure~\ref{fig:b2_vmag}, but for the B8 models.}
    \label{fig:b8_all_vmag}
    \end{figure}

\section{Supplementary H$\alpha$ figures}\label{secA2}

For completeness, we also include additional figures for the spectral types not included in the main results shown in Section~\ref{sec:ha_ew_results}. 

\begin{figure}
    \centering
    \begin{minipage}[b]{0.22\textwidth} \centering\includegraphics[width=\textwidth]{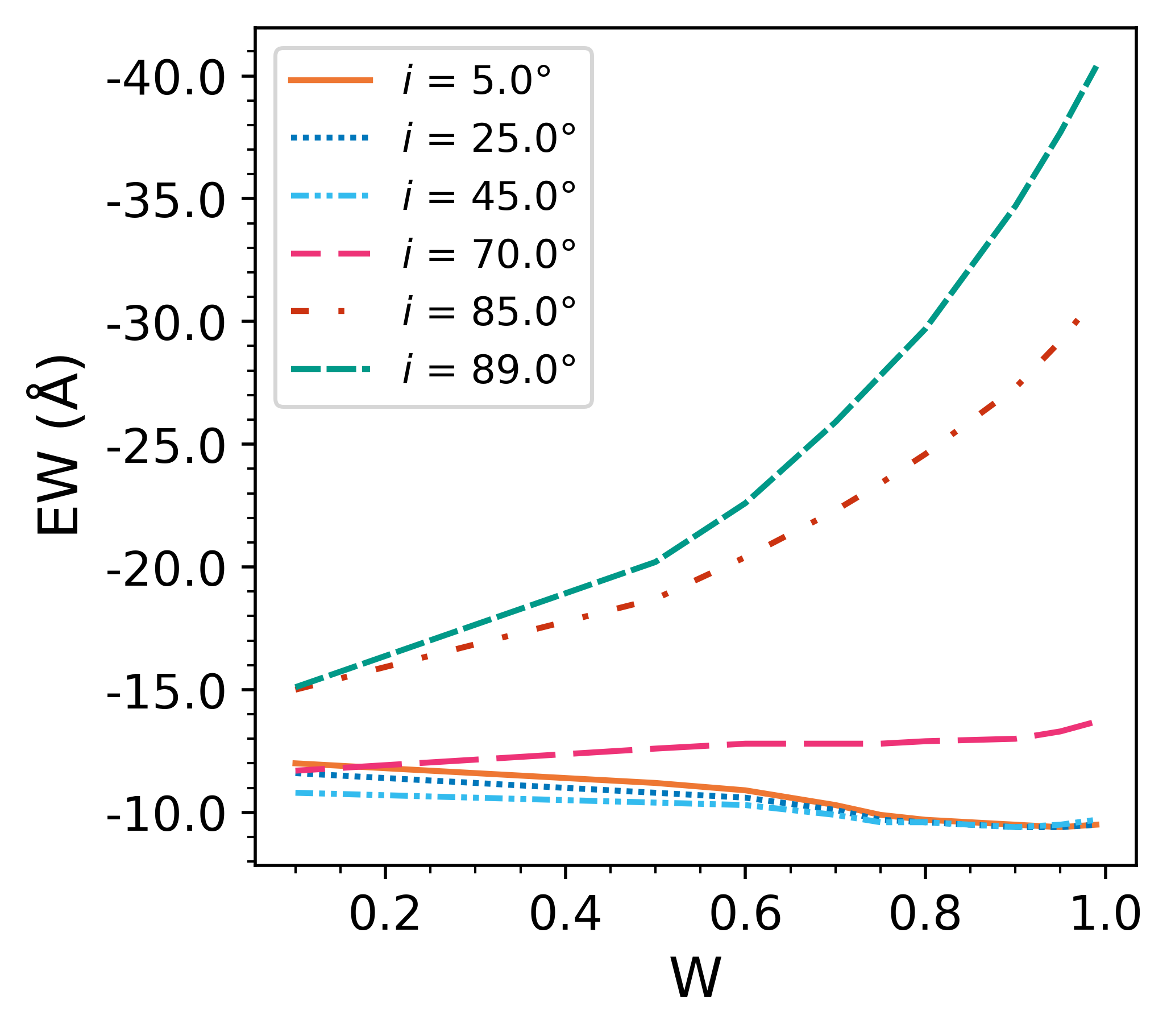}\par\vspace{0.5em}
    \includegraphics[width=\textwidth  ]{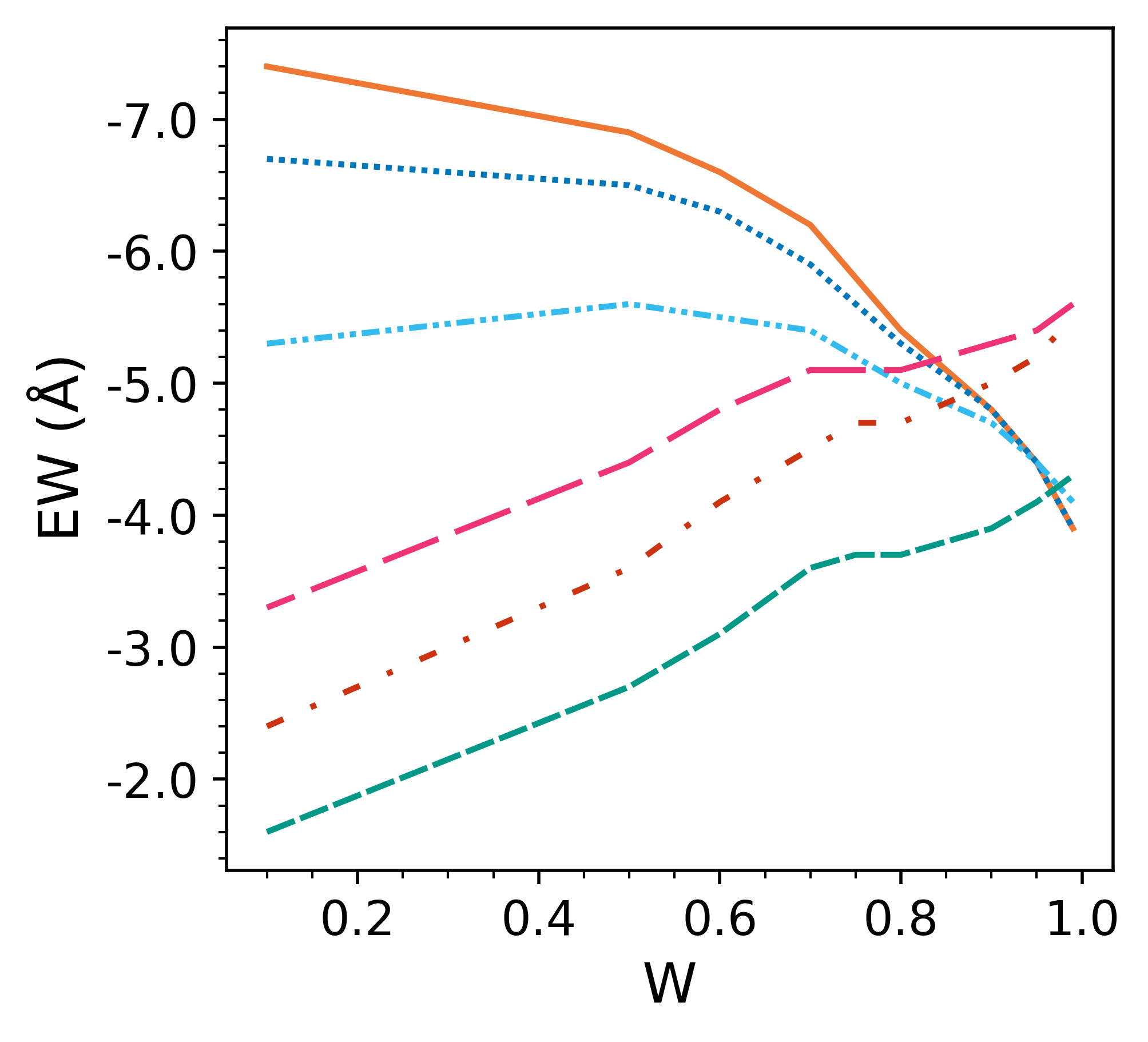}\par\vspace{0.5em}
    \includegraphics[width=\textwidth]{b0/b0_rho0_1e-12_ew_comp.png}
    \end{minipage}%
    \hfill
    \begin{minipage}[b]{0.25\textwidth}  
        \centering
    \includegraphics[width=\textwidth]{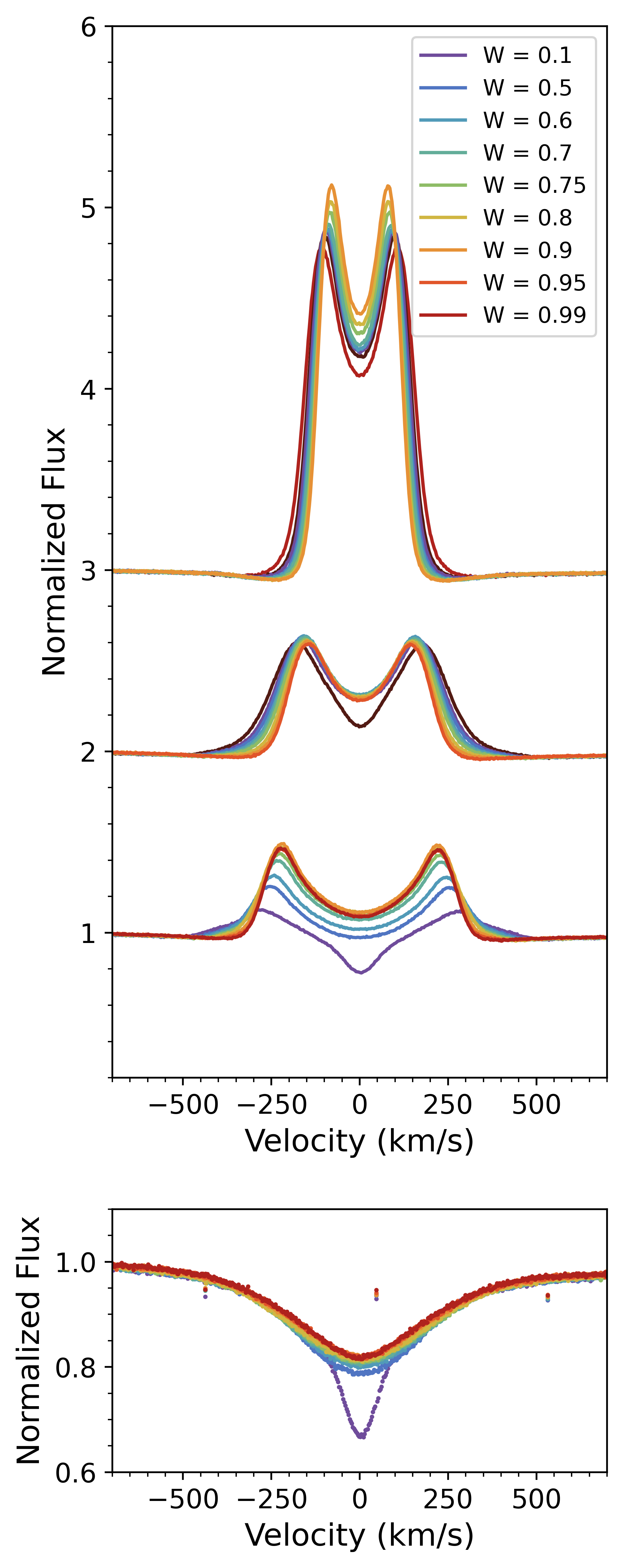}
    \end{minipage}
    \caption{Same as Figure~\ref{fig:B2_ew}, but for the B0 models.}
    \label{fig:B0_ew}
\end{figure}

\begin{figure}
    \centering
    \begin{minipage}[b]{0.22\textwidth} \centering\includegraphics[width=\textwidth]{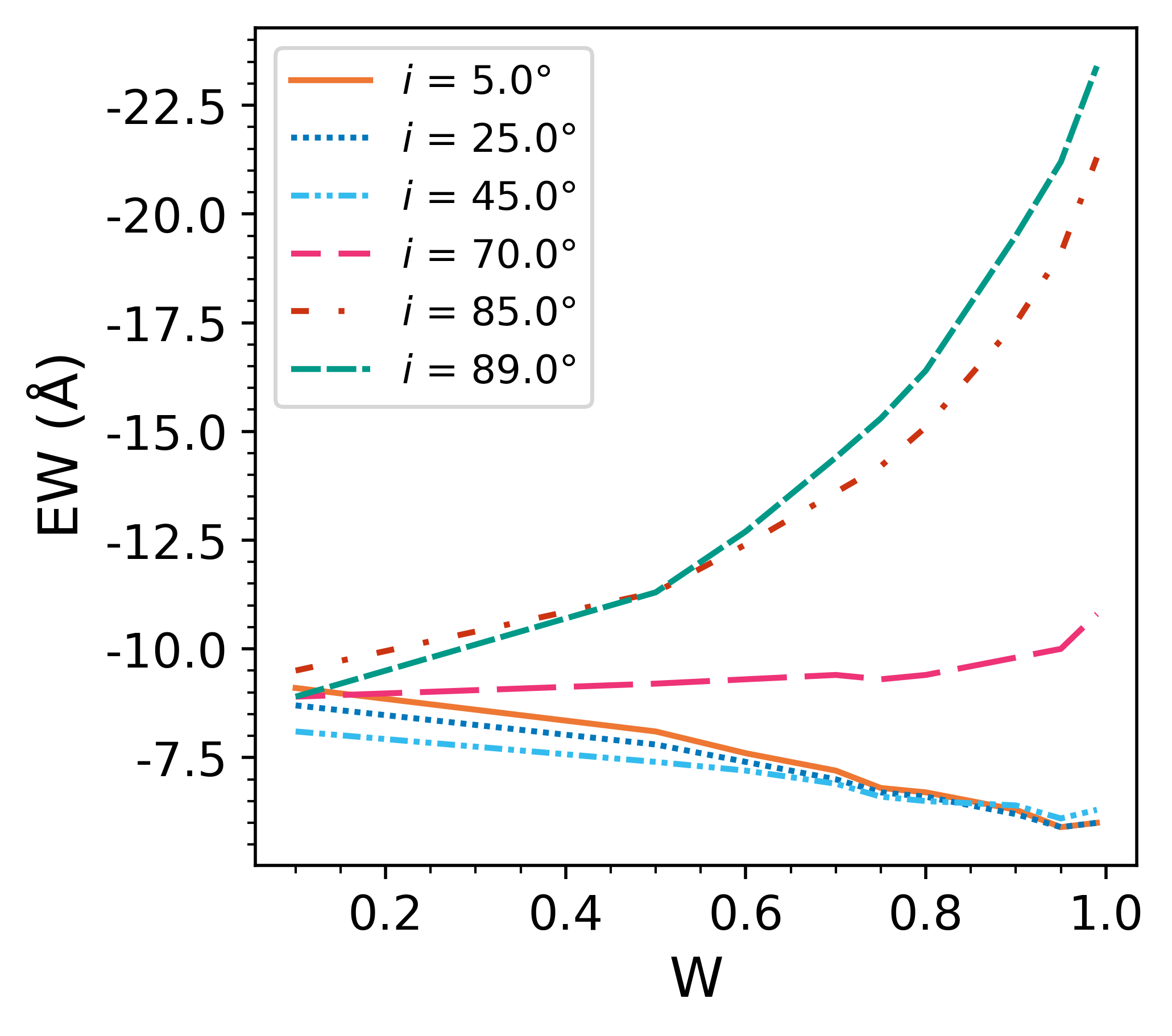}\par\vspace{0.5em}
    \includegraphics[width=\textwidth  ]{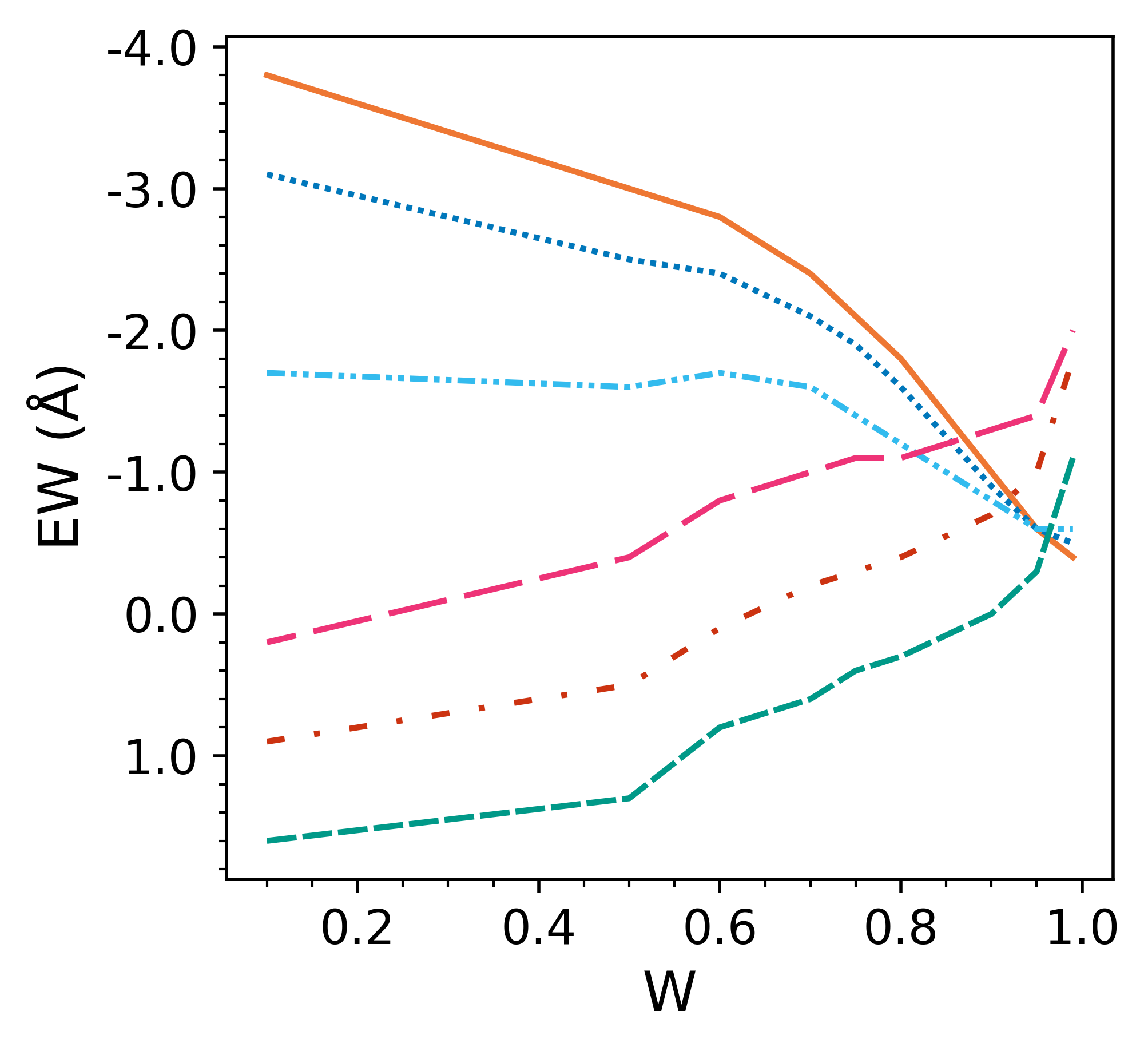}\par\vspace{0.5em}
    \includegraphics[width=\textwidth]{b5/b5_rho0_5e-13_ew_comp.png}
    \end{minipage}%
    \hfill
    \begin{minipage}[b]{0.25\textwidth}  
        \centering
    \includegraphics[width=\textwidth]{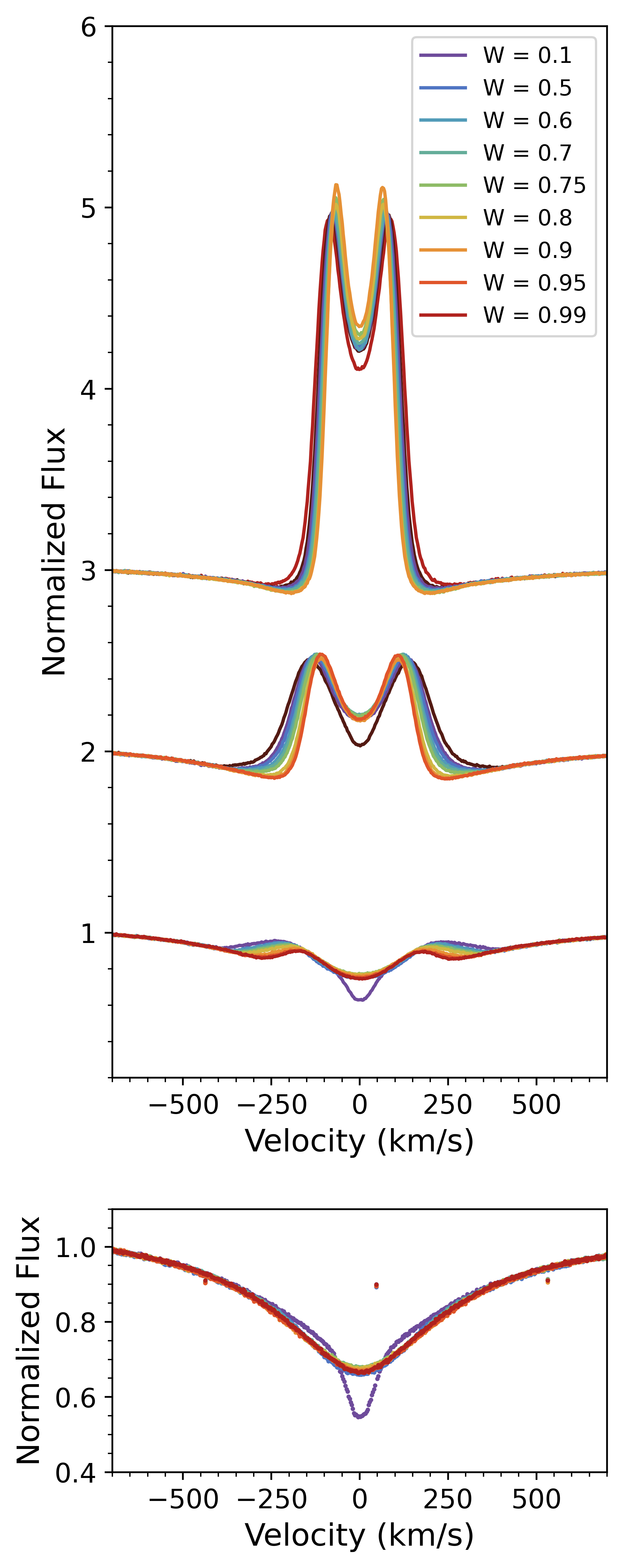}
    \end{minipage}
    \caption{Same as Figure~\ref{fig:B2_ew}, but for the B5 models.}
    \label{fig:B5_ew}
\end{figure}

\section{Supplementary polarization figures}\label{secA3}

This section features figures for the spectral types not included in the main results shown in Section~\ref{sec:pol_results}. 

\begin{figure}
    \centering
    \begin{minipage}[b]{0.23\textwidth} \centering\includegraphics[width=\textwidth]{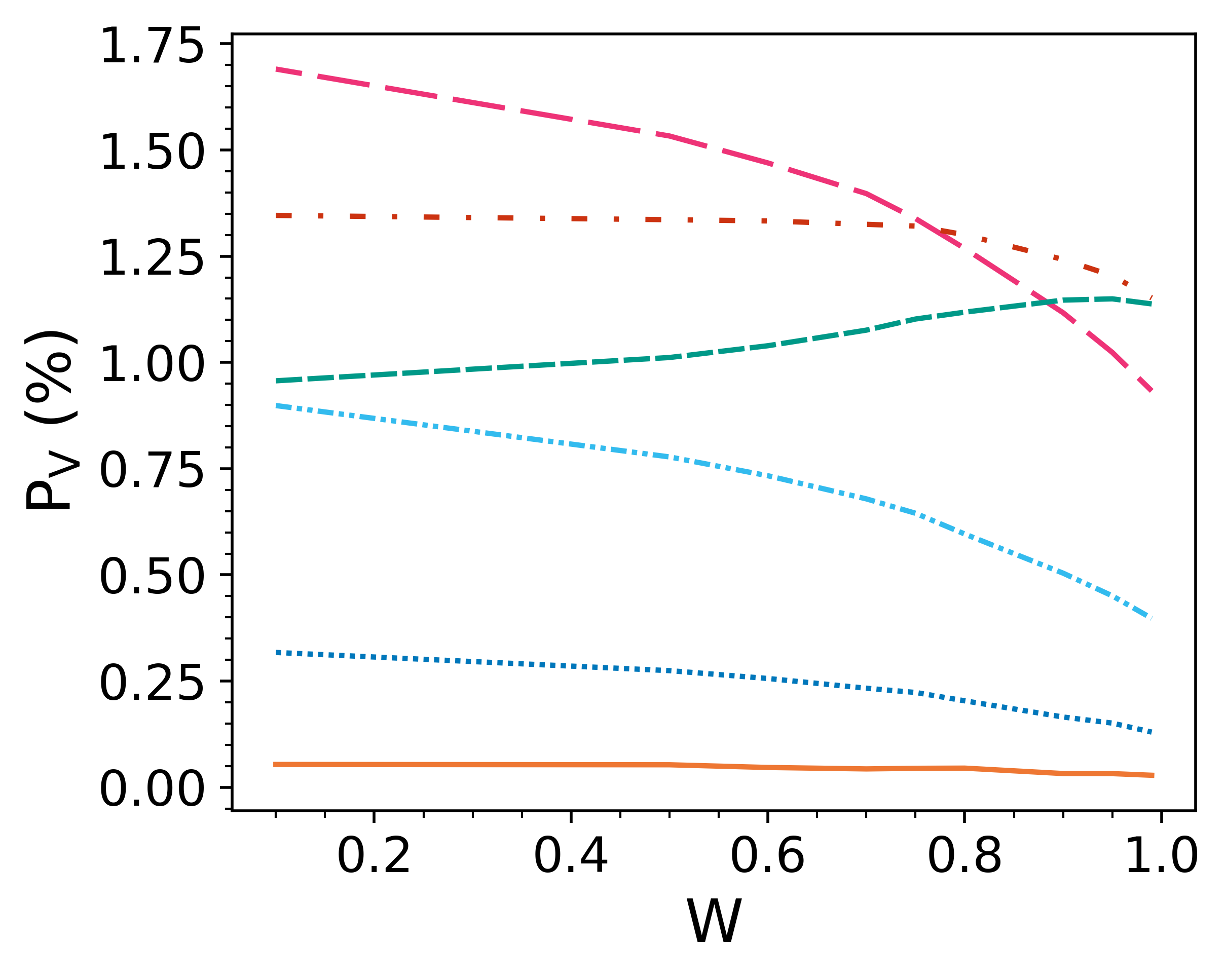}\par\vspace{0.5em}
    \includegraphics[width=\textwidth  ]{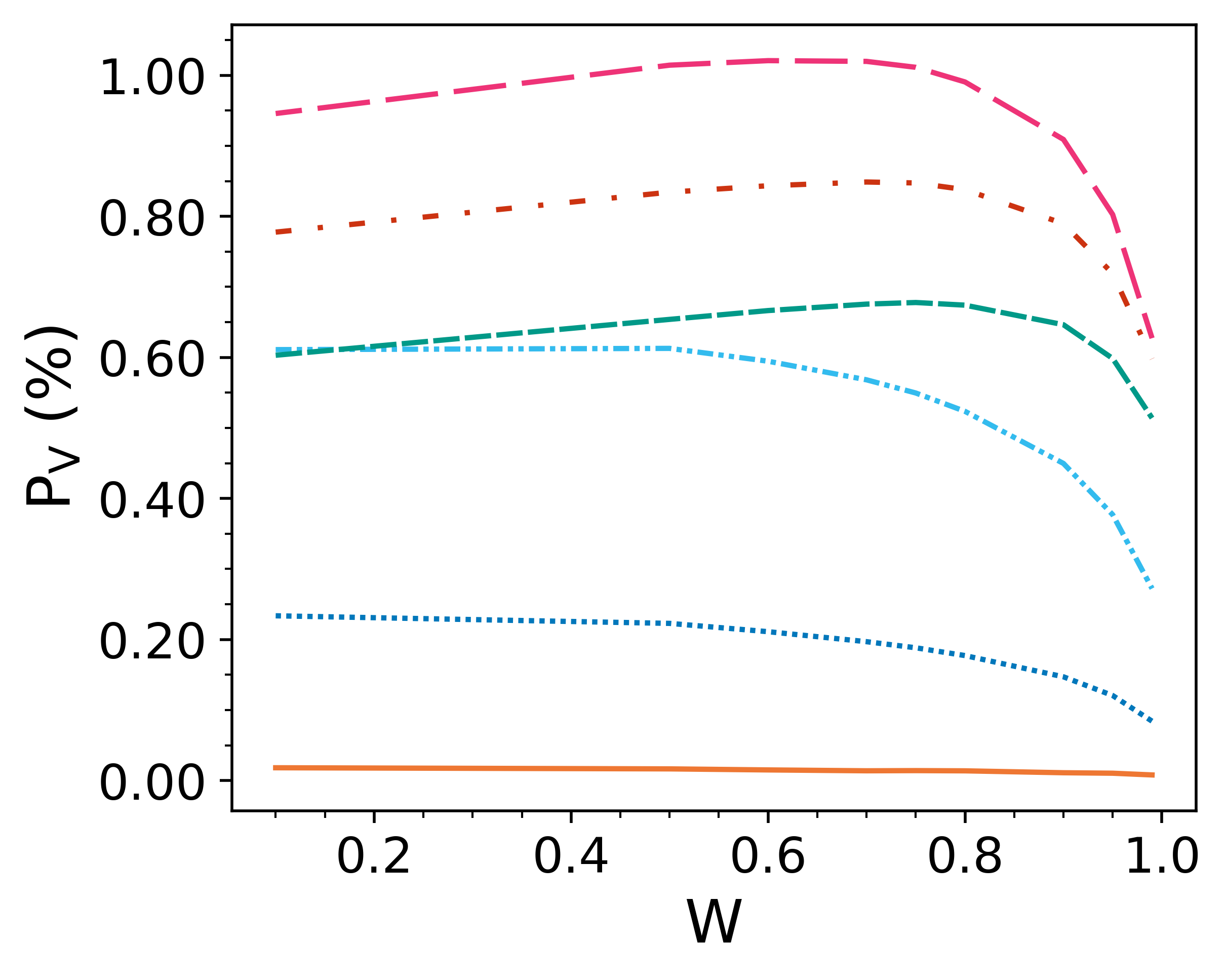}\par\vspace{0.5em}
    \includegraphics[width=\textwidth]{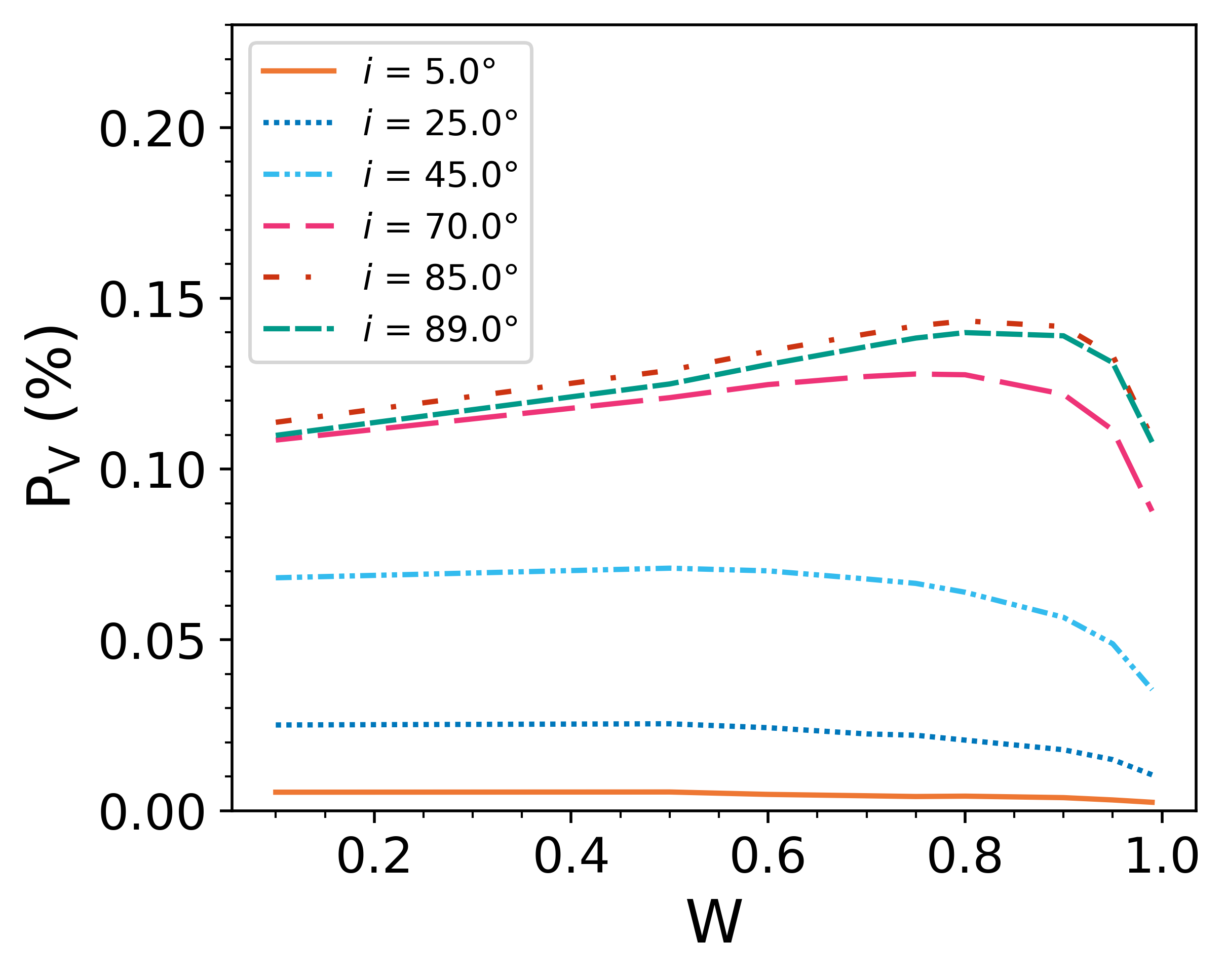}
    \end{minipage}%
    \hfill
    \begin{minipage}[b]{0.23\textwidth} \centering\includegraphics[width=\textwidth]{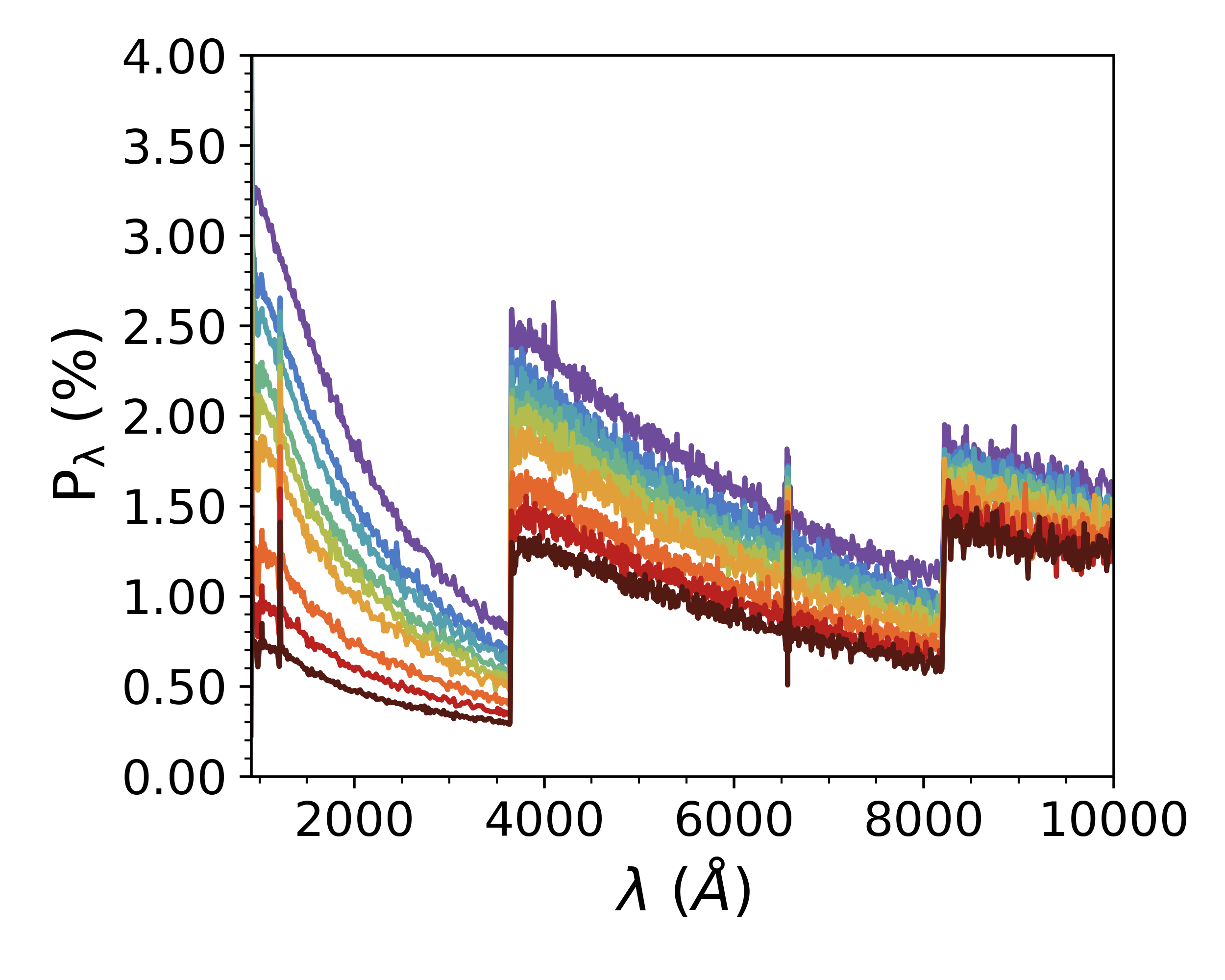}\par\vspace{0.5em}
    \includegraphics[width=\textwidth  ]{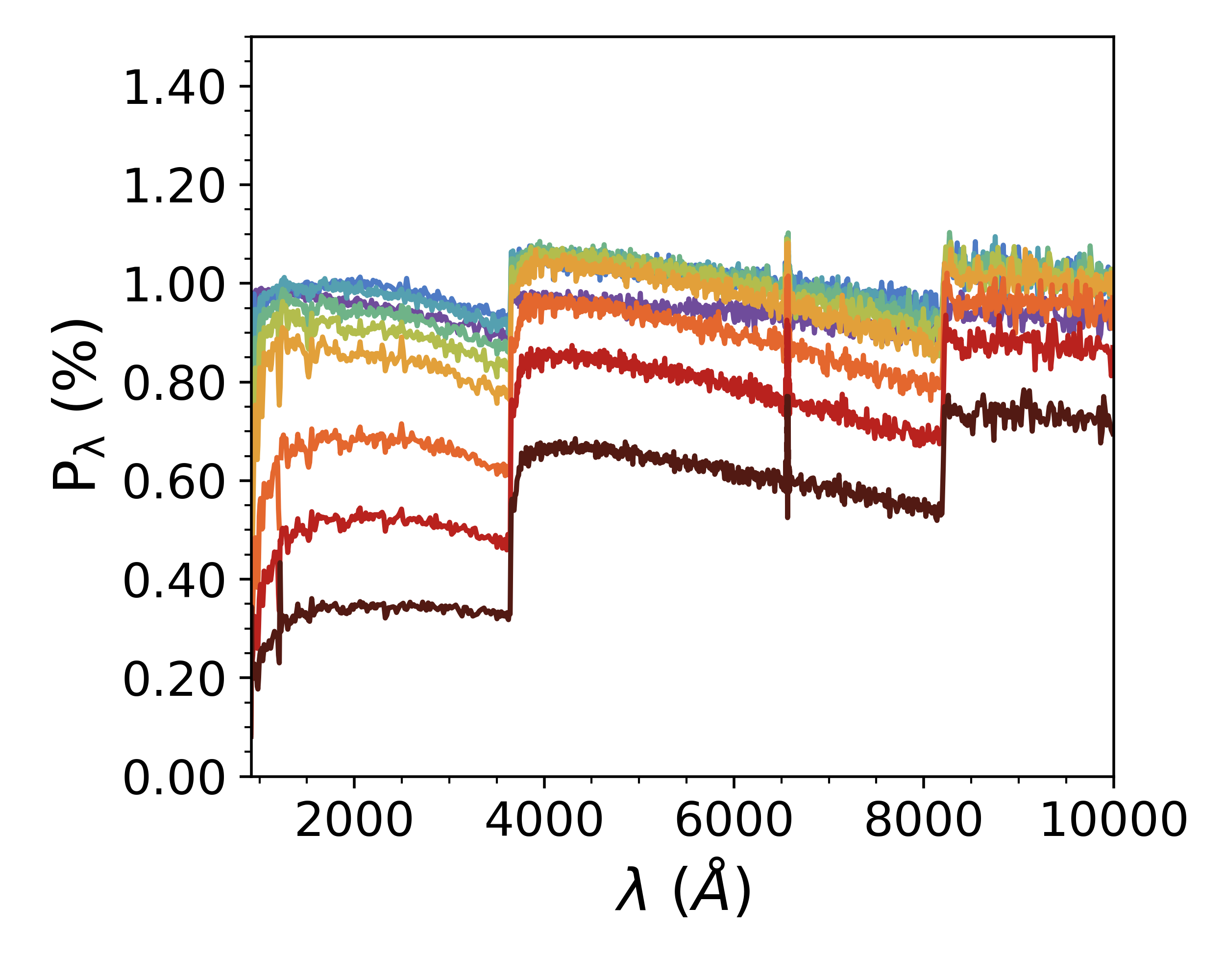}\par\vspace{0.5em}
    \includegraphics[width=\textwidth]{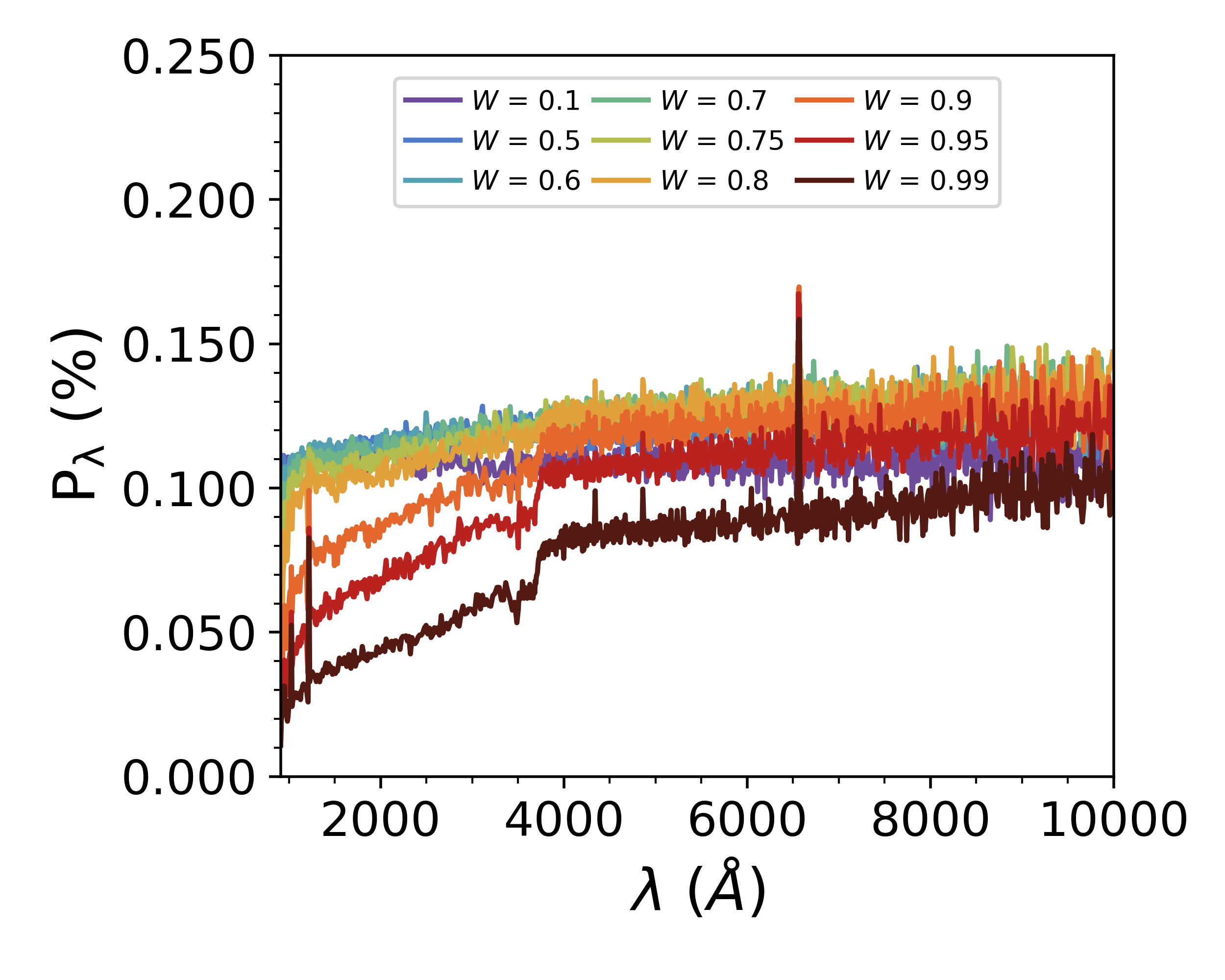}
    \end{minipage}%
    \hfill
    \caption{Same as Figure~\ref{fig:B2_pol}, but for the B0 models.}
    \label{fig:B0_pol}
\end{figure}

\begin{figure}
    \centering
    \begin{minipage}[b]{0.23\textwidth} \centering\includegraphics[width=\textwidth]{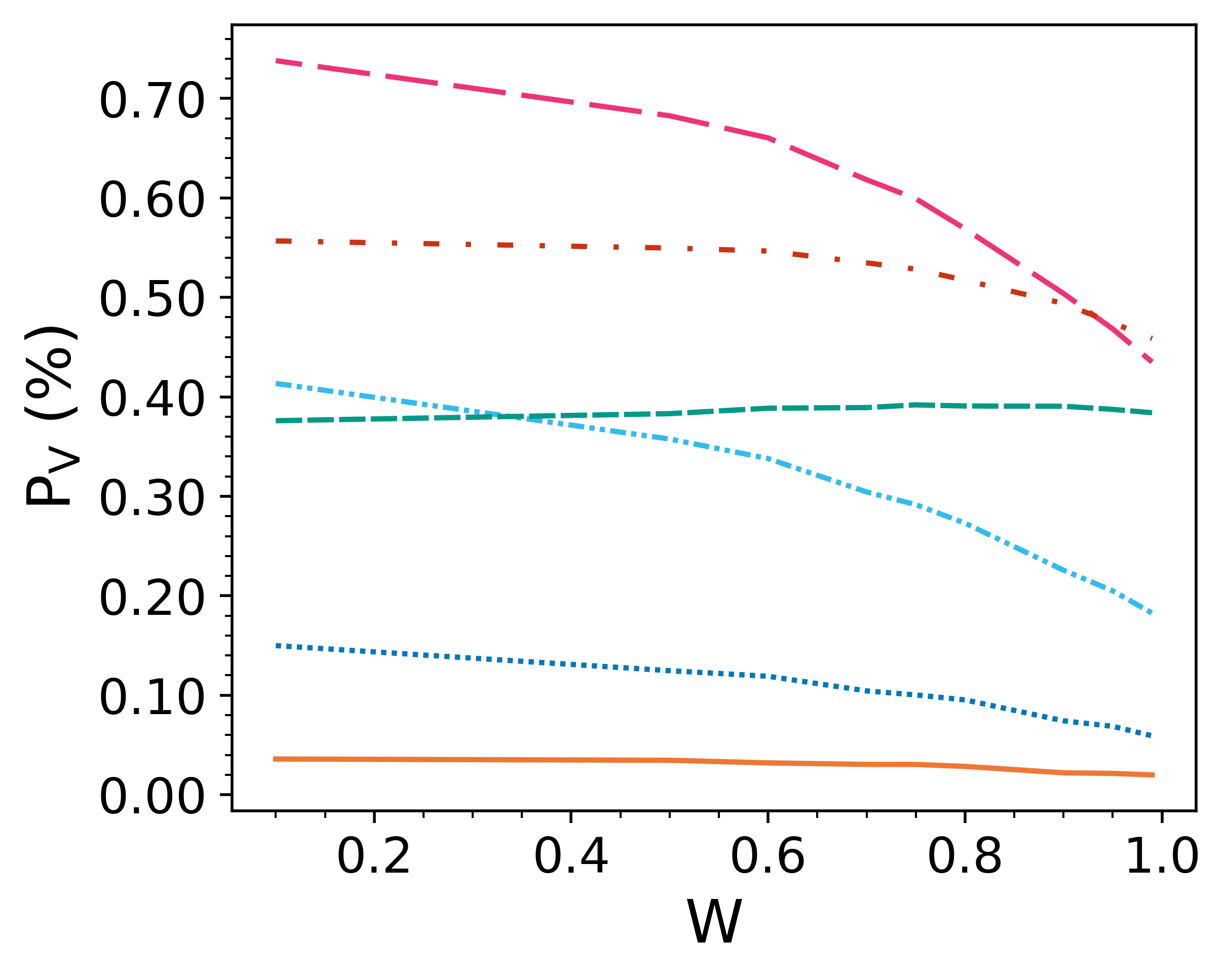}\par\vspace{0.5em}
    \includegraphics[width=\textwidth  ]{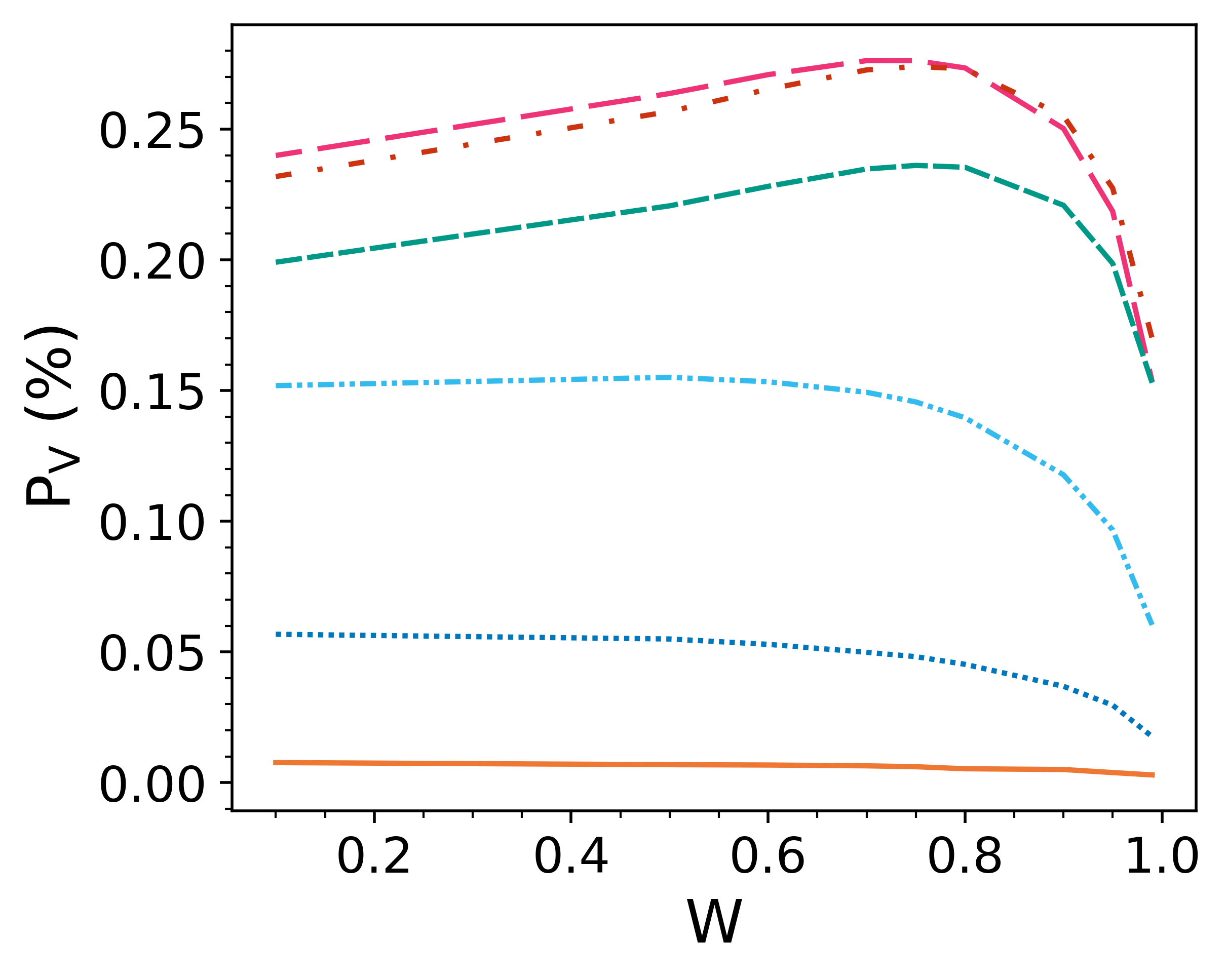}\par\vspace{0.5em}
    \includegraphics[width=\textwidth]{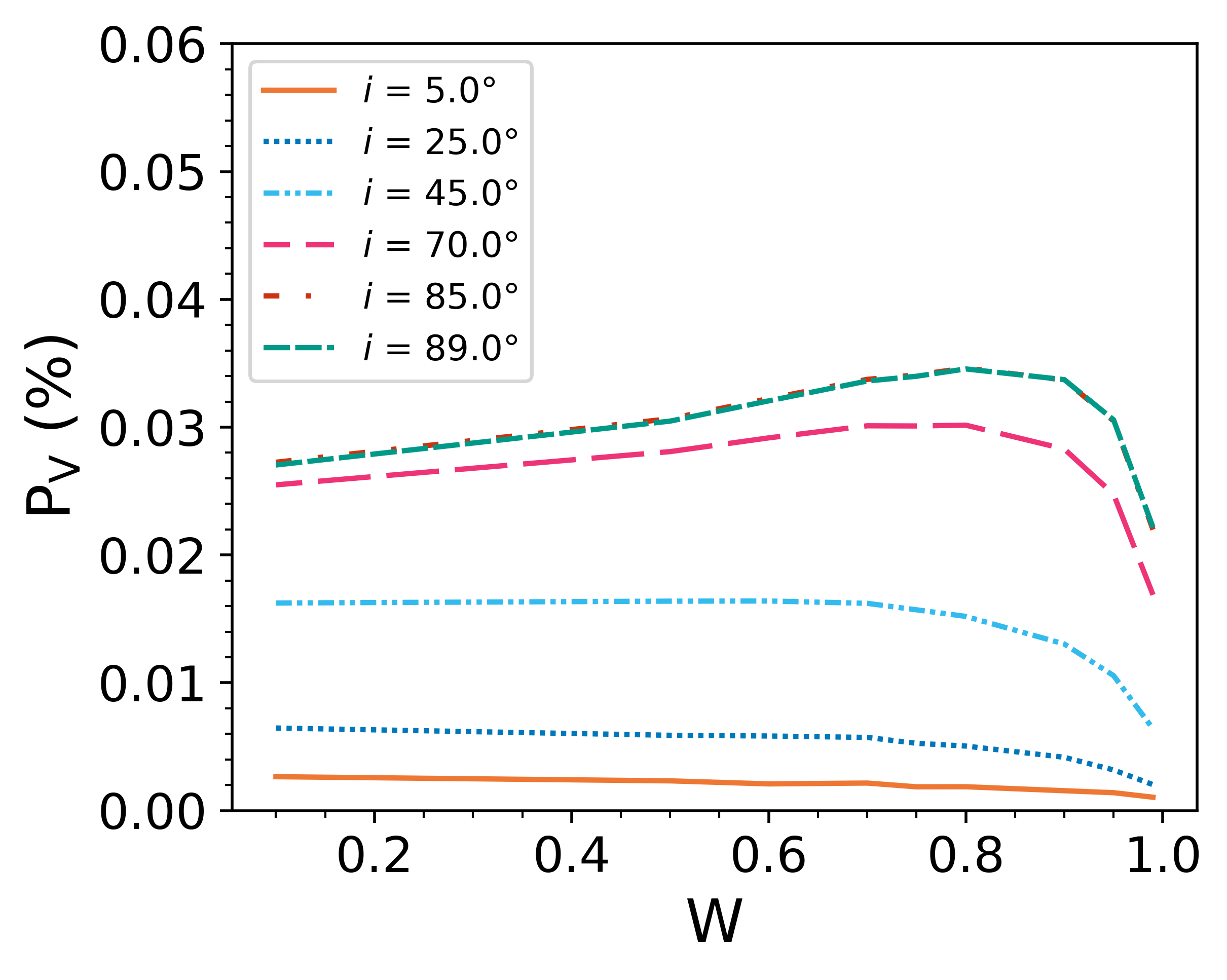}
    \end{minipage}%
    \hfill
    \begin{minipage}[b]{0.23\textwidth} \centering\includegraphics[width=\textwidth]{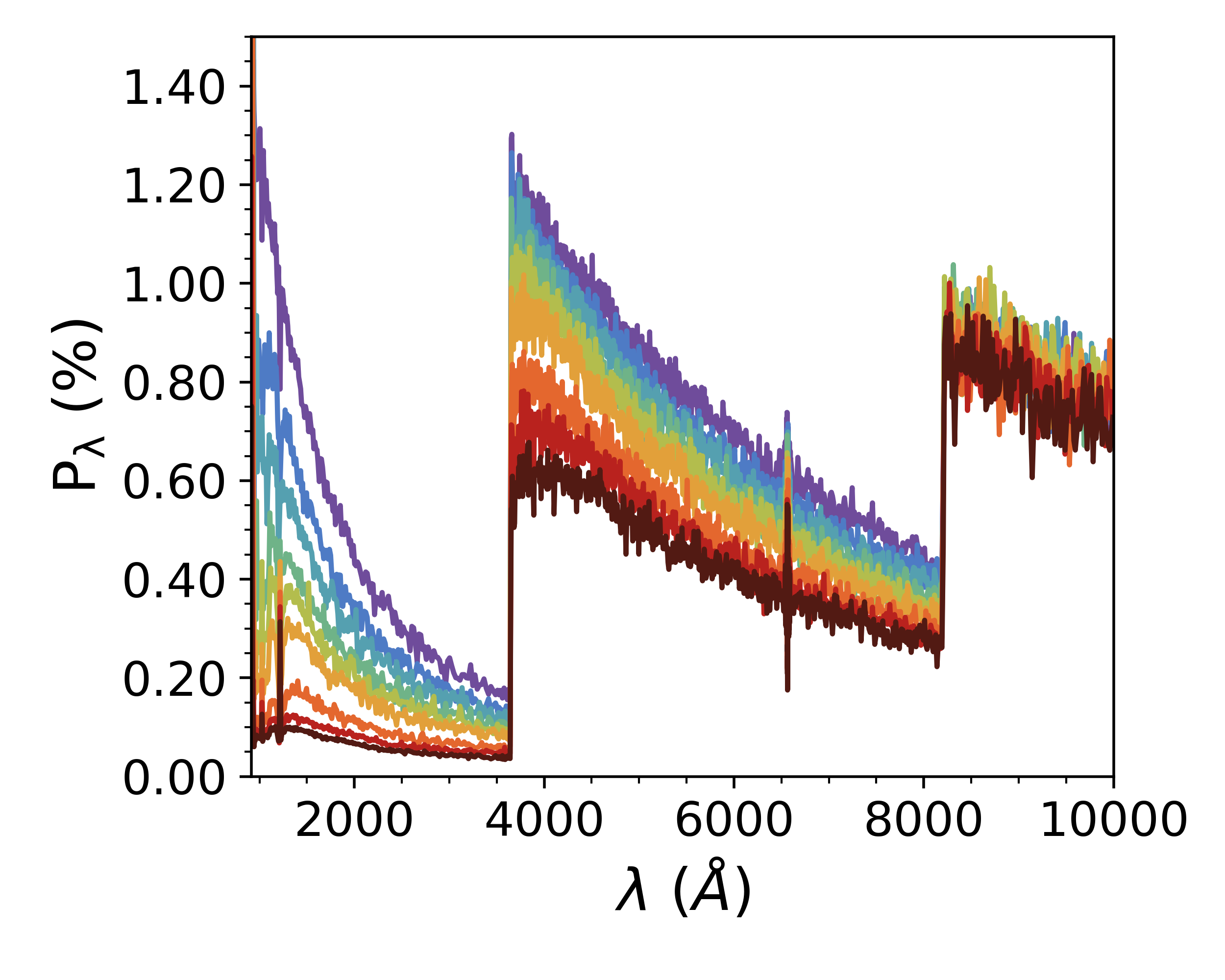}\par\vspace{0.5em}
    \includegraphics[width=\textwidth  ]{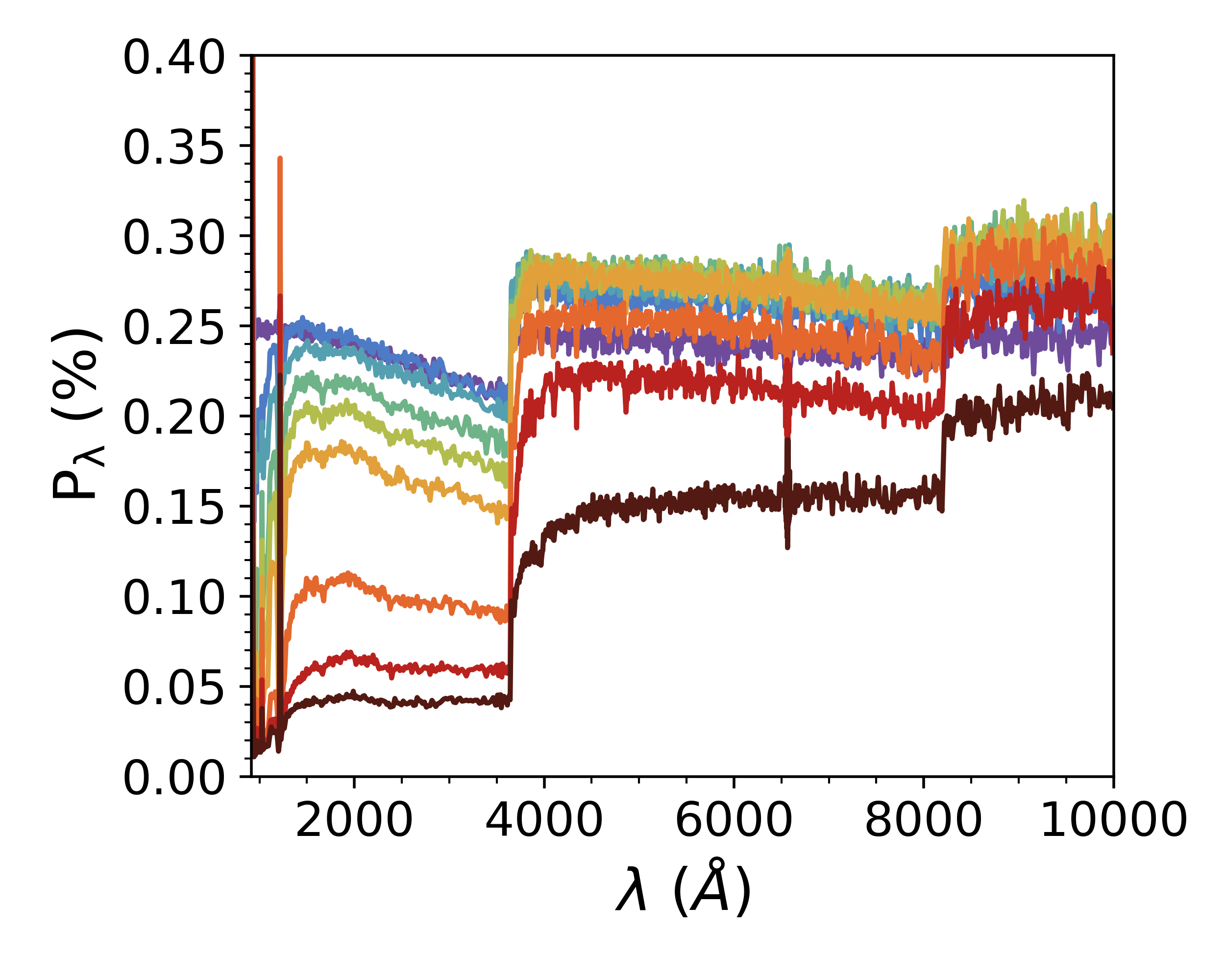}\par\vspace{0.5em}
    \includegraphics[width=\textwidth]{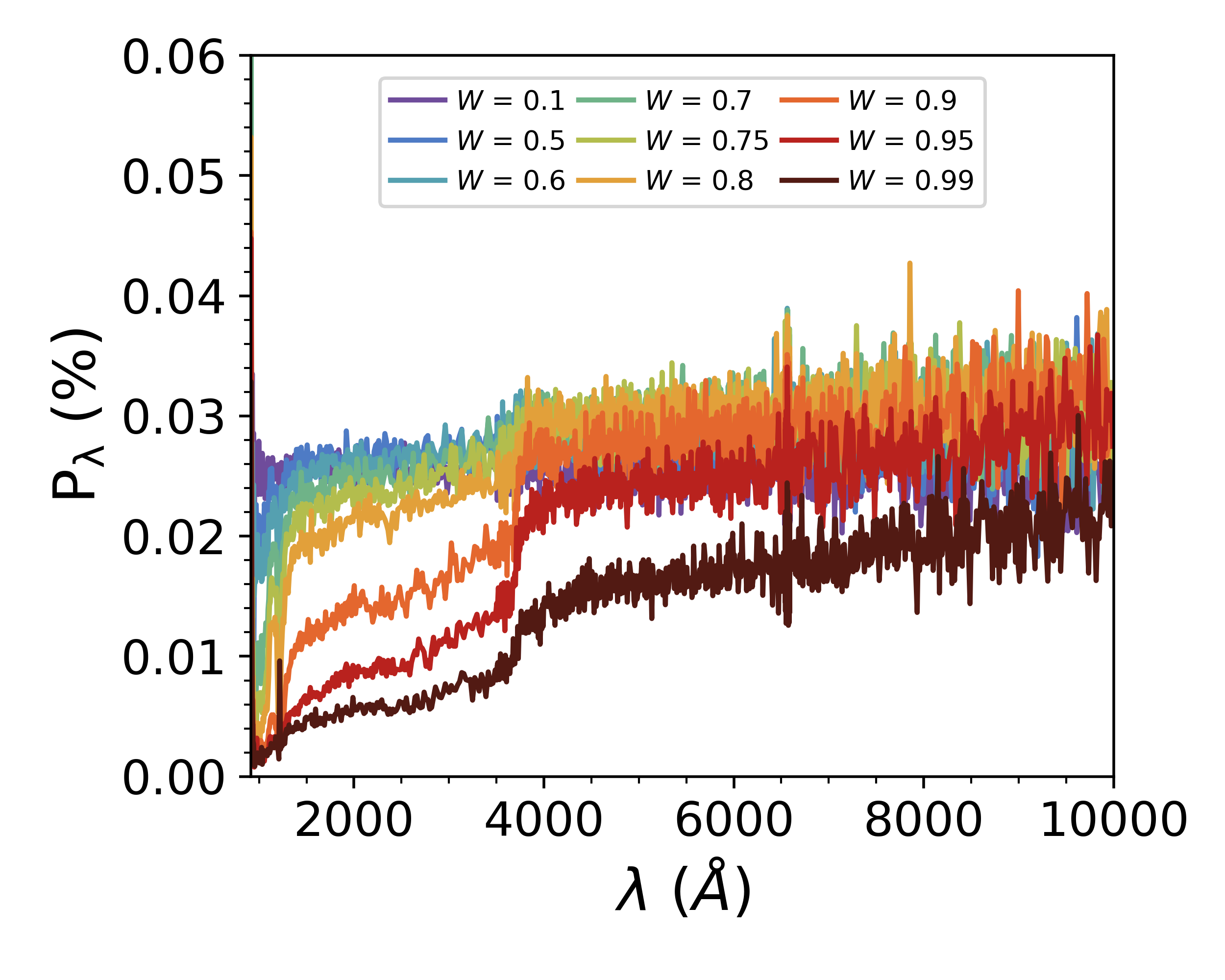}
    \end{minipage}%
    \hfill
    \caption{Same as Figure~\ref{fig:B2_pol}, but for the B5 models.}
    \label{fig:B5_pol}
\end{figure}

\section{Supplementary temperature cross-sections}\label{secA4}

Here, we provide temperature cross-sections for the disk in each of the models presented in this work. 

\begin{figure}
    \centering
    \includegraphics[width=0.49\textwidth]{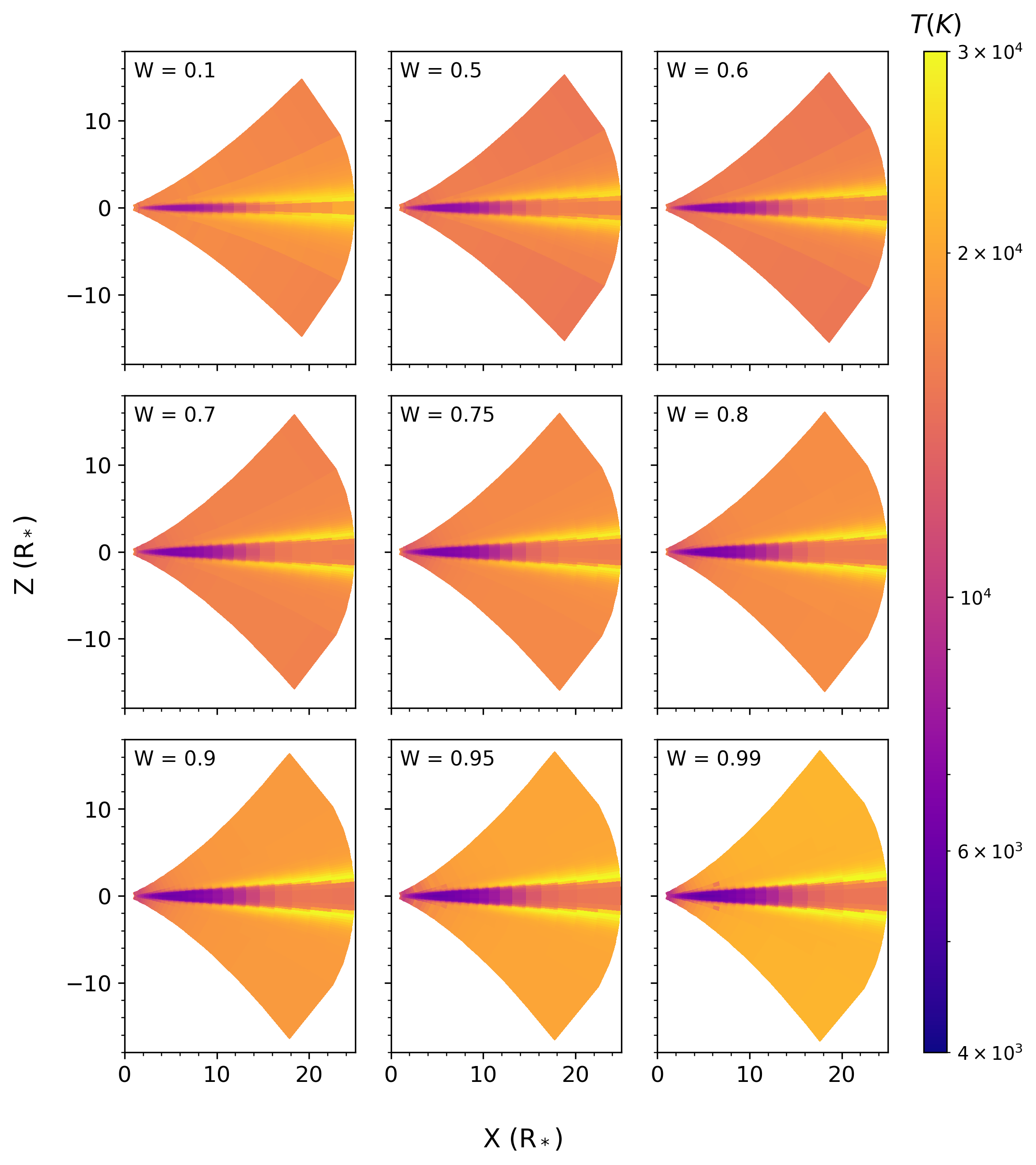}
    \caption{Temperature structure of a cross-section of the B0 models with high density disks. Each panel represents a different rotation rate, indicated in the top left corner, and the color represents temperature, as indicated by the color bar on the right.}
    \label{fig:B0_highdens_temps}
\end{figure}

\begin{figure}
    \centering
    \includegraphics[width=0.49\textwidth]{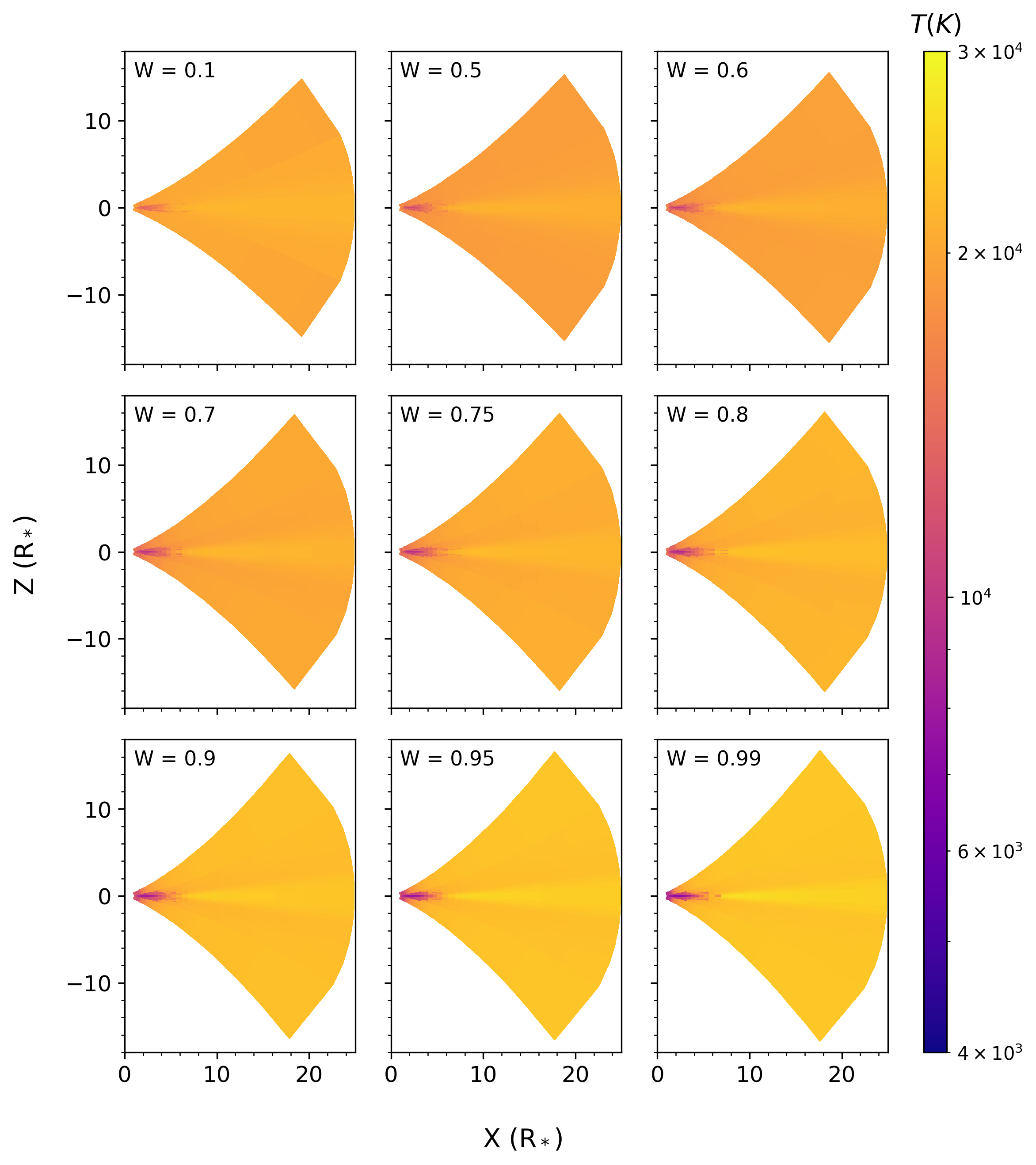}
    \caption{Same as Figure~\ref{fig:B0_highdens_temps}, but for the B0 moderate density disks.}
    \label{fig:B0_moddens_temps}
\end{figure}

\begin{figure}
    \centering
    \includegraphics[width=0.49\textwidth]{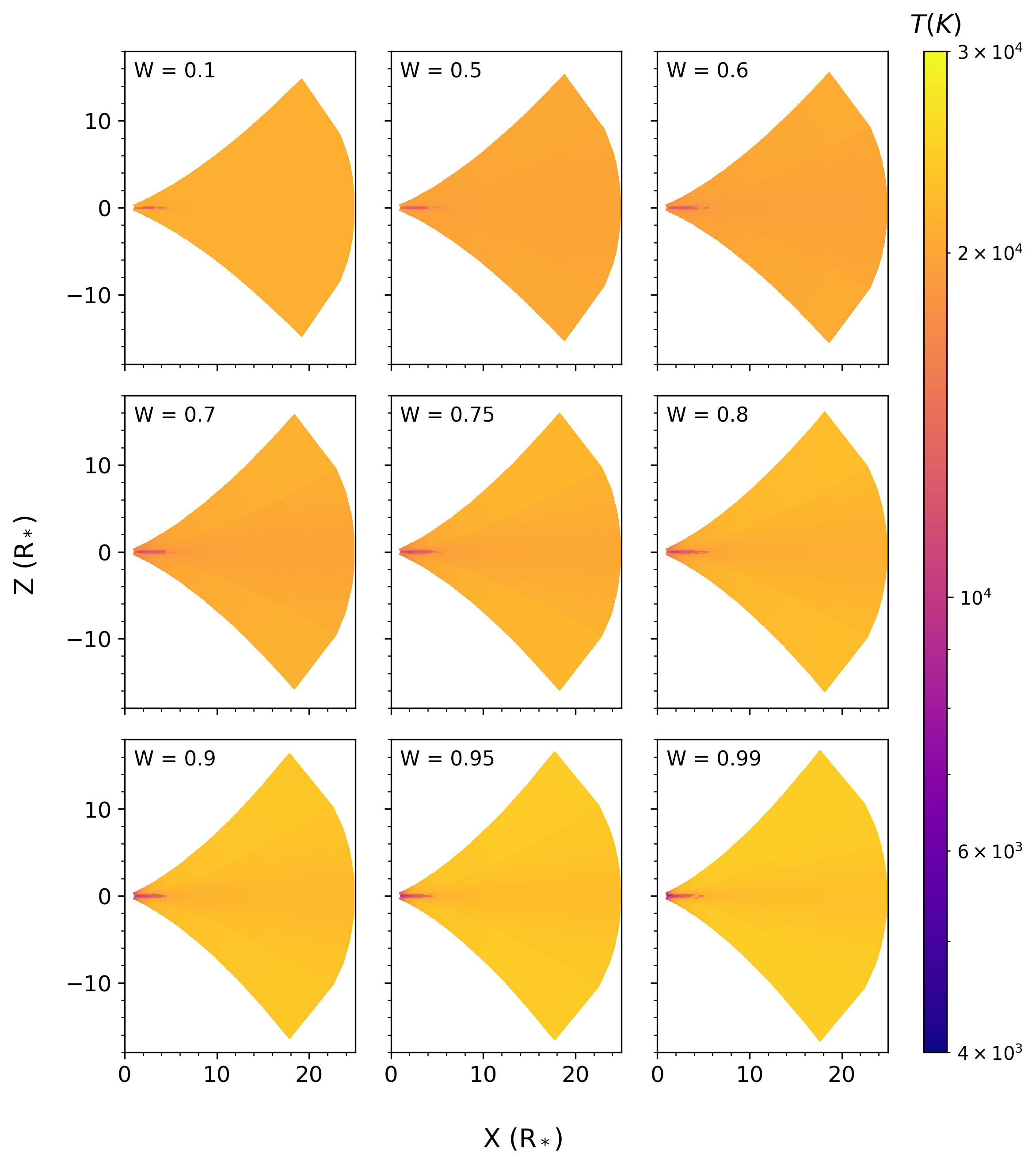}
    \caption{Same as Figure~\ref{fig:B0_highdens_temps}, but for the B0 low density disks.}
    \label{fig:B0_lowdens_temps}
\end{figure}

\begin{figure}
    \centering
    \includegraphics[width=0.49\textwidth]{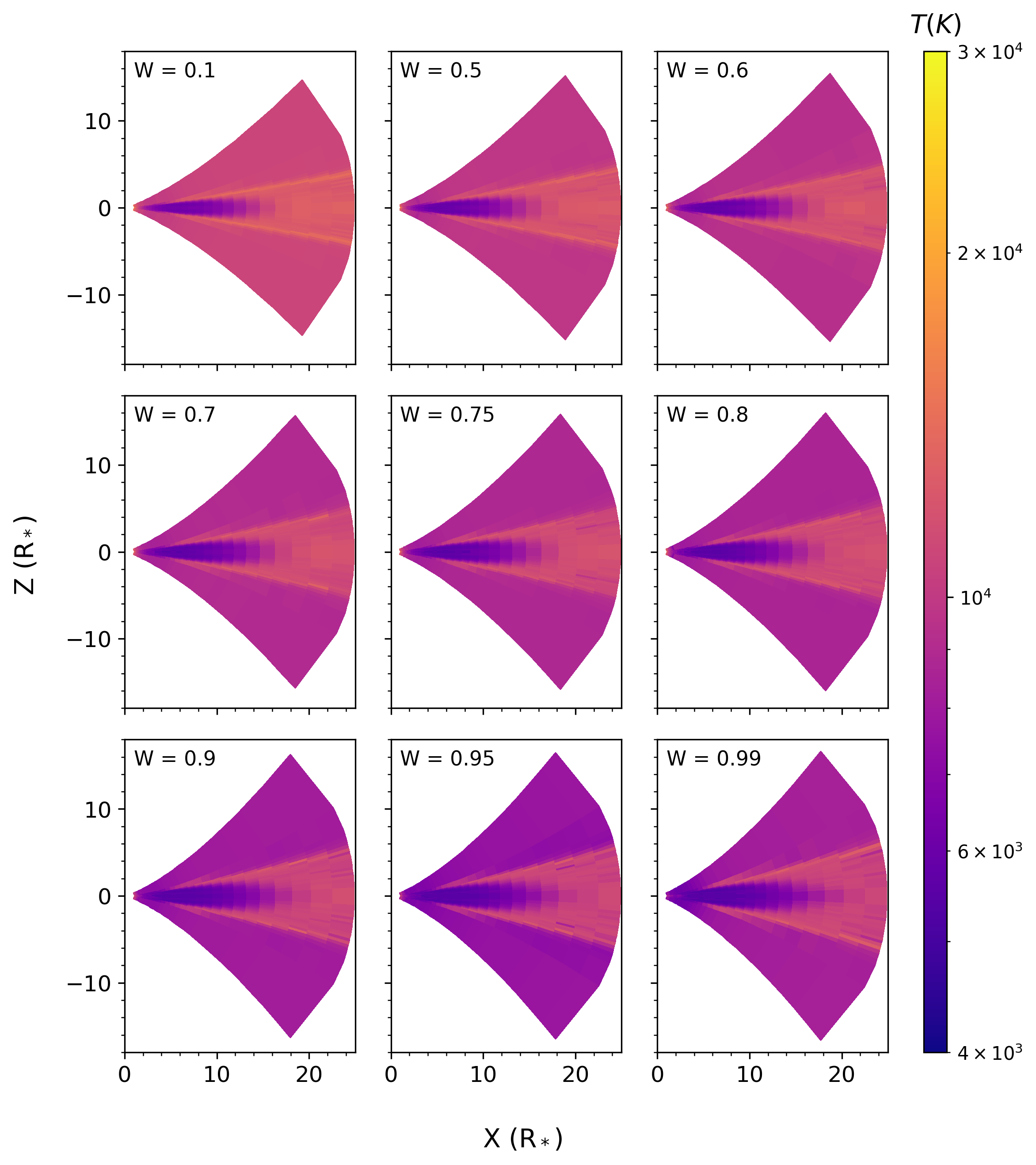}
    \caption{Same as Figure~\ref{fig:B0_highdens_temps}, but for the B2 high density disks.}
    \label{fig:B2_highdens_temps}
\end{figure}

\begin{figure}
    \centering
    \includegraphics[width=0.49\textwidth]{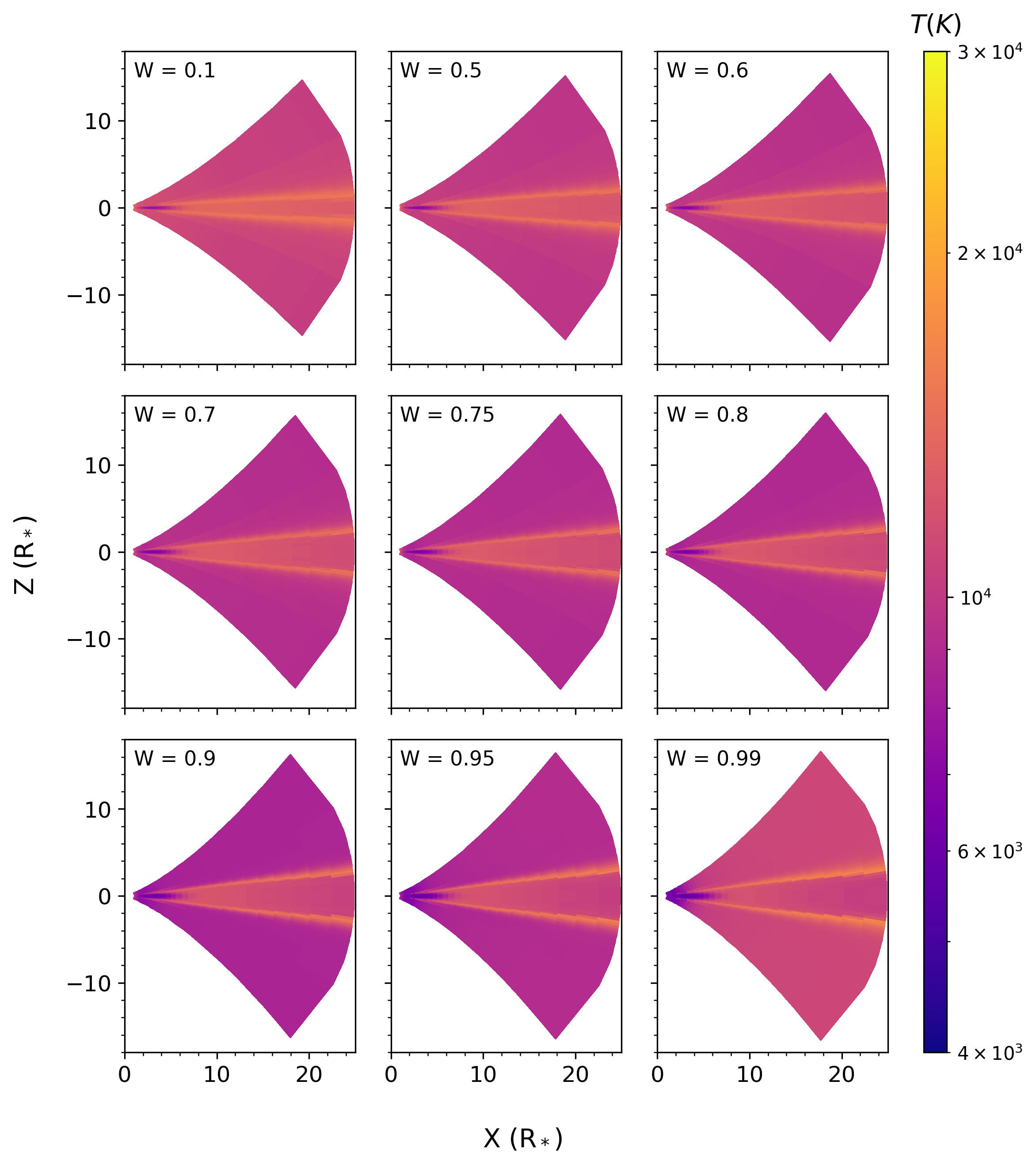}
    \caption{Same as Figure~\ref{fig:B0_highdens_temps}, but for the B2 moderate density disks.}
    \label{fig:B2_moddens_temps}
\end{figure}

\begin{figure}
    \centering
    \includegraphics[width=0.49\textwidth]{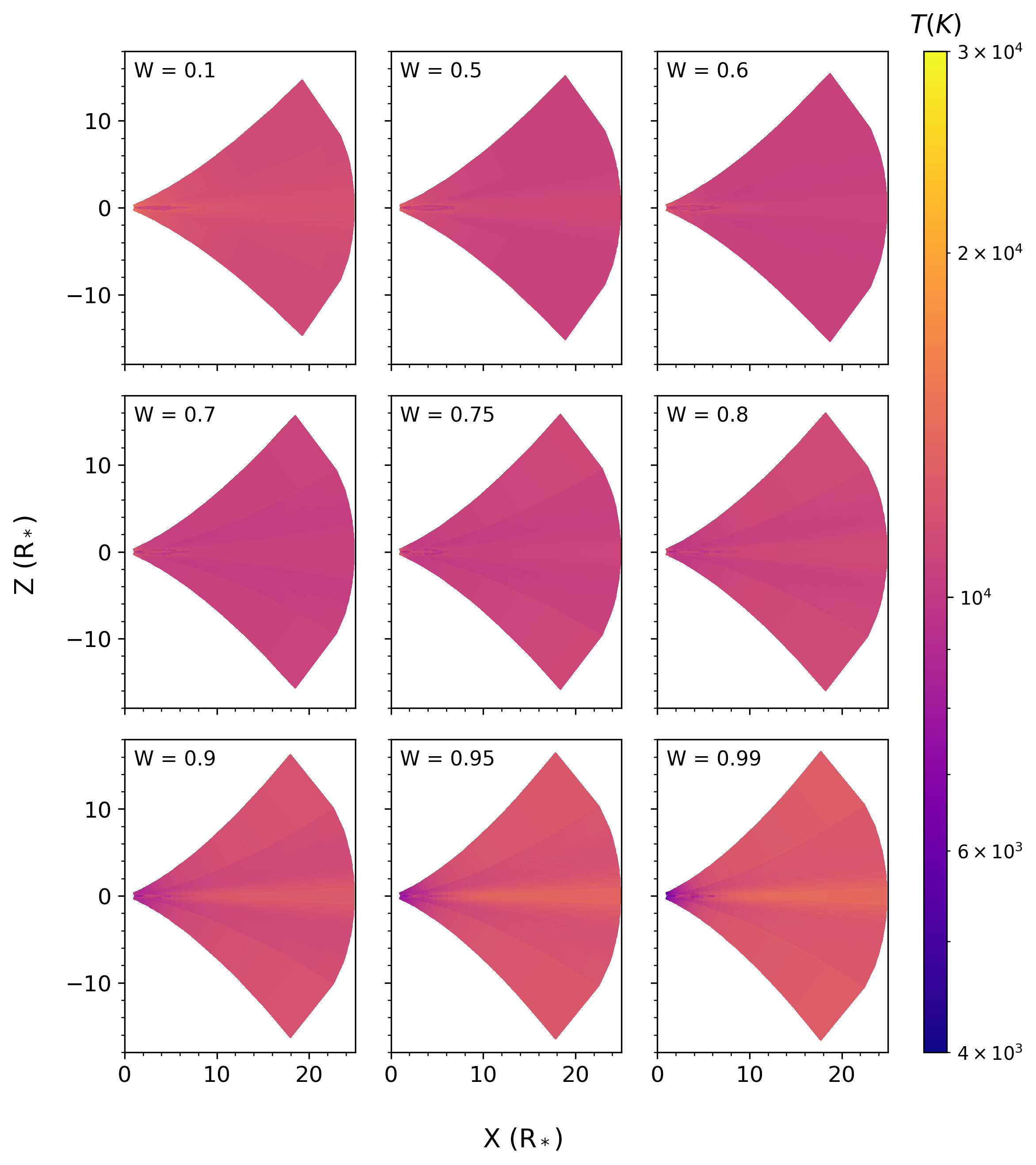}
    \caption{Same as Figure~\ref{fig:B0_highdens_temps}, but for the B2 low density disks.}
    \label{fig:B2_lowdens_temps}
\end{figure}

\begin{figure}
    \centering
    \includegraphics[width=0.49\textwidth]{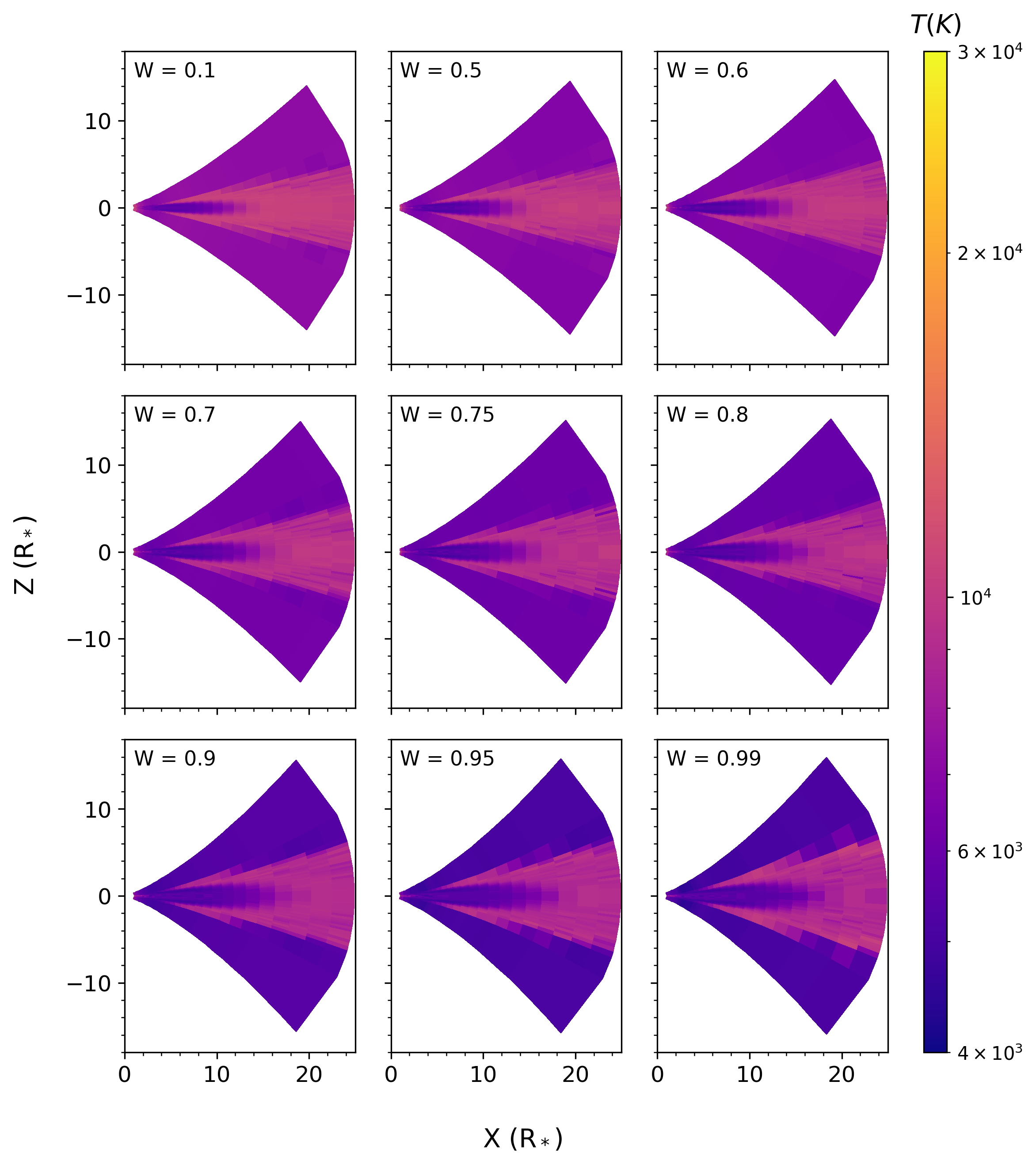}
    \caption{Same as Figure~\ref{fig:B0_highdens_temps}, but for the B5 high density disks.}
    \label{fig:B5_highdens_temps}
\end{figure}

\begin{figure}
    \centering
    \includegraphics[width=0.49\textwidth]{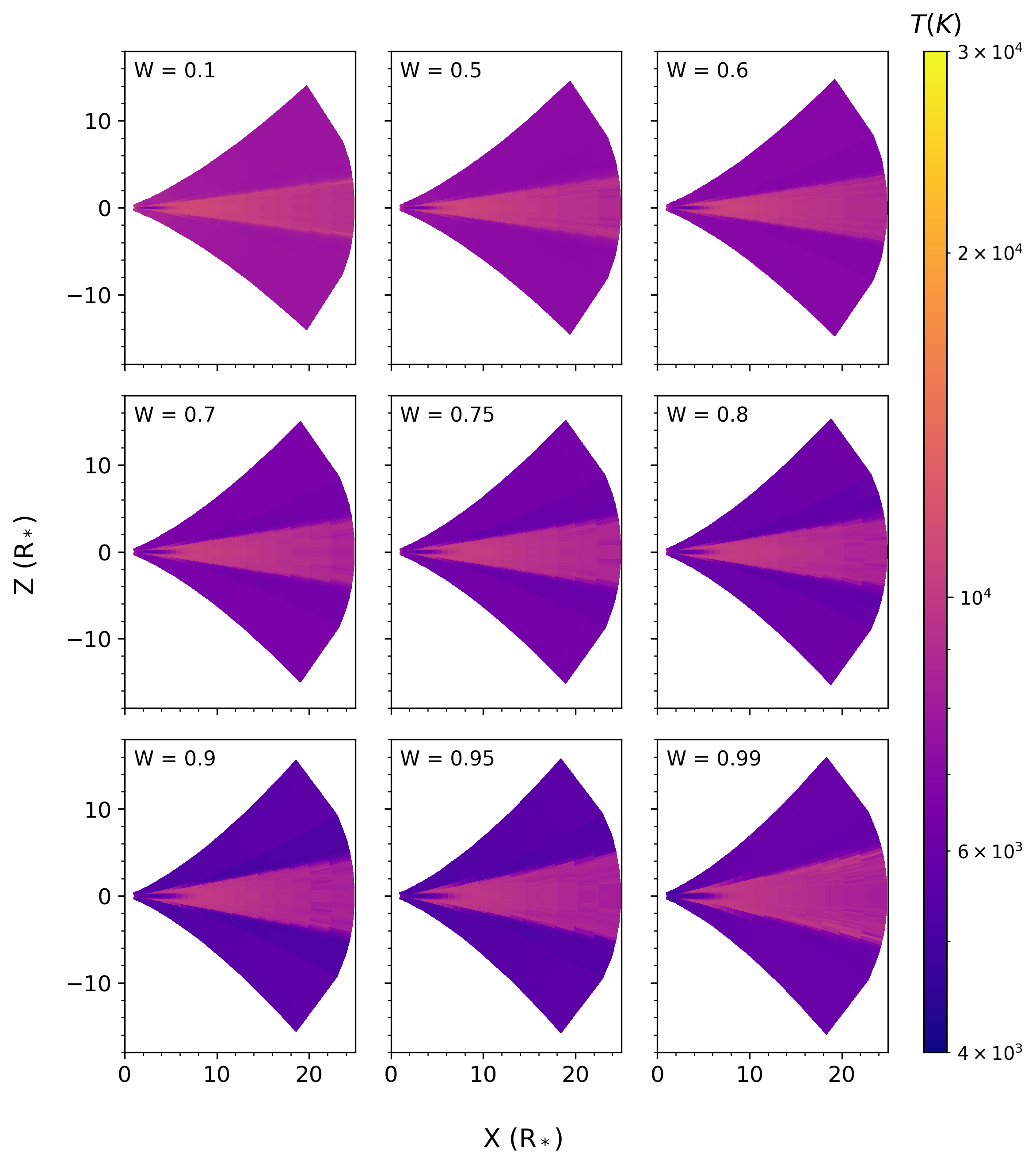}
    \caption{Same as Figure~\ref{fig:B0_highdens_temps}, but for the B5 moderate density disks.}
    \label{fig:B5_moddens_temps}
\end{figure}

\begin{figure}
    \centering
    \includegraphics[width=0.49\textwidth]{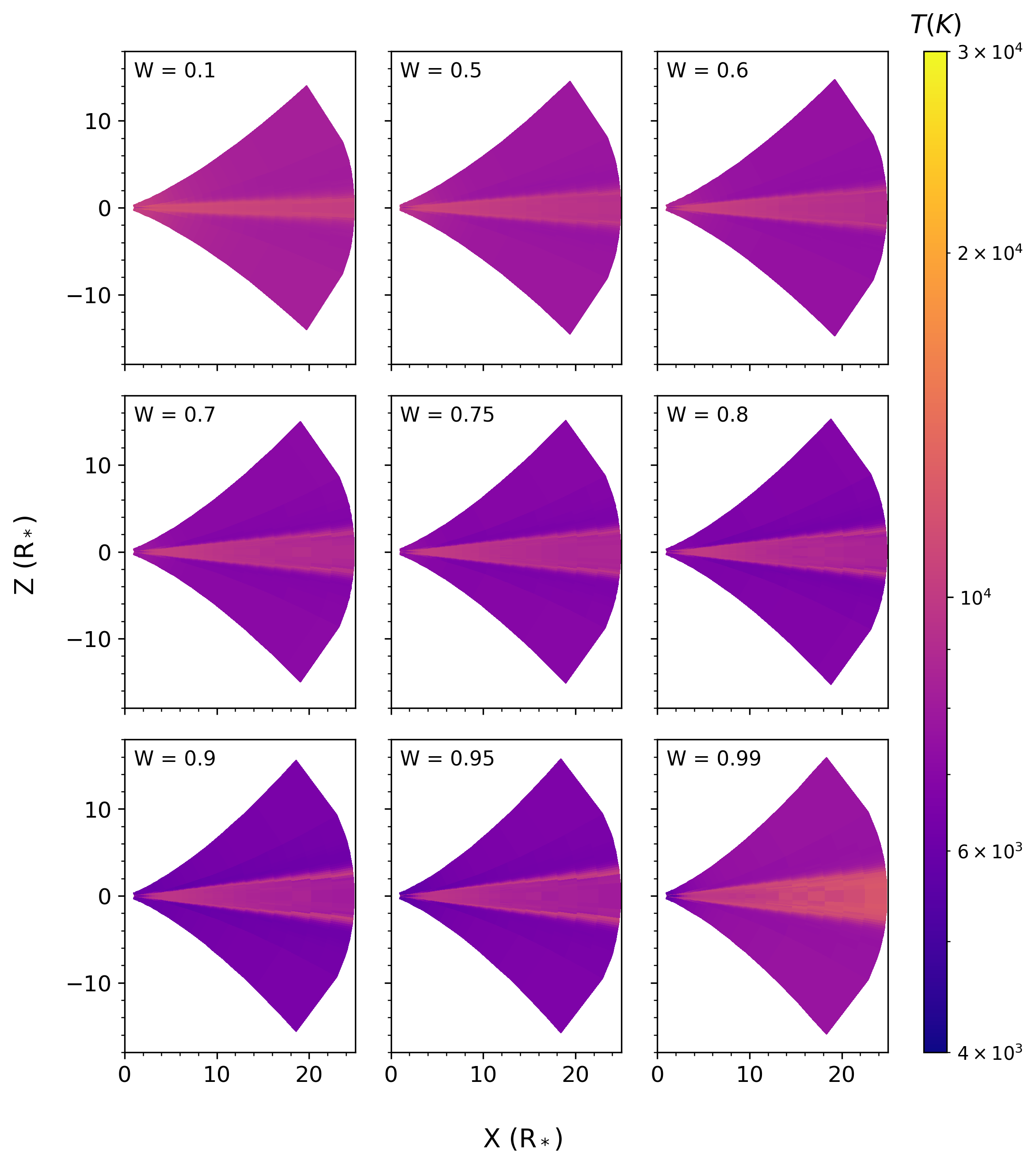}
    \caption{Same as Figure~\ref{fig:B0_highdens_temps}, but for the B5 low density disks.}
    \label{fig:B5_lowdens_temps}
\end{figure}

\begin{figure}
    \centering
    \includegraphics[width=0.49\textwidth]{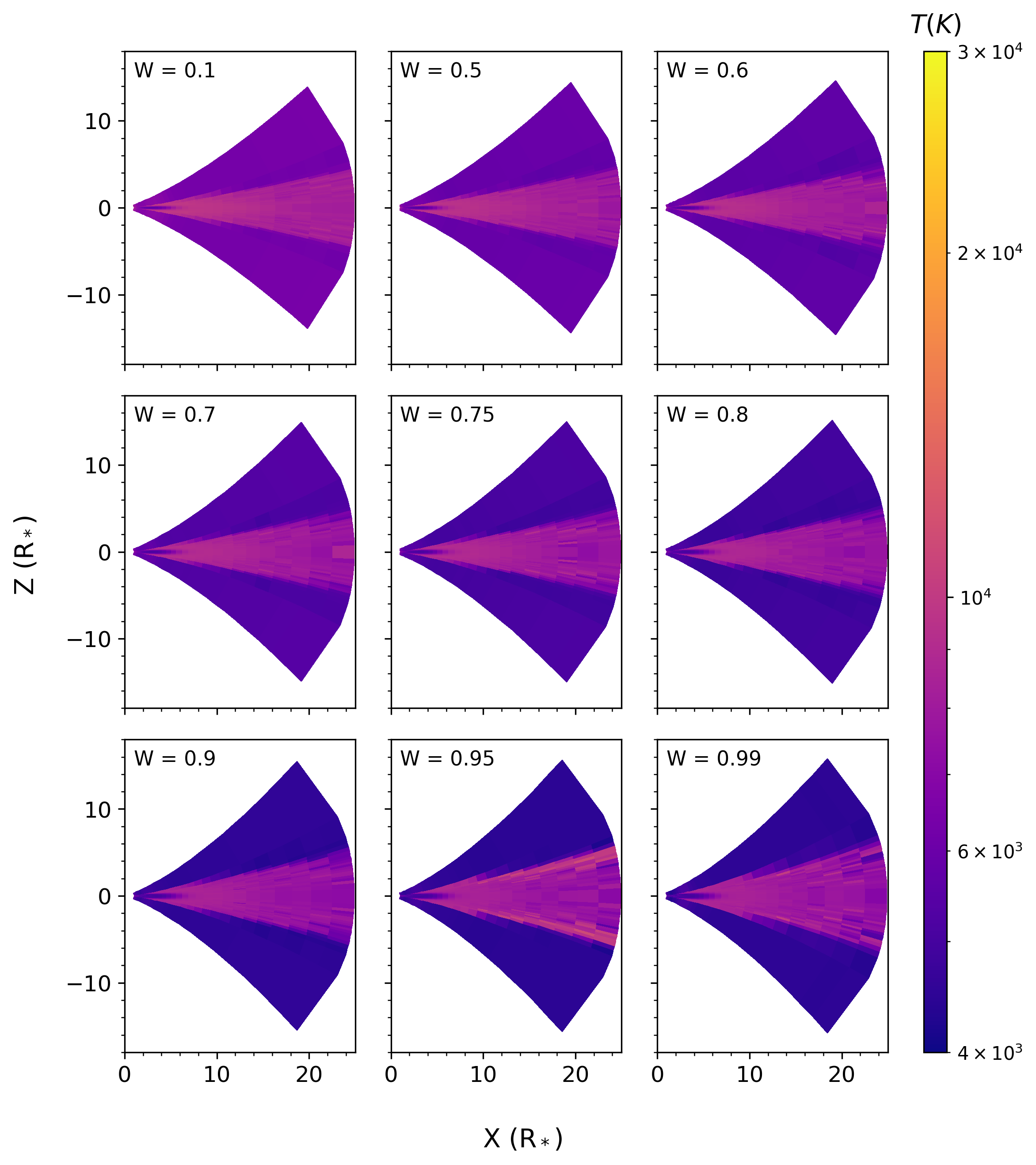}
    \caption{Same as Figure~\ref{fig:B0_highdens_temps}, but for the B8 high density disks.}
    \label{fig:B8_highdens_temps}
\end{figure}

\begin{figure}
    \centering
    \includegraphics[width=0.49\textwidth]{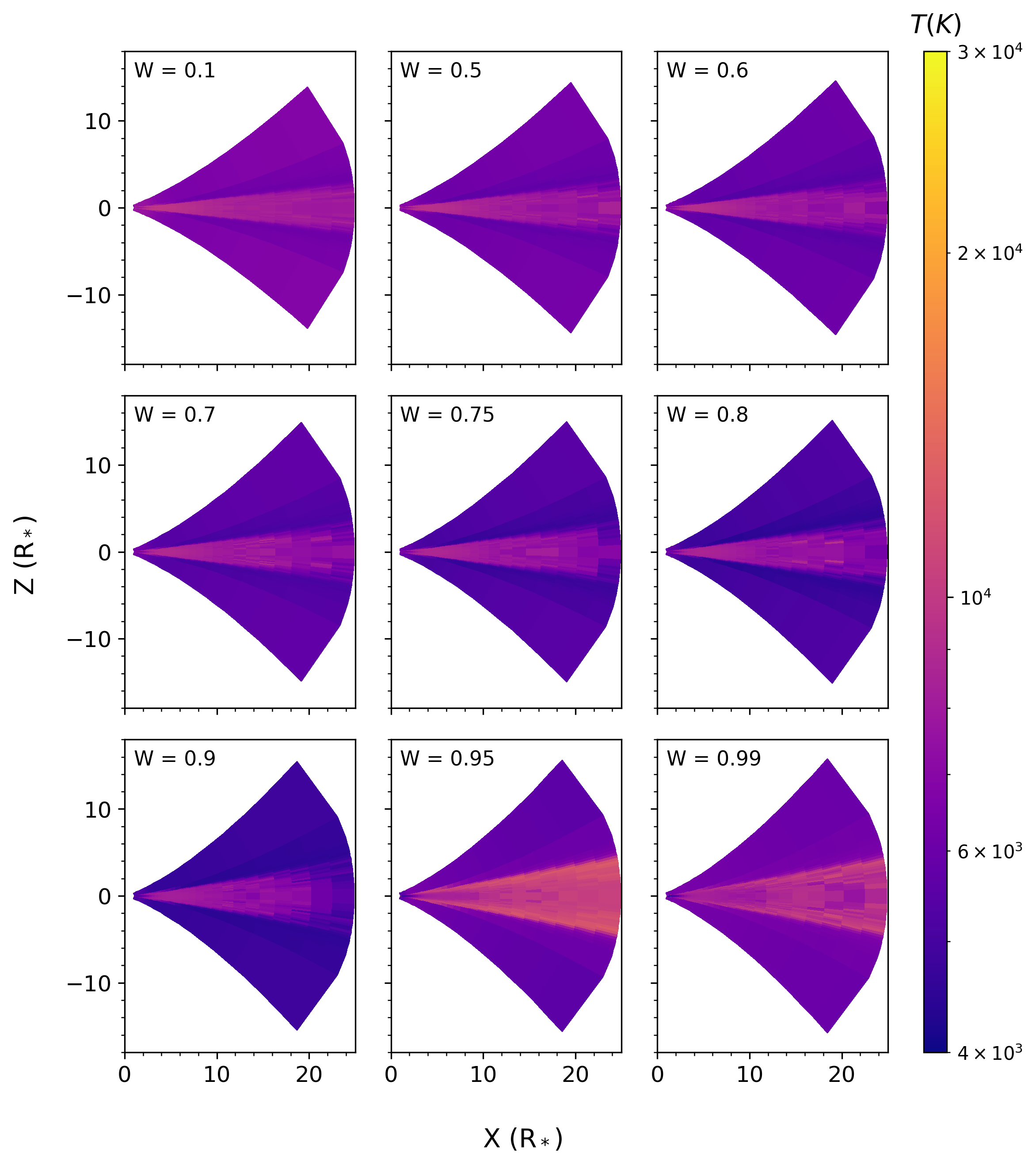}
    \caption{Same as Figure~\ref{fig:B0_highdens_temps}, but for the B8 moderate density disks.}
    \label{fig:B8_moddens_temps}
\end{figure}

\begin{figure}
    \centering
    \includegraphics[width=0.49\textwidth]{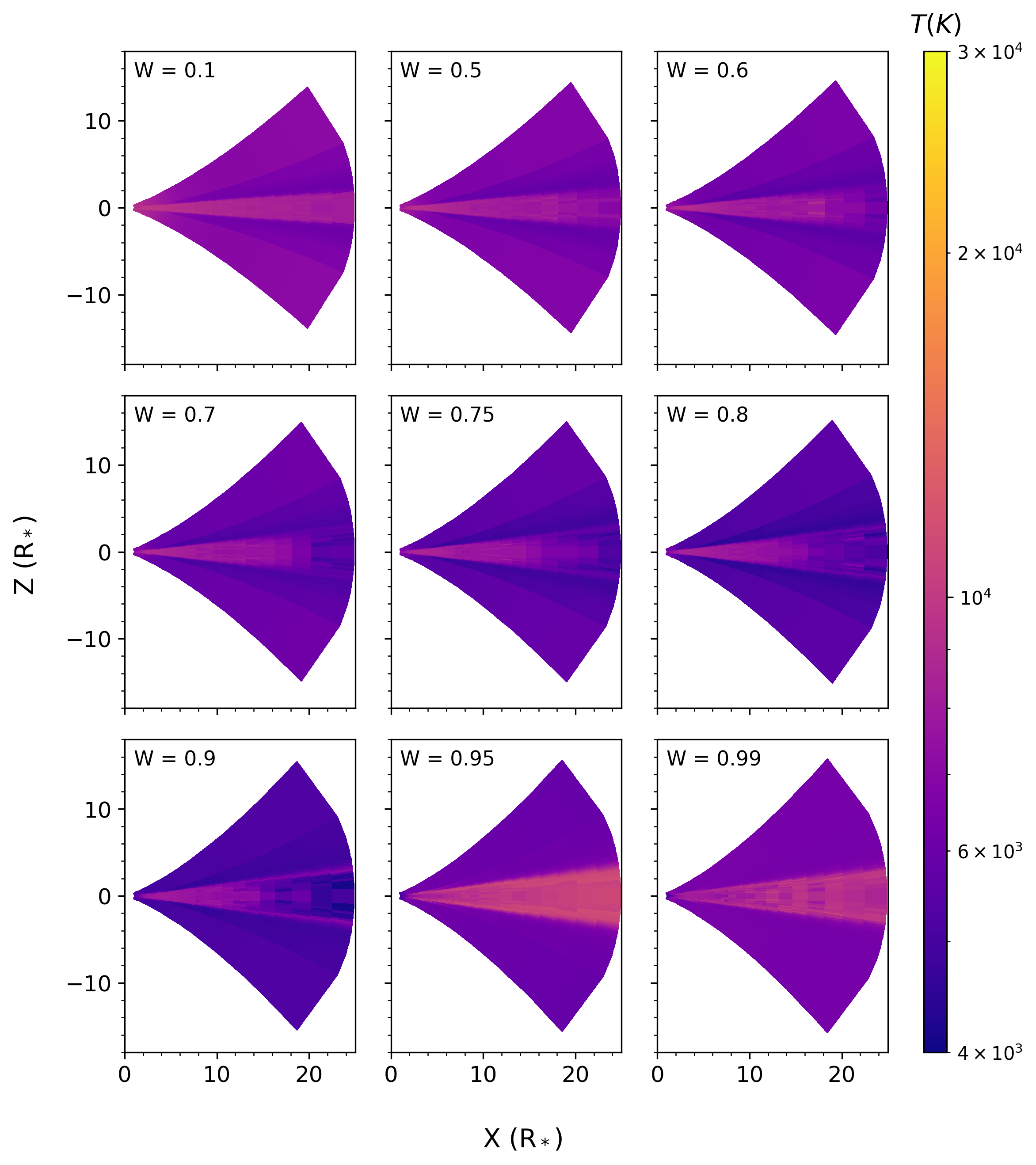}
    \caption{Same as Figure~\ref{fig:B0_highdens_temps}, but for the B8 low density disks.}
    \label{fig:B8_lowdens_temps}
\end{figure}

\end{appendices}

\bibliography{references.bib}

\section*{Declarations}
RGR received funding from the Natural Sciences and Engineering Research Council of Canada (NSERC) Postgraduate Scholarship - Doctoral program. CEJ received support from the NSERC Discovery Research Program. AuD has received research support from NASA through Chandra Award number TM4-25001A issued by the Chandra X-ray Observatory 27 Center, which is operated by the Smithsonian Astrophysical Observatory for and on behalf of NASA under contract NAS8-03060. JLB received funding from the European Union (ERC, MAGNIFY, Project 101126182). ACC acknowledges support from the Conselho Nacional de Desenvolvimento Cient\'ifico e Tecnol\'ogico (CNPq, grant 314545/2023-9) and the S\~ao Paulo Research Foundation (FAPESP, grants 2018/04055-8 and 2019/13354-1). The authors have no relevant financial or non-financial interests to disclose. 

All authors contributed to the study conception and design. RGR led the effort, including running the majority of the simulations and conducting all analysis, in addition to contributing the text related to the Abstract, Methods, Results, Discussion and Conclusions sections. CEJ contributed the first draft of the Introduction, provided feedback on the simulations and their analysis, and offered detailed comments on the manuscript that shaped the structure of the paper. PQ contributed compute time to run a subset of the simulations with the adjusted density parameter. ACC contributed special expertise to the discussion of the polarization, and greatly assisted in the interpretation of the polarization color as a diagnostic for the stellar rotation rate. All authors commented on previous versions of the manuscript, and have read and approved the final manuscript.

Ethics declarations: Not applicable.

\end{document}